\RequirePackage{fix-cm}
\documentclass[smallextended]{svjour3}
\smartqed  
\usepackage{graphicx}
\usepackage{fancybox, ascmac, multirow, hhline, atbegshi, lscape}
\usepackage{authblk}
\usepackage{natbib}
\usepackage{latexsym}
\usepackage{amsmath}
\usepackage{amssymb}
\usepackage[linesnumbered, ruled]{algorithm2e}
\usepackage{bm}
\usepackage{subfig}
\usepackage{url}
\usepackage{color}

\newtheorem{thm}{Theorem}
\newtheorem{lem}{Lemma}

\newtheorem{rem}{Remark}

\newcommand{\argmax}{\mathop{\rm arg~max}\limits}

\begin{document}

\title{Selective Inference via Marginal Screening for High Dimensional Classification}
\author{Yuta Umezu         \and
        Ichiro Takeuchi 
}

\institute{Yuta Umezu \at
              Nagoya Institute of Technology,  Aichi, Japan \\
           \and
           Ichiro Takeuchi \at
              Nagoya Institute of Technology,  Aichi, Japan/
              RIKEN Center for Advanced Intelligence Project, Tokyo, Japan/
              Center for Materials Research by Information Integration, National Institute for Materials Science, Ibaraki, Japan \\
              \email{ichiro.takeuchi@nitech.ac.jp}           
}

\date{Received: date / Accepted: date}

\maketitle

\begin{abstract}
Post-selection inference is a statistical technique for determining salient variables after model or variable selection. Recently, selective inference, a kind of post-selection inference framework, has garnered  the attention in the statistics and machine learning communities. By conditioning on a specific variable selection procedure, selective inference can properly control for so-called selective type I error, which is a type I error conditional on a variable selection procedure, without imposing excessive additional computational costs. While selective inference can provide a valid hypothesis testing procedure, the main focus has hitherto been on Gaussian linear regression models. In this paper, we develop a selective inference framework for binary classification problem. We consider a logistic regression model after variable selection based on marginal screening, and derive the high dimensional statistical behavior of the post-selection estimator. This enables us to asymptotically control for selective type I error for the purposes of hypothesis testing after variable selection. We conduct several simulation studies to confirm the statistical power of the test, and compare our proposed method with data splitting and other methods.

\end{abstract}

\section{Introduction}
\label{sec:intro}
Discovering statistically significant variables in high dimensional data is an important problem for many applications such as bioinformatics, materials informatics, and econometrics, to name a few.
To achieve this, for example in a regression model, data analysts often attempt to reduce the dimensionality of the model by utilizing a particular {\it model selection} or {\it variable selection} method.
For example, the Lasso \citep{Tib96} and marginal screening \citep{FanLv08} are frequently used in model selection contexts.
In many applications, data analysts conduct statistical inference based on the selected model as if it is known a priori, but this practice has been referred to as ``a quiet scandal in the statistical community'' in \cite{Bre92}.
If we select a model based on the available data, then we have to pay heed to the effect of model selection when we conduct a statistical inference.
This is because the selected model is no longer deterministic, i.e., random, and statistical inference after model selection is affected by {\it selection bias}.
In hypothesis testing of the selected variables, the validity of the inference is compromised when a test statistic is constructed without taking account of the model selection effect.
This means that, as a consequence, we can no longer effectively control type I error or the false positive rate.
This kind of problem falls under the banner of {\it post-selection inference} in the statistical community and is recently attracted a lot of attention \citep[see, e.g.,][]{Ber13, Efr14, Bar16, Lee16}.

Post-selection inference consists of the following two steps: 
\begin{description}
\item[Selection:] The analyst chooses a model or subset of variables and constructs hypothesis, based on the data.
\item[Inference:] The analyst tests the hypothesis by using the  selected model.
\end{description}
Broadly speaking, the selection step determines what issue to address, i.e., a hypothesis selected from the data, and the inference step conducts hypothesis testing to enable a conclusion to be drawn about the issue under consideration.
To navigate the issue of selection bias, there are several approaches for conducting the inference step.

{\it Data splitting} is the most common procedure for selection bias correction.
In a high dimensional linear regression model, \cite{WasRoe09} and \cite{Mei09} succeed in assigning a $p$-value for each selected variable by splitting the data into two subsets.
Specifically, they first reduce the dimensionality of the model using the first subset, and then make the final selection using the second subset of the data, by assigning a $p$-value based on a classical least square estimation.
While such a data splitting method is mathematically valid straightforward to implement, it leads to low power for extracting truly significant variables because only sub-samples, whose size is obviously smaller than that of the full sample, can be used in each of the selection and inference steps.

As an alternative, {\it simultaneous inference}, which takes account all possible subsets of variables, has been developed for correcting selection bias.
\cite{Ber13} showed that the type I error can be successfully controlled even if the full sample is used in both the selection and inference steps by adjusting multiplicity of model selection.
Since the number of all possible subsets of variables increases exponentially, computational costs associated with this method become excessive when the dimension of parameters is greater than 20.

On the other hand, {\it selective inference}, which only takes the selected model into account, is another approach for post-selection inference, and provides a new framework for combining selection and hypothesis testing.
Since hypothesis testing is conducted only for the selected model, it makes sense to condition on an event that ``a certain model is selected''.
This event is referred to as a {\it selection event}, and we conduct hypothesis testing conditional on the event.
Thus, we can avoid having to compare coefficients across two different models.
Recently, \cite{Lee16} succeeded in using this method to conduct hypothesis testing through constructing confidence intervals for selected variables by the Lasso in s linear regression modeling context.
When a specific confidence interval is constructed, the corresponding hypothesis testing can be successfully conducted
They also show that the type I error, which is also conditioned on the selection event and is called {\it selective type I error}, can be appropriately controlled.
It is noteworthy that by conditioning on the selection event in a certain class, we can construct exact $p$-values in the meaning of conditional inference based on a truncated normal distribution.

Almost all studies which have followed since the seminal work by \cite{Lee16}, however, focus on linear regression models.
Particularly, normality of the noise is crucial to control selective type I error.
To relax this assumption, \cite{TiaTay15} developed an asymptotic theory for selective inference in a generalized linear modeling context.
Although their results can be available for high dimensional and low sample size data, we can only test a global null hypothesis, that is, a hypothesis that all regression hypothesis is zero, just like with covariance test \citep{Loc14}.
On the other hand, \cite{Tay16} proposed a procedure to test individual hypotheses in a logistic regression model with the Lasso.
By debiasing the Lasso estimator for both the active and inactive variables, they require a joint asymptotic distribution of the debiased Lasso estimator and conduct hypothesis testing for regression coefficients individually.
However, the method is justified only for low dimensional scenarios since they exploit standard fixed dimensional asymptotics.

Our main contribution is that, by utilizing marginal screening as a variable selection method, we can show that the selective type I error rate for logistic regression model is appropriately controlled even in a high dimensional asymptotic scenario.
In addition, our method is applicable not only with respect to testing the global null hypothesis but also hypotheses pertaining to individual regression coefficients.
Specifically, we first utilize marginal screening for the selection step in a similar way to \cite{LeeTay14}.
Then, by considering a logistic regression model for the selected variables, we derive a high dimensional asymptotic property of a maximum likelihood estimator.
Using the asymptotic results, we can conduct selective inference of a high dimensional logistic regression, i.e., valid hypothesis testing for the selected variables from high dimensional data.

The rest of the paper is organized as follows.
Section \ref{sec:SI_and_related} briefly describes the notion of selective inference and intruduces several related works.
In Section \ref{sec:setting}, the model setting and assumptions are described.
An asymptotic property of the maximum likelihood estimator of our model is discussed in Section \ref{sec:propose}.
In Section \ref{sec:simulation}, we conduct several simulation studies to explore the performance of the proposed method before application to real world empirical data sets in Section \ref{sec:real}.
Theorem proofs are relegated to Section \ref{sec:proof}.
Finally, Section \ref{sec:conclusion} offers concluding remarks and suggestions for future research in this domain.

\subsection*{Notation}
Throughout the paper, row and column vectors of $X\in \mathbb{R}^{n\times d}$ are denoted by $\bm{x}_i~(i=1,\ldots, n)$ and $\tilde{\bm{x}}_j,~(j=1,\ldots, d)$, respectively.
An $n\times n$ identity matrix is denoted by $I_n$.
The $\ell_2$-norm of a vector is denoted by $\|\cdot\|$ provided there is no confusion.
For any subset $J\subseteq\{1,\ldots, d\}$, its complement is denoted by $J^\bot=\{1,\ldots,d\}\backslash S$.
We also denote $\bm{v}_J=(v_i)_{i\in J}\in \mathbb{R}^{|J|}$ and $X_J=(\bm{x}_{J,1},\ldots,\bm{x}_{J,n})^\top\in\mathbb{R}^{n\times |J|}$ as a sub-vector of $\bm{v}$ and a sub-matrix of $X$, respectively.
For a differentiable function $f$, we denote $f'$ and $f''$ as the first and second derivatives and so on.

\section{Selective Inference and Related Works}
\label{sec:SI_and_related}
In this section, we overview fundamental notion of selective inference through a simple linear regression model \citep{Lee16}.
We also review related existing works on selective inference.

\subsection{Selective Inference in Linear Regression Model}
\label{subsec:SI_in_LR}
Let $\bm{y}\in\mathbb{R}^n$ and $X\in\mathbb{R}^{n\times d}$ be a response and non-random regressor, respectively, and let us consider a linear regression model 
\begin{align*}
\bm{y}=X\bm{\beta}^*+\bm{\varepsilon},
\end{align*}
where $\bm{\beta}^*$ is the true regression coefficient vector and $\bm{\varepsilon}$ is distributed according to ${\rm N}(\bm{0},\sigma^2 I_n)$ with known variance $\sigma^2$.
Suppose that a subset of variables $S$ is selected in the selection step (e.g., Lasso or marginal screening as in \citet{Lee16, LeeTay14}) and let us consider hypothesis testing for $j\in \{1,\ldots, |S|\}$:
\begin{align}
\text{H}_{0,j}:\beta_{S, j}^*=0
\qquad
\text{vs.}
\qquad
\text{H}_{1,j}:\beta_{S, j}^*\neq 0.
\label{eq;stest}
\end{align}
If $S$ is non-random, a maximum likelihood estimator $\hat{\bm{\beta}}_S=(X_S^\top X_S)^{-1}X_S^\top\bm{y}$ is distributed according to ${\rm N}(\bm{\beta}_S^*,\sigma^2(X_S^\top X_S)^{-1})$, as is well-known.
However, we cannot use this sampling distribution when $S$ is selected based on the data, and the selected variable $S$ is also random.

If a subset of variables, i.e., the active set, $\hat{S}$ is selected by the Lasso or marginal screening, the event $\{\hat{S}=S\}$ can be written as an affine set with respect to $\bm{y}$, that is, in the form of $\{\bm{y};A\bm{y}\leq \bm{b}\}$ for some non-random matrix $A$ and vector $\bm{b}$ \citep{Lee16, LeeTay14}, in which the event $\{\hat{S}=S\}$ is called a {\it selection event}.
\cite{Lee16} showed that if $\bm{y}$ follows a normal distribution and the selection event can be written as an affine set, the following lemma holds:

\begin{lem}[Polyhedral Lemma; \cite{Lee16}]
\label{lem1}
Suppose $\bm{y}\sim{\rm N}(\bm{\mu},\Sigma)$.
Let $\bm{c}=\Sigma\bm{\eta}(\bm{\eta}^\top\Sigma\bm{\eta})^{-1}$ for any $\bm{\eta}\in\mathbb{R}^n$, and let $\bm{z}=(I_n-\bm{c}\bm{\eta}^\top)\bm{y}$.
Then we have
\begin{align*}
\{\bm{y};A\bm{y}\leq \bm{b}\}=\{\bm{y}; L(\bm{z})\leq \bm{\eta}^\top\bm{y}\leq U(\bm{z}),\;N(\bm{z})\geq 0\},
\end{align*}
where
\begin{align*}
L(\bm{z})
=\max_{j:(A\bm{c})_j<0}\frac{b_j-(A\bm{z})_j}{(A\bm{c})_j},~~~~~
U(\bm{z})
=\min_{j:(A\bm{c})_j>0}\frac{b_j-(A\bm{z})_j}{(A\bm{c})_j}
\end{align*}
and $N(\bm{z})=\max_{j:(A\bm{c})_j=0}b_j-(A\bm{z})_j$.
In addition, $(L(\bm{z}),U(\bm{z}),N(\bm{z}))$ is independent of $\bm{\eta}^\top\bm{y}$.
\end{lem}
\noindent
By using the lemma, we can find that the distribution of the pivotal quantity for $\bm{\eta}^\top\bm{\mu}$ is given by a truncated normal distribution.
Specifically, let $F^{[L,U]}_{\mu,\sigma^2}$ be a cumulative distribution function of a truncated normal distribution ${\rm TN}(\mu, \sigma^2, L, U)$, that is,
\begin{align*}
F^{[L,U]}_{\mu,\sigma^2}(x)
=\frac{\Phi((x-\mu)/\sigma)-\Phi((L-\mu)/\sigma)}{\Phi((U-\mu)/\sigma)-\Phi((L-\mu)/\sigma)},
\end{align*}
where $\Phi$ is a cumulative distribution function of a standard normal distribution.
Then, for any value of $\bm{z}$, we have
\begin{align*}
\left[
F^{[L(\bm{z}),U(\bm{z})]}_{\bm{\eta}^\top\bm{\mu},\bm{\eta}^\top\Sigma\bm{\eta}}(\bm{\eta}^\top\bm{y})
\mid A\bm{y}\leq \bm{b}
\right]
 \sim{\rm Unif}(0,1),
\end{align*}
where $L(\bm{z})$ and $U(\bm{z})$ are defined in the above lemma.
This pivotal quantity allows us to construct a so-called {\it selective $p$-value}.
Precisely, by choosing $\bm{\eta}=X_S(X_S^\top X_S)^{-1}\bm{e}_j$, we can construct a right-side selective $p$-value as
\begin{align*}
P_j
=1-F^{[L(\bm{z}_0), U(\bm{z}_0)]}_{0,\bm{\eta}^\top\Sigma\bm{\eta}}(\bm{\eta}^\top\bm{y}),
\end{align*}
where $\bm{e}_j\in\mathbb{R}^{|S|}$ is a unit vector whose $j$-th element is 1 and 0 otherwise, and $\bm{z}_0$ is a realization of $\bm{z}$.
Note that the value of $P_j$ represents a right-side $p$-value conditional on the selection event under the null hypothesis $\text{H}_{0,j}:\beta_{S,j}^*=\bm{\eta}^\top \bm{\mu}=0$ in (\ref{eq;stest}).
In addition, for the $j$-th test in (\ref{eq;stest}), a two-sided selective $p$-value can be defined as 
\begin{align*}
\tilde{P}_j=2\min\{P_j, 1-P_j \},
\end{align*}
which also follows from standard uniform distribution under the null hypothesis.
Therefore, we reject the $j$-th null hypothesis at level $\alpha$ when $\tilde{P}_j\leq \alpha$, and the probability
\begin{align}
\text{P}(\text{H}_{0,j}~\text{is falsely rejected}\mid \hat{S}=S)
=\text{P}(\tilde{P}_j\leq \alpha\mid \hat{S}=S)
\label{eq:sFPR}
\end{align}
is referred to as a {\it selective type I error}.

\subsection{Related Works}
In selective inference, we use the same data in variable selection and statistical inference.
Therefore, the selected model is not deterministic and we can not apply classical hypothesis testing due to selection bias.

To navigate this problem, {\it data splitting} has been commonly utilized.
In data splitting, the data are randomly divided into two disjoint sets, and one of them is used for variable selection and the other is used for hypothesis testing.
This is a particularly versatile method and is widely applicable if we can divide the data randomly \citep[see e.g.,][]{Cox75, WasRoe09, Mei09}.
Since the data are split randomly, i.e., independent of the data, we can conduct hypothesis testing in the inference step independent of the selection step.
Thus, we do not need to concerned with selection bias. 
It is noteworthy that data splitting can be viewed as a method of selective inference because the inference is conducted only for the selected variables in the selection step.
However, a drawback of data splitting is that only a part of the data are available for each split, precisely because the essence of this approach involves rendering some data available for the selection step and the remainder for the inference step.
Because only a subset of the data can be used in variable selection, the risk of failing to select truly important variables increases.
Similarly, the power of hypothesis testing would decrease since inference proceeds on the basis of a subset of the total data.
In addition, since data splitting is executed at random, it is possible and plausible that the final results and conclusions will vary non-trivially depending on exactly how this split is manifested.

On the other hand, in the traditional statistical community, {\it simultaneous inference} has been developed for correcting selection bias \citep[see e.g.,][]{Ber13, Dic14}.
In simultaneous inference, type I error is controlled at level $\alpha$ by considering all possible subsets of variables.
Specifically, let $\hat{S}\subseteq \{1,\ldots, d\}$ be the set of variables selected by a certain variable selection method and $P_j(\hat{S})$ be a $p$-value for the $j$-th selected variable in $\hat{S}$.
Then, in simultaneous inference, the following type I error should be adequately controlled:
\begin{align}
{\rm P}(P_j(\hat{S})\leq \alpha ~\text{for any}~\hat{S}\subseteq\{1,\ldots,d\})\leq \alpha.
\label{eq:FPR}
\end{align}
To examine the relationship between selective inference and simultaneous inference, note that the left-hand side in (\ref{eq:FPR}) can be rewritten as
\begin{align*}
&{\rm P}(P_j(\hat{S})\leq \alpha ~\text{for any}~\hat{S}\subseteq\{1,\ldots,d\}) \\
&=\sum_{S\subseteq\{1,\ldots,d\}}{\rm P}(P_j(S)\leq \alpha \mid \hat{S}=S){\rm P}(\hat{S}=S).
\end{align*}
The right-hand side in the above equality is simply a weighted sum of selective type I errors over all possible subsets of variables.
Therefore, if we control selected type I errors for all possible subsets of variables, we can also control type I errors in the sense of simultaneous inference.
However, because the number of all possible subsets of variables is $2^d$, it becomes overly cumbersome to compute the left-hand side in (\ref{eq:FPR}) even for $d=20$.
In contrast to simultaneous inference, selective inference only considers the selected variables, and thus the computational cost is low compared to simultaneous inference.
 
Following the seminal work of \cite{Lee16}, selective inference for variable selection has been intensively studied \citep[e.g.,][]{FitSunTay14, LeeTay14, taylor2016inference, tian2018selective}.
All these methods, however, rely on the assumption of normality of the data.

\subsection{Beyond Normality}
\label{subsec:non-normality}
It is important to relax the assumption of the normality for applying selective inference to more general cases such as generalized linear models.
To the best of our knowledge, there is death of research into selective inference in such a generalized setting.
Here, we discuss the few studies which do exist in this respect.

\cite{FitSunTay14} derived an exact post-selection inference for a natural parameter of exponential family, and obtained the uniformly most powerful unbiased test in the framework of selective inference.
However, as suggested in their paper, the difficulty in constructing exact inference in generalized linear models emanates from the discreteness of the response distribution.

Focusing on an asymptotic behavior in a generalized linear model context with the Lasso penalty, \cite{TiaTay15} directly considered the asymptotic property of a pivotal quantity.
Although their work can be applied in high dimensional scenarios, we can only test a global null, that is, ${\rm H}_0:\bm{\beta}^*=\bm{0}$, except for the linear regression model case.
This is because that, when we conduct selective inference for individual coefficient, the selection event does not form a simple structure such as an affine set.

On the other hand, \cite{Tay16} proposed a procedure to test individual hypotheses fin logistic regression model context based on the Lasso.
Their approach is fundamentally based on solving the Lasso by approximating the log-likelihood up to the second order, and on debiasing the Lasso estimator.
Because the objective function now becomes quadratic as per the linear regression model, the selection event reduces to a relatively simple affine set.
After debiasing the Lasso estimator, they derive an asymptotic joint distribution of active and inactive estimators.
However, since they required $d$ dimensional asymptotics, high dimensional scenarios can not be supported in their theory.

In this paper, we extend selective inference for logistic regression in \cite{Tay16} to high dimensional settings in the case where variable selection is conducted by marginal screening. 
We do not consider asymptotics for a $d$ dimensional original parameter space, but for a $K$ dimensional selected parameter space. 
Unfortunately, however, we cannot apply this asymptotic result directly to the polyhedral lemma (Lemma \ref{lem1}) in \cite{Lee16}. 
To tackle this problem, we consider a score function for constructing a test statistic for our selective inference framework.
We first define a function $\bm{T}_n(\bm{\beta}_{S}^*)$ based on a score function as a ``source'' for constructing a test statistic.
To apply the polyhedral lemma to $\bm{T}_n(\bm{\beta}_{S}^*)$, we need to asymptotically ensure that 
i) the selection event is represented by affine constraints with respect to $\bm{T}_n(\bm{\beta}_{S}^*)$, and 
ii) the function in the form of $\bm \eta^\top \bm{T}_n(\bm{\beta}_{S}^*)$ is independent of the truncation points. 
Our main technical contribution herein is that, by carefully analyzing problem configuration and by introducing reasonable additional assumptions, we can show that those two requirements for the polyhedral lemma are satisfied asymptotically.

\begin{figure*}[t]
\begin{center}
\includegraphics[scale=0.45, bb=0 0 360 360]{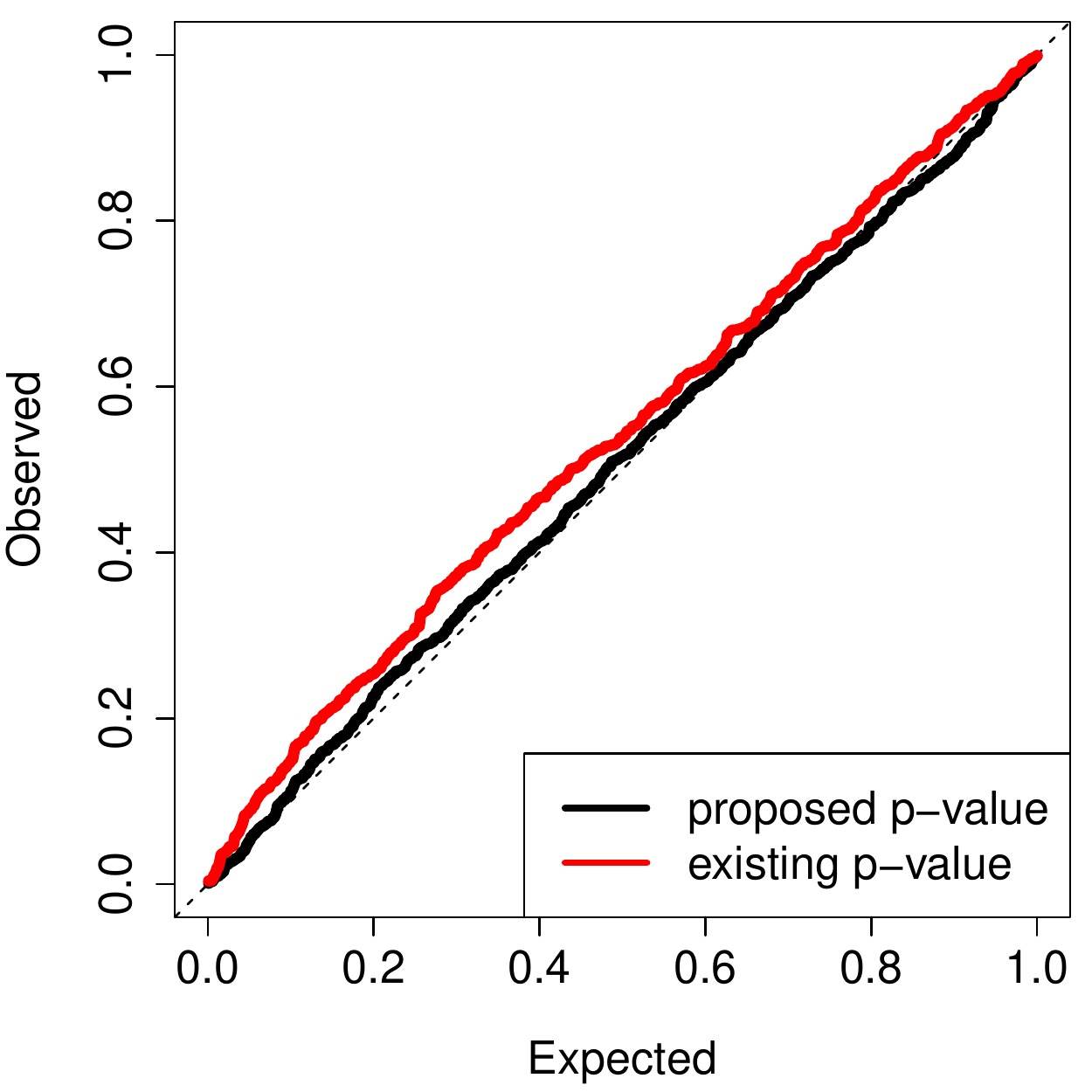}
\end{center}
\caption{Comparison between empirical distributions of selective $p$-values in (\ref{eq;pval}) (black solid line) and \cite{Tay16} (red solid line).
The dashed line shows the cumulative distribution function of the standard uniform distribution.
Data were simulated for $n=50$ and $d=3{,}000$ under the global null and $x_{ij}$ was independently generated from a normal distribution $\text{N}(0, 1)$.
Our proposed method appears to offer superior approximation accuracy compared to the extant alternative.
}
\label{fig:selective_p}
\end{figure*}

Figure \ref{fig:selective_p} shows the asymptotic distribution of selective $p$-values in our setting and in \cite{Tay16} based on 1,000 Monte-Carlo simulation.
While the theory in \cite{Tay16} does not support high dimensionality, their selective $p$-value (red solid line) appears to effective in high dimensional scenarios, although it is slightly mode conservative compared to the approach developed in this paper (black solid line).
Our high dimensional framework means that the number of selected variables grows with the sample size in an appropriate order, and a proposed method allows us to test (\ref{eq;stest}) individually even in high dimensional contexts.

\section{Setting and Assumptions}
\label{sec:setting}
As already noted, our objective herein is to develop a selective inference approach applicable to logistic regression models when the variables are selected by marginal screening.
Let $(y_i,\bm{x}_i)$ be the $i$-th pair of the response and regressor.
We assume that the $y_i$'s are independent random variables which take values in $\{0,1\}$, and the $\bm{x}_i$'s are a $d$ dimensional vector of known constants.
Further, let $X=(\bm{x}_1,\ldots,\bm{x}_n)^\top\in\mathbb{R}^{n\times d}$ and $\bm{y}=(y_1,\ldots,y_n)^\top \in\{0,1\}^n$.
Unlike \cite{Tay16}, we do not require that the dimension $d$ be fixed, that is, $d$ may increase, as well as the sample size $n$.

\subsection{Marginal Screening and Selection Event}
\label{subsec:MS}

In this study, we simply select variables based on a score between the regressor and response $\bm{z}=X^\top\bm{y}$ as per a linear regression problem.
Specifically, we select the top $K$ coordinates of absolute values in $\bm{z}$, that is, 
\begin{align*}
\hat{S}=\{j;|z_j|~ \text{is among the first}~ K~ \text{largest of all}\}.
\end{align*}
To avoid computational issues, we consider the event $\{(\hat{S}, s_{\hat{S}})=(S, s_S)\}$ as a selection event (see, e.g., \cite{LeeTay14, TiaTay15, Lee16}).
Here, $\bm{s}_S$ is a vector of sign $z_j~(j\in S)$.
Then, the selection event $\{(\hat{S}, \bm{s}_{\hat{S}})=(S, \bm{s}_S)\}$ can be rewritten as
\begin{align*}
|z_j|\geq |z_k|,
\qquad
\forall (j,k)\in S\times S^\bot,
\end{align*}
which is equivalent to
\begin{align*}
-s_jz_j\leq z_k\leq s_jz_j,
\quad
s_j z_j\geq 0,
\qquad
\forall(j,k)\in S\times S^\bot.
\end{align*}
Therefore, $\{(\hat{S}, s_{\hat{S}})=(S, s_S)\}$ is reduced to an affine set $\{\bm{z};\;A\bm{z}\leq \bm{0}\}$ for an appropriate $\{2K(d-K)+K\}\times d$ dimensional matrix $A$.

In the following, we assume that a sure screening property holds.
This is desirable property for variable selection \citep[see e.g.,][]{FanLv08, FanSon10} and the statement is as follows:
\begin{description}
\item[(C0)] For the true active set $S^*=\{j;\beta_j^*\neq 0\}$, the probability ${\rm P}(\hat{S}\supset S^*)$ converges to 1 as $n$ goes to infinity.
\end{description}
In the above assumption, we denote $\bm{\beta}^*\in\mathbb{R}^d$ as a true value of the coefficient vector.
This assumption requires that the set of selected variables contain the set of true active variables with probability tending to 1.
In the linear regression model, (C0) holds under some regularity conditions in high dimensional settings \cite[see, e.g.,][]{FanLv08}.
The sufficient condition concerning about high dimensionality for (C0) is $\log d=O(n^\xi)$ for some $\xi\in(0,1/2)$, and thus we allow $d$ to be exponentially large.
Because (C0) is not directly related in selective inference, we do not further discuss it.

\subsection{Selective Test}
For a subset of variables $\hat{S}~(=S)$ selected by marginal screening, we consider $K$ selective tests (\ref{eq;stest}) for each variable $\beta_j^*,~j\in S$.
Let us define the loss function of logistic regression with the selected variables as follows:
\begin{align}
\ell_n(\bm{\beta}_S)
=\sum_{i=1}^n\{y_i\bm{x}_{S,i}^\top\bm{\beta}_S-\psi(\bm{x}_{S,i}^\top\bm{\beta}_S)\},
\label{eq;loss}
\end{align}
where $\psi(\bm{x}_{S,i}^\top\bm{\beta}_S)=\log(1+\exp(\bm{x}_{S,i}^\top\bm{\beta}_S))$ is a cumulant generating function.
Observe that $\ell_n(\bm{\beta}_S)$ is concave with respect to $\bm{\beta}_S$.
Thus we can define the maximum likelihood estimator of $\bm{\beta}_S$ as the optimal solution that attains the maximum of the following optimization problem:
\begin{align}
\hat{\bm{\beta}}_S
=\argmax_{\bm{\beta}_S\in{\cal B}}\ell_n(\bm{\beta}_S),
\label{eq;estimator}
\end{align}
where ${\cal B}\subseteq\mathbb{R}^K$ is a parameter space.
\begin{rem}
Suppose that $S~(\supset S^*)$ is fixed.
Then, it holds that
\begin{align*}
\psi'(\bm{x}_{S,i}^\top\bm{\beta}_S^*)=\psi'(\bm{x}_{S^*,i}^\top\bm{\beta}_{S^*}^*),~~~
\psi''(\bm{x}_{S,i}^\top\bm{\beta}_S^*)=\psi''(\bm{x}_{S^*,i}^\top\bm{\beta}_{S^*}^*),
\end{align*}
and thus, we have
\begin{align*}
{\rm P}(y_i=1)={\rm E}[y_i]=\psi'(\bm{x}_{S^*,i}^\top\bm{\beta}_{S^*}^*),~~~
{\rm V}[y_i]=\psi''(\bm{x}_{S^*,i}^\top\bm{\beta}_{S^*}^*).
\end{align*}
\end{rem}

We construct test statistics for (\ref{eq;stest}) by deriving an asymptotic distribution of $\hat{\bm{\beta}}_S$.
To develop our asymptotic theory, we further assume the following conditions in addition to (C0) for a fixed $S$ with $|S|=K$:
\begin{description}
\item[(C1)]
$\max_{i}\|\bm{x}_{S,i}\|={\rm O}(\sqrt{K})$.
In addition, for a $K\times K$ dimensional matrix
\begin{align*}
\Xi_{S,n}=\frac{1}{n}X_S^\top X_S=\frac{1}{n}\sum_{i=1}^{n}\bm{x}_{S,i}\bm{x}_{S,i}^\top \in\mathbb{R}^{K\times K},
\end{align*}
the following holds:
\begin{align*}
0<C_1<\lambda_{\rm min}(\Xi_{S,n})\leq \lambda_{\rm max}(\Xi_{S,n})<C_2<\infty,
\end{align*}
where $C_1$ and $C_2$ are constants that depend on neither $n$ nor $K$.
\item[(C2)]
There exists a constant $\xi\;(<\infty)$ such that $\max_i|\bm{x}_{S,i}^\top\bm{\beta}_S^*|<\xi$.
In addition, parameter space ${\cal B}$ is 
\begin{align*}
{\cal B}=\{\bm{\beta}_S\in\mathbb{R}^{K};\max_i|\bm{x}_{S,i}^\top\bm{\beta}_S|<\tilde{\xi}\}
\end{align*}
for some constant $\tilde{\xi}\;(\in(\xi,\infty))$.
\item[(C3)]
$K^3/n={\rm o}(1)$.
\item[(C4)]
For any $p\times q$ dimensional matrix $A$, we denote the spectral norm of $A$ by $\|A\|=\sup_{\bm{v}\neq \bm{0}}\|A\bm{v}\|/\|\bm{v}\|$.
Then the following holds:
\begin{align*}
\Bigl\|\frac{1}{\sqrt{n}}X_{S^\bot}^\top X_S\Bigr\|={\rm O}(K).
\end{align*}
\end{description}
The condition (C1) pertains to the design matrix.
Note that we only consider a high dimensional and small sample size setting for the original data set, and not for selected variables.
This assumption is reasonable for high dimensional and large sample scenarios.
(C2) requires that ${\rm P}(y_i=1)$ not converge to 0 or 1 for any $i=1,\ldots,n$.
Observe that the parameter space ${\cal B}$ is an open and convex set with respect to $\bm{\beta}_S$.
This assumption naturally holds when the space of regressors is compact and $\bm{\beta}_S$ does not diverge.
In addition, if the maximum likelihood estimator $\hat{\bm{\beta}}_S$ is $\sqrt{n/K}$-consistent, then $\hat{\bm{\beta}}_S$ lies in ${\cal B}$ with probability converging to 1.
The condition (C3) represents the relationship between the sample size and the number of selected variables for high dimensional asymptotics in our model.
As related conditions, \cite{FanPen04} employs $K^5/n\to0$, and \cite{DasKhaGho14} employs $K^{6+\delta}/n\to0$ for some $\delta>0$ to derive an asymptotic expansion of a posterior distribution in a Bayesian setting.
Furthermore, \cite{Hub73} employs the same condition as in (C3) in the scenario for $M$-estimation.
Finally, (C4) requires that regressors of selected variables and those of unselected variables be only weakly correlated.
A similar assumption is required in \cite{Hua08} for deriving an asymptotic distribution for a bridge estimator.
This type of assumption, e.g., a restricted eigenvalue condition \citep{bickel2009simultaneous}, is essential for handling high dimensional behavior of the estimator.

\section{Proposed Method}
\label{sec:propose}
In this section, we present the proposed method for selective inference for high dimensional logistic regression with marginal screening.
We first consider a subset of features $\hat{S} = S (\supset S^*)$ as a fixed set, and derive an asymptotic distribution of $\hat{\bm{\beta}}_S$ under the assumptions (C1) -- (C3). 
Then, we introduce the ``source'' of the test statistic $\bm{T}_n(\bm{\beta}_S^*)$, which is defined by a score function, and apply it to the polyhedral lemma, where we will show that the truncation points are independent of the $\bm{\eta}^\top \bm T_n(\bm{\beta}_S^*)$ with the assumption (C4).

To extend the selective inference framework to logistic regression, we first consider a subset of variables $\hat{S}=S~(\supset S^*)$ as a fixed set.
From (\ref{eq;loss}), let us define a score function and observed information matrix by
\begin{align*}
\bm{s}_n(\bm{\beta}_S)
&=\frac{1}{\sqrt{n}}\ell'_n(\bm{\beta}_S) 
=\frac{1}{\sqrt{n}}\sum_{i=1}^{n}\bm{x}_{S,i}(y_i-\psi'(\bm{x}_{S,i}^\top\bm{\beta}_S))
\intertext{and}
\Sigma_n(\bm{\beta}_S)
&=-\frac{1}{n}\ell''_n(\bm{\beta}_S)
=\frac{1}{n}\sum_{i=1}^{n}\psi''(\bm{x}_{S,i}^\top\bm{\beta}_S)\bm{x}_{S,i}\bm{x}_{S,i}^\top,
\end{align*}
respectively.
To simplify the notation, we denote $\bm{s}_n(\bm{\beta}_S^*)$ and $\Sigma_n(\bm{\beta}_S^*)$ by $\bm{s}_n$ and $\Sigma_n$, respectively, for the true value of $\bm{\beta}_S^*$.
Because $\psi''(\bm{x}_{S,i}^\top\bm{\beta}_S^*)$ is uniformly bounded on ${\cal B}$ from (C2), $\Sigma_n$ is a symmetric and positive definite matrix when (C1) holds.
Then, by the same argument as in \cite{FanPen04}, if $K^2/n\to 0$, we have
\begin{align}
\|\hat{\bm{\beta}}_S-\bm{\beta}_S^*\|={\rm O}_{{\rm p}}(\sqrt{K/n}).
\label{eq;consistency}
\end{align}
By using Taylor's theorem, we have
\begin{align*}
\bm{0}
=\ell'_n(\hat{\bm{\beta}}_S)
\approx \sqrt{n}\bm{s}_n-n\Sigma_n(\hat{\bm{\beta}}_S-\bm{\beta}_S^*),
\end{align*}
and thus
\begin{align*}
\sqrt{n}(\hat{\bm{\beta}}_S-\bm{\beta}_S^*)
\approx \Sigma_n^{-1}\bm{s}_n.
\end{align*}
As per Remark 1, $S\supset S^*$ implies
\begin{align*}
{\rm E}[\bm{s}_n]
=\frac{1}{\sqrt{n}}\sum_{i=1}^{n}\bm{x}_{S,i}({\rm E}[y_i]-\psi'(\bm{x}_{S,i}^\top\bm{\beta}_S^*))
=\bm{0}.
\end{align*}
In addition, because the $y_i$'s are independent of each other, we observe that
\begin{align*}
{\rm V}[\bm{s}_n]
=\frac{1}{n}\sum_{i=1}^{n}{\rm V}[y_i]\bm{x}_{S,i}\bm{x}_{S,i}^\top
=\Sigma_n.
\end{align*}
Therefore, by recalling asymptotic normality of the score function, we expect that a distribution of $\Sigma_n^{-1}\bm{s}_n$ can be approximated by a normal distribution with mean $\bm{0}$ and covariance matrix $\Sigma_n^{-1}$.
Indeed, if $S$ is fixed, this approximation is true under the conditions (C1) -- (C3):
\begin{thm}
\label{thm1}
Suppose that the conditions (C1) -- (C3) hold.
Then, for any fixed $S~(\supset S^*)$ and $\bm{\eta}\in\mathbb{R}^K$ with $\|\bm{\eta}\|<\infty$, we have
\begin{align}
\sqrt{n}\sigma_n^{-1}\bm{\eta}^\top(\hat{\bm{\beta}}_S-\bm{\beta}^*_S)
=\sigma_n^{-1}\bm{\eta}^\top\Sigma_n^{-1}\bm{s}_n+{\rm o}_{\rm p}(1) \stackrel{{\rm d}}{\to}
{\rm N}(0,1),
\label{eq;adist}
\end{align}
where $\sigma_n^2=\bm{\eta}^\top\Sigma_n^{-1}\bm{\eta}$ and ${\rm o}_{{\rm p}}(1)$ is a term that converges to 0 in probability uniformly with respect to $\bm{\eta}$ and $S$.
\end{thm}

Note that, under the conditions (C1), (C2) and $d^3/n\to 0$, Theorem \ref{thm1} also holds when we do not enforce variable selection (see e.g., \cite{FanPen04}).
To formulate a selective test, let us consider 
\begin{align}
\bm{T}_n(\bm{\beta}_S^*)
=\Sigma_n^{-1}\bm{s}_n
=\Sigma_n^{-1}\left\{
\frac{1}{\sqrt{n}}X_S^\top(\bm{y}-\bm{\psi}'(\bm{\beta}_S^*))
\right\}
\label{eq;test_stat}
\end{align}
as a ``source" of a test statistic, where $\bm{\psi}'(\bm{\beta}_S^*)=(\psi'(\bm{x}_{S,i}^\top\bm{\beta}_S^*))_{i=1,\ldots, n}$.
The term ``source" means that we cannot use it as a test statistic directly because $\bm{T}_n(\bm{\beta}_S^*)$ depends on $\bm{\beta}_S^*$.
In the following, for notational simplicity, we denote $\bm{T}_n(\bm{\beta}_S^*)$ and $\bm{\psi}'(\bm{\beta}_S^*)$ by $\bm{T}_n$ and $\bm{\psi}'$, respectively.

As noted in Section \ref{subsec:MS}, by using an appropriate non-random matrix $A\in\mathbb{R}^{K(2d-2K+1)\times d}$, the marginal screening selection event can be expressed as an affine constraint with respect to $\bm{z}=X^\top\bm{y}$, that is, $\{\bm{z};A\bm{z}\leq \bm{0}\}$.
Then, by appropriately dividing $A$ and $X$ based on the selected $S$, we can rewrite it as follows:
\begin{align*}
A\bm{z}\leq \bm{0}
\quad
\Leftrightarrow 
\quad
A_SX_S^\top\bm{y}+A_{S^\bot}X_{S^\bot}^\top\bm{y}\leq \bm{0}
\quad
\Leftrightarrow 
\quad
\tilde{A}\bm{T}_n\leq \tilde{\bm{b}}.
\end{align*}
The last inequality is an affine constraint with respect to $\bm{T}_n$, where
\begin{align*}
\tilde{A}=A_S\Sigma_n
\qquad
\text{and}
\qquad
\tilde{\bm{b}}=-\frac{1}{\sqrt{n}}(A_SX_S^\top\bm{\psi}'+A_{S^\bot}X_{S^\bot}^\top\bm{y}).
\end{align*}
Unlike the polyhedral lemma in Section \ref{subsec:SI_in_LR}, $\tilde{\bm{b}}$ depends on $\bm{y}$ and so is a random vector.
By using (C4), we can prove that $\tilde{\bm{b}}$ is asymptotically independent of $\bm{\eta}^\top\bm{T}_n$, which implies the polyhedral lemma holds asymptotically.
\begin{thm}
\label{thm2}
Suppose that (C1) -- (C4) all hold.
Let $\bm{c}=\Sigma_n^{-1}\bm{\eta}/\sigma_n^2$ for any $\bm{\eta}\in\mathbb{R}^K$ with $\|\bm{\eta}\|<\infty$, and $\bm{w}=(I_K-\bm{c}\bm{\eta}^\top)\bm{T}_n$, where $\sigma_n^2=\bm{\eta}^\top\Sigma_n^{-1}\bm{\eta}$.
Then, for any fixed $S~(\supset S^*)$, the selection event can be expressed as
\begin{align*}
\{\bm{T};\tilde{A}\bm{T}\leq \tilde{\bm{b}}\}
=\{\bm{T};L_n\leq \bm{\eta}^\top\bm{T}\leq U_n,N_n=0\},
\end{align*}
where
\begin{align}
L_n
=\max_{l:(\tilde{A}\bm{c})_l<0}\frac{\tilde{b}_l-(\tilde{A}\bm{w})_l}{(\tilde{A}\bm{c})_l},
\qquad
U_n
=\min_{l:(\tilde{A}\bm{c})_l>0}\frac{\tilde{b}_l-(\tilde{A}\bm{w})_l}{(\tilde{A}\bm{c})_l},
\label{eq;truncation}
\end{align}
and $N_n=\max_{l:(\tilde{A}\bm{c})_l=0}\tilde{b}_l-(\tilde{A}\bm{w})_l$.
In addition, $(L_n,U_n,N_n)$ is asymptotically independent of $\bm{\eta}^\top\bm{T}_n$.
\end{thm}

As a result of Theorem \ref{thm1}, Theorem \ref{thm2} and (C0), we can asymptotically identify a pivotal quantity as a truncated normal distribution, that is, by letting $\bm{\eta}=\bm{e}_j\in\mathbb{R}^K$, 
\begin{align*}
\left[F^{[L_n,U_n]}_{0,\sigma_n^2}(\bm{\eta}^\top\bm{T}_n) \mid \tilde{A}\bm{T}_n\leq\tilde{\bm{b}}\right]
\stackrel{{\rm d}}{\to}{\rm Unif}(0,1)
\end{align*}
for any $\bm{w}$, under ${\rm H}_{0,j}$.
Therefore, we can define an asymptotic selective $p$-value for selective test (\ref{eq;stest}) under ${\rm H}_{0,j}$ as follows:
\begin{align}
P_{n,j}=2\min\Bigl\{F^{[L_n,U_n]}_{0,\sigma_n^2}(\bm{\eta}^\top\bm{T}_n), 1-F^{[L_n,U_n]}_{0,\sigma_n^2}(\bm{\eta}^\top\bm{T}_n) \Bigr\},
\label{eq;pval}
\end{align}
where $L_n$ and $U_n$ are evaluated at the realization of $\bm{w}=\bm{w}_0$.
Unfortunately, because $\bm{T}_n$, $\Sigma_n$, $L_n$ and $U_n$ are still dependent on the true value of $\bm{\beta}_S^*$, we construct a test statistic by introducing a maximum likelihood estimator (\ref{eq;estimator}), which is a consistent estimator of $\bm{\beta}_S^*$.

\subsection{Computing Truncation Points}
In practice, we need to compute truncation points in (\ref{eq;truncation}).
When we utilize marginal screening for variable selection, it becomes difficult to compute $L_n$ and $U_n$ because $\tilde{A}$ becomes a $\{2K(d-K)+K\}\times K$ dimensional matrix.
For example, even when $d=1{,}000$ and $K=20$, we need to handle a 39,220 dimensional vector.
To reduce the computational burden, we derive a simple form of (\ref{eq;truncation}) in this section.

We first derive $A_S$.
As notedd in Section \ref{subsec:MS}, selection event $\{(\hat{S}, s_{\hat{S}})=(S, s_S)\}$ can be rewritten as
\begin{align*}
-s_jz_j\leq z_k\leq s_jz_j,~s_j z_j\geq 0,
~~~~~
\forall (j,k)\in S\times S^\bot,
\end{align*}
where $s_j={\rm sgn}(z_j)$ is the sign of the $j$-th element of $\bm{z}=X^\top\bm{y}$.
Let $S=\{j_1,\ldots, j_K\}$ and $q=2(d-K)+1$.
Then, by a simple calculation, we have
\begin{align*}
A_S
=\left( \begin{array}{ccc}
-s_{j_1}\bm{1}_{q}&&O \\
&\ddots& \\
O&& -s_{j_K}\bm{1}_{q}
\end{array} \right)
=-J\otimes\bm{1}_{q},
\end{align*}
where $J$ is a $K\times K$ dimensional diagonal matrix whose $j$-th diagonal element is $s_j$ and $\otimes$ denotes a Kronecker product.
Since $\tilde{A}=A_S\Sigma_n$ and $\bm{c}=\Sigma_n^{-1}\bm{\eta}/\sigma_n^2$, the denominator in (\ref{eq;truncation}) reduces to $\tilde{A}\bm{c}=A_S\bm{\eta}//\sigma_n^2$.
For $\bm{\eta}=\bm{e}_j$, we can further evaluate $A_S\bm{\eta}$ as
\begin{align*}
A_S\bm{\eta}=-s_j(\bm{0}_{(j-1)q}^\top,\bm{1}_{q}^\top,\bm{0}_{(K-j)q}^\top)^\top\in\mathbb{R}^{Kq}.
\end{align*}
Further, by the definition of $\tilde{A},~\tilde{\bm{b}}$, and $\bm{w}$, we have
\begin{align*}
\tilde{\bm{b}}-\tilde{A}\bm{w}
=\tilde{\bm{b}}-\tilde{A}\bm{T}_n+(\bm{\eta}^\top\bm{T}_n)\tilde{A}\bm{c}
=-\frac{1}{\sqrt{n}}A\bm{z}+T_{n,j}\tilde{A}\bm{c}.
\end{align*}
Because $\sigma_n^2$, the $j$-th diagonal element of $\Sigma_n^{-1}$, is positive, it is straightforward to observe that
\begin{align*}
\{l:(\tilde{A}\bm{c})_l<0\}
=
\begin{cases}
\{(j-1)q+1,\ldots,jq\}, & \text{if}~s_j=1 \\
\emptyset, & \text{otherwise}
\end{cases}
\end{align*}
for $j=1,\ldots,K$.
Note that, for each $j=1,\ldots,K$, $(A\bm{z})_{l=(j-1)q+1,\ldots,jq}$ consists of $q$ elements of $z_j$ and $z_j\pm z_k$ for any $k\in S^\bot$.
Therefore, for each $j=1,\ldots,K$, we have
\begin{align*}
\max_{l=(j-1)q+1,\ldots,jq}(A\bm{z})_l
=\max_{k\in S^\bot}\{z_j,z_j\pm z_k\}
=z_j+\max_{k\in S^\bot}|z_k|.
\end{align*}
As a consequence, we obtain
\begin{align}
L_n
&=\max_{l:(\tilde{A}\bm{c})_l<0}\frac{\tilde{b}_l-(\tilde{A}\bm{w})_l}{(\tilde{A}\bm{c})_l} \nonumber \\
&=\max_{l:(\tilde{A}\bm{\eta})_l<0}\frac{-(A\bm{z})_l/\sqrt{n}}{(\tilde{A}\bm{\eta})_l/\sigma_n^2}+T_{n,j} \nonumber \\
&=\frac{\sigma_n^2}{\sqrt{n}} \max_{l=(j-1)q+1,\ldots,jq}(A\bm{z})_l+T_{n,j} \nonumber \\
&=\frac{\sigma_n^2}{\sqrt{n}}(|z_j|+\max_{k\in S^\bot}|z_k|)+T_{n,j},
\label{eq;lower}
\end{align}
if $s_j=1$, and $L_n=-\infty$, otherwise.
Similarly, we obtain
\begin{align}
U_n
&=\min_{l:(\tilde{A}\bm{c})_l>0}\frac{\tilde{b}_l-(\tilde{A}\bm{w})_l}{(\tilde{A}\bm{c})_l} \nonumber \\
&=\min_{l:(\tilde{A}\bm{\eta})_l>0}\frac{-(A\bm{z})_l/\sqrt{n}}{(\tilde{A}\bm{\eta})_l/\sigma_n^2}+T_{n,j} \nonumber \\
&=-\frac{\sigma_n^2}{\sqrt{n}} \max_{l=(j-1)q+1,\ldots,jq}(A\bm{z})_l+T_{n,j} \nonumber \\
&=\frac{\sigma_n^2}{\sqrt{n}}(|z_j|-\max_{k\in S^\bot}|z_k|)+T_{n,j},
\label{eq;upper}
\end{align}
if $s_j=-1$, and $U_n=\infty$, otherwise.
Because of this simple form, we can calculate truncation points efficiently.
We summarize the algorithm to compute selective $p$-values of the $K$ selective test in Algorithm \ref{alg1}.

\begin{algorithm}[t]
    \SetKwInOut{Input}{Input}
    \SetKwInOut{Output}{Output}
    \Input{Data $(\bm{y},X)\in\{0,1\}^n\times\mathbb{R}^{n\times d}$, \# of selected variables $K$}
    \Output{Selective $p$-value for $K$ selective test (\ref{eq;stest})}
    {
    $\bm{z}\leftarrow X^\top\bm{y}$\;
    $S\leftarrow \{j;|z_j|~ \text{is among the first}~ K~ \text{largest of all}\}$\;
    $\hat{\bm{\beta}}_S\leftarrow \argmax_{\bm{\beta}_S\in{\cal B}}\ell_n(\bm{\beta}_S)$\;
    $\bm{p}\leftarrow \bm{0}$\;
    }
    \For{$j=1,\ldots,K$}
      {
        $\bm{\eta}\leftarrow \bm{e}_j$\; 
        Compute $\bm{\eta}^\top\bm{T}_n,\sigma_n^2, L_n$ and $U_n$ based on (\ref{eq;lower}) and (\ref{eq;upper})\;
        $p_j\leftarrow2\min\Bigl\{F^{[L_n,U_n]}_{0,\sigma_n^2}(\bm{\eta}^\top\bm{T}_n), 1-F^{[L_n,U_n]}_{0,\sigma_n^2}(\bm{\eta}^\top\bm{T}_n) \Bigr\}$
      }
      {\bf Return} $\bm{p}\in[0,1]^K$
    \caption{Selective Inference for Classification}
    \label{alg1}
\end{algorithm}

\subsection{Controlling Family-wise Error Rate}
\label{subsec:FWER}
Since selective test (\ref{eq;stest}) consists of $K$ hypotheses, we may be concerned about multiplicity when $K>1$.
In this case, instead of selective type I error, we control the {\it family-wise error rate} (FWER) in the sense of selective inference and we term it selective FWER.

For the selected variable $\hat{S}=S$, let us denote a family of true null by ${\cal H}=\{{\rm H}_{0,j}:{\rm H}_{0,j}(j\in S)$ is true null$\}$.
Then, let us define the selective FWER by
\begin{align}
{\rm sFWER}
={\rm P}(\text{at least one}~{\rm H}_{0,j}\in{\cal H}~\text{is rejected}\mid \hat{S}=S)
\label{eq;FWER}
\end{align}
in the same way as the classic FWER.
Next, we asymptotically control the selective FWER at level $\alpha$ by utilizing Bonferroni correction for $K$ selective tests.
Specifically, we adjust selective $p$-values (\ref{eq;pval}) as follows.
Let us define $\tilde{\alpha}=\alpha/K$.
Since selective $p$-value $P_{n,j}$ is asymptotically distributed according to ${\rm Unif}(0,1)$, we have that a limit superior of (\ref{eq;FWER}) can be bounded as follows:
\begin{align*}
\limsup_{n\to\infty}{\rm P}\Bigl(\bigcup_{j:{\rm H}_{0,j}\in{\cal H}}\{P_{n,j}\leq \tilde{\alpha} \} \mid \hat{S}=S\Bigr)
&\leq \limsup_{n\to\infty}\sum_{j:{\rm H}_{0,j}\in{\cal H}}{\rm P}(P_{n,j}\leq \tilde{\alpha} \mid \hat{S}=S) \\
&\leq \sum_{j:{\rm H}_{0,j}\in{\cal H}}\limsup_{n\to\infty}{\rm P}(P_{n,j}\leq \tilde{\alpha} \mid \hat{S}=S) \\
&\leq |{\cal H}|\tilde{\alpha}
\leq \alpha.
\end{align*}
In the last inequality, we simply use $|{\cal H}|\leq K$.
Accordingly, letting $p_{n,j}$ be a realization of (\ref{eq;pval}), we reject a null hypothesis when $\{p_{n,j}\leq \tilde{\alpha} \}$.
In the following, we refer to $\tilde{p}_{n,j}=\min\{1,Kp_{n,j}\}$ as an {\it adjusted selective $p$-value}.
Note that we can utilize not only Bonferroni's method but also other methods for correcting multiplicity such as Scheff$\acute{\text{e}}$'s method, Holm's method, and so on.
We use Bonferroni's method for expository purposes.

\section{Simulation Study}
\label{sec:simulation}
Through simulation studies, we explore the performance of the proposed method in Section \ref{sec:propose}, which we term ASICs (Asymptotic Selective Inference for Classification) here.

We first identify if the ASICs can control selective type I error.
We also check the selective type I error when data splitting (DS) and nominal test (NT) methods are used.
In DS, we first randomly divide the data into two disjoint sets.
Then, after selecting $\hat{S}=S$ with $|S|=K$ by using one of these sets, we construct a test statistic $\bm{T}_n(\hat{\bm{\beta}}_S)$ based on the other sets and reject the $j$-th selective test (\ref{eq;stest}) when $|T_{n,j}/\sigma_n|\geq z_{\alpha/2}$, where $z_{\alpha/2}$ is an upper $\alpha/2$-percentile of a standard normal distribution.
In NT, we cannot control type I errors since selection bias is ignored:
it selects $K$ variables by marginal screening first, then rejects the $j$-th selective test (\ref{eq;stest}) when $|T_{n,j}/\sigma_n|\geq z_{\alpha/2}$, where the entire data set is used for both selection and inference steps.
Finally, we explore whether the ASICs can effectively control selective FWER, and at the same time, confirm its statistical power by comparing it with that of DS.

The simulation settings are as follows.
As $d$ dimensional regressor $\bm{x}_i$ ($i=1,\ldots,n$), we used vectors obtained from ${\rm N}(\bm{0},\Sigma)$, where $\Sigma$ is a $d\times d$ dimensional covariance matrix whose $(j,k)$-th element is set to $\rho^{|j-k|}$.
We set $\rho=0$ or $0.5$ in Case 1 and Case 2, respectively.
Note that each element of $\bm{x}_i$ is independent in Case 1 but correlated in Case 2.
Then, for each $\bm{x}_i$, we generate $y_i$ from ${\rm Bi}(\psi'(\bm{x}_i^\top\bm{\beta}^*))$, where $\bm{\beta}^*$ is a $d$ dimensional true coefficient vector and ${\rm Bi}(p)$ is a Bernoulli distribution with parameter $p$.
In the following, we conduct simulations using 1,000 Monte-Carlo runs.
We use the {\tt glm} package in {\tt R} for parameter estimation.

\subsection{Controlling Selective Type I Error}
\label{subsec:selective_type_I_error}

To check if ASICs can control selective type I error, we consider a selective test (\ref{eq;stest}).
Specifically, we first select $K=1$ variable by marginal screening and then conduct a selective test at the 5\% level.
By setting $\bm{\beta}^*=\bm{0}\in\mathbb{R}^d$, we can confirm selective type I error because the selective null is always true.
Therefore, we assess the following index as an estimator of the selective type I error:
letting $\beta$ be the selected variable in each simulation, we evaluate an average and standard deviation of
\begin{align}
I\{{\rm H}_0~\text{is rejected}\},
\label{eq;FP}
\end{align}
where $I$ is an indicator function and ${\rm H}_0:\beta^*=0$ is a selective null.
We construct a selective test at the 5\% level in all simulations.
In the same manner as classical type I error, it is desirable when the above index is less than or equal to 0.05, with particularly small values indicating that the selective test is overly conservative.

Table \ref{tab:FPR} presents averages and standard deviations of (\ref{eq;FP}) based on 1,000 runs.
It is clear that NT cannot control selective type I error;
it becomes larger as the dimension $d$ increases.
In addition, NT does not improve even if the sample size becomes large, because there exist selection bias in the selection step.
On the other hand, both ASICs and DS adequately control selective type I error, although the latter appears slightly more conservative than the former.
Moreover, unlike NT, these two methods can adequately control selective type I error, even when the covariance structure of $\bm{x}_i$ and the number of dimensions change.

\begin{table}[t]
\caption{Method comparison using simulated data based on 1,000 Monte-Carlo runs. 
Each cell denotes an average with standard deviations of (\ref{eq;FP}) in parentheses.}
\begin{tabular}{@{\extracolsep{-9pt}}ccrrrrrrr}
&&&\multicolumn{6}{c}{sample size} \\ \cline{4-9}
&$d$&method&50&100&200&500&1,000&1,500 \\ \hline\hline
Case 1&200&ASICs&.029 {\footnotesize (.168)}&.049 {\footnotesize (.216)}&.038 {\footnotesize (.191)}&.031 {\footnotesize (.173)}&.028 {\footnotesize (.165)}&.033 {\footnotesize (.179)} \\
&&DS&.012 {\footnotesize (.109)}&.015 {\footnotesize (.122)}&.004 {\footnotesize (.063)}&.004 {\footnotesize (.063)}&.011 {\footnotesize (.104)}&.011 {\footnotesize (.104)} \\
&&NT&.184 {\footnotesize (.388)}&.226 {\footnotesize (.418)}&.219 {\footnotesize (.414)}&.261 {\footnotesize (.439)}&.255 {\footnotesize (.436)}&.256 {\footnotesize (.437)} \\ \cline{2-9}
&500&ASICs&.028 {\footnotesize (.165)}&.043 {\footnotesize (.203)}&.039 {\footnotesize (.194)}&.039 {\footnotesize (.194)}&.032 {\footnotesize (.176)}&.036 {\footnotesize (.186)} \\
&&DS&.012 {\footnotesize (.109)}&.006 {\footnotesize (.077)}&.008 {\footnotesize (.089)}&.009 {\footnotesize (.094)}&.005 {\footnotesize (.071)}&.008 {\footnotesize (.089)} \\
&&NT&.267 {\footnotesize (.044)}&.273 {\footnotesize (.446)}&.304 {\footnotesize (.460)}&.301 {\footnotesize (.459)}&.326 {\footnotesize (.469)}&.325 {\footnotesize (.469)} \\ \cline{2-9}
&1,000&ASICs&.041 {\footnotesize (.198)}&.044 {\footnotesize (.205)}&.023 {\footnotesize (.150)}&.032 {\footnotesize (.176)}&.038 {\footnotesize (.191)}&.044 {\footnotesize (.205)} \\
&&DS&.006 {\footnotesize (.077)}&.011 {\footnotesize (.104)}&.010 {\footnotesize (.100)}&.009 {\footnotesize (.094)}&.013 {\footnotesize (.113)}&.010 {\footnotesize (.100)} \\
&&NT&.294 {\footnotesize (.456)}&.345 {\footnotesize (.476)}&.390 {\footnotesize (.488)}&.402 {\footnotesize (.491)}&.411 {\footnotesize (.492)}&.405 {\footnotesize (.491)} \\ \hline
Case 2&200&ASICs&.038 {\footnotesize (.191)}&.038 {\footnotesize (.191)}&.040 {\footnotesize (.196)}&.032 {\footnotesize (.176)}&.028 {\footnotesize (.165)}&.031 {\footnotesize (.173)} \\
&&DS&.012 {\footnotesize (.109)}&.007 {\footnotesize (.083)}&.012 {\footnotesize (.109)}&.010 {\footnotesize (.100)}&.012 {\footnotesize (.109)}&.004 {\footnotesize (.063)} \\
&&NT&.177 {\footnotesize (.382)}&.207 {\footnotesize (.405)}&.234 {\footnotesize (.424)}&.211 {\footnotesize (.408)}&.219 {\footnotesize (.414)}&.210 {\footnotesize (.408)} \\ \cline{2-9}
&500&ASICs&.049 {\footnotesize (.216)}&.038 {\footnotesize (.191)}&.030 {\footnotesize (.171)}&.030 {\footnotesize (.171)}&.039 {\footnotesize (.194)}&.034 {\footnotesize (.181)} \\
&&DS&.007 {\footnotesize (.083)}&.006 {\footnotesize (.077)}&.010 {\footnotesize (.100)}&.009 {\footnotesize (.094)}&.007 {\footnotesize (.083)}&.007 {\footnotesize (.083)} \\
&&NT&.247 {\footnotesize (.431)}&.269 {\footnotesize (.443)}&.291 {\footnotesize (.454)}&.295 {\footnotesize (.456)}&.309 {\footnotesize (.462)}&.318 {\footnotesize (.466)} \\ \cline{2-9}
&1,000&ASICs&.049 {\footnotesize (.216)}&.047 {\footnotesize (.212)}&.031 {\footnotesize (.173)}&.034 {\footnotesize (.181)}&.024 {\footnotesize (.153)}&.046 {\footnotesize (.210)} \\
&&DS&.009 {\footnotesize (.094)}&.008 {\footnotesize (.089)}&.013 {\footnotesize (.113)}&.006 {\footnotesize (.077)}&.006 {\footnotesize (.077)}&.010 {\footnotesize (.100)} \\
&&NT&.290 {\footnotesize (.454)}&.350 {\footnotesize (.477)}&.375 {\footnotesize (.484)}&.396 {\footnotesize (.489)}&.407 {\footnotesize (.492)}&.414 {\footnotesize (.493)} \\ \hline\hline
\end{tabular}
\label{tab:FPR}
\end{table}

\subsection{FWER and Power}
\label{subsec:FWER_Power}
Here, we explore selective FWER and statistical power with respect to ASICs and DS for $K$ selective tests (\ref{eq;stest}), where we set $K=5, 10, 15$, and $20$.
Note that, as discussed in the above section, NT is disregarded here because it does no adequately control selective type I error.
We adjust multiplicity by utilizing Bonferroni's method as noted in Section \ref{subsec:FWER}.

The true coefficient vector is set to be $\bm{\beta}^*=(2\times\bm{1}_5^\top,\bm{0}_{d-5}^\top)^\top$ and $\bm{\beta}^*=(2\times\bm{1}_5^\top,-2\times\bm{1}_5^\top,\bm{0}_{d-10}^\top)^\top$ in Model 1 and Model 2, respectively.
In the following, we assess the indices as an estimator of selective FWER and power.
Letting $\hat{S}=S$ be the subset of selected variables for each simulation, we evaluate an average of
\begin{align}
I\{\text{at least one}~{\rm H}_{0,j}\in{\cal H}~\text{is rejected}\}
\label{eq;FWER.est}
\end{align}
and
\begin{align}
\frac{1}{|S^*|}\sum_{j\in S}I\{{\rm H}_{0,j}\not\in{\cal H}~\text{is rejected}\},
\label{eq;TPR}
\end{align}
where, for each $j\in S$, ${\rm H}_{0,j}:\beta_j^*=0$ is the selective null and ${\cal H}$ is a family of true nulls.
Note that, by using Bonferroni's method, we use $\tilde{\alpha}=\alpha/K$ as an adjusted significance level for $\alpha=0.05$.
Similar to the selective type I error, it is desirable when (\ref{eq;FWER.est}) is less than or equal to $\alpha$.
In addition, higher values of (\ref{eq;TPR}) are desiable in the same manner as per classical power.
We evaluate (\ref{eq;TPR}) as the proportion of rejected hypotheses for false nulls to that of true active variables.
We employ this performance index because it is important to identify how many truly active variables are extracted in practice.

\begin{figure*}[t]
\begin{center}
Case 1
\end{center}
\vspace{-35pt}
\begin{center}
\subfloat[$d=200$]{\includegraphics[scale=0.16, bb=0 0 720 720]{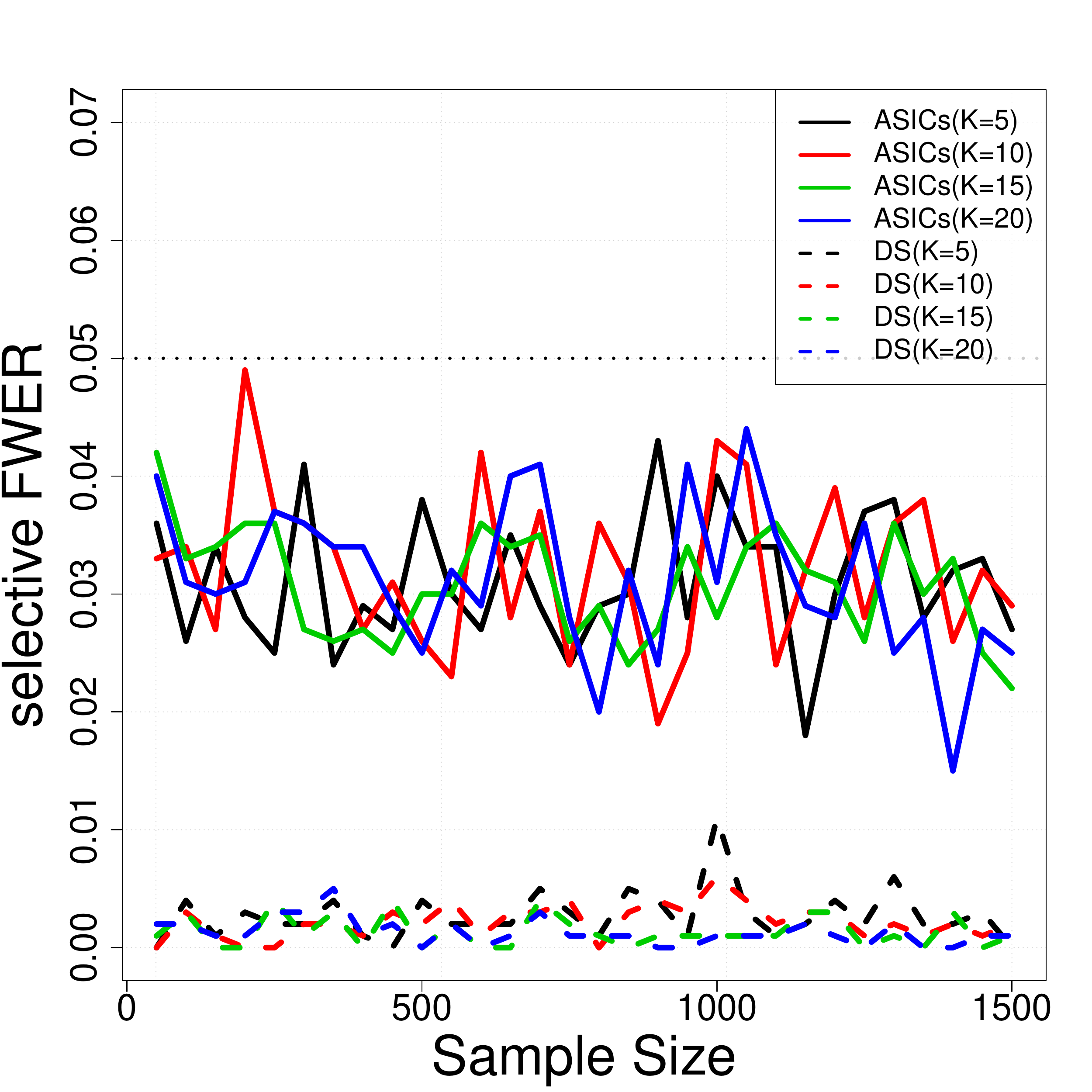}}
\subfloat[$d=500$]{\includegraphics[scale=0.16, bb=0 0 720 720]{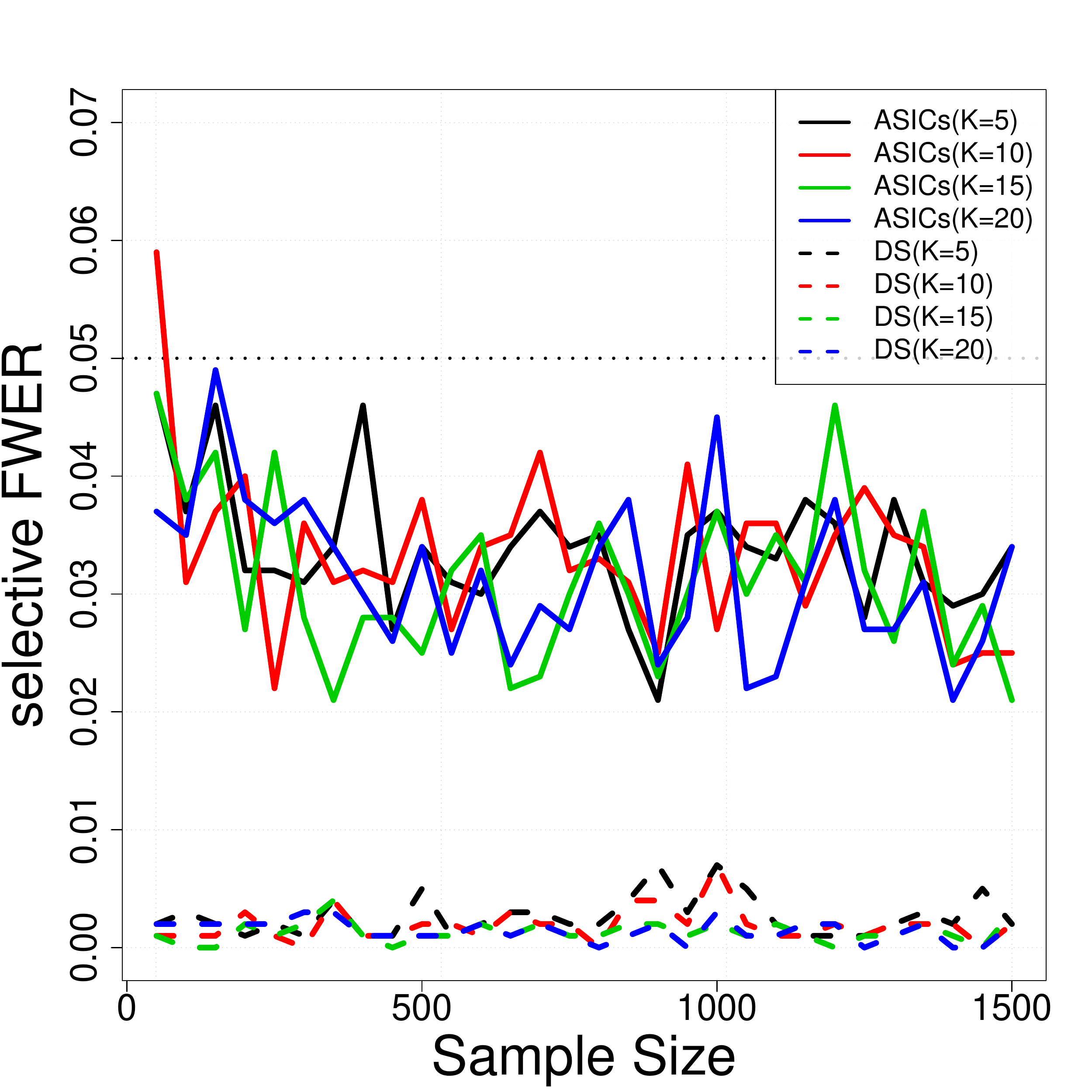}} 
\subfloat[$d=1,000$]{\includegraphics[scale=0.16, bb=0 0 720 720]{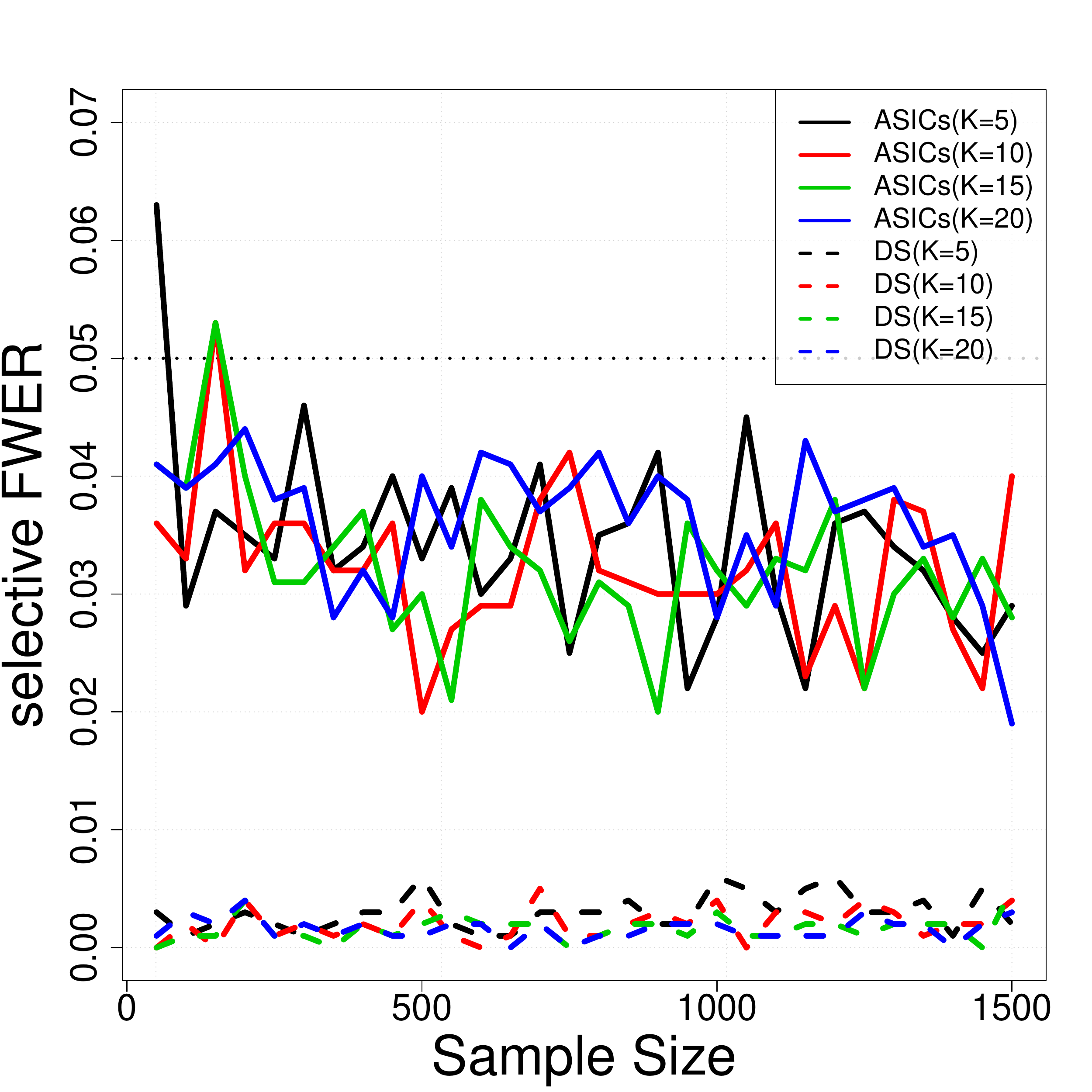}}
\end{center}
\begin{center}
Case 2
\end{center}
\vspace{-35pt}
\begin{center}
\subfloat[$d=200$]{\includegraphics[scale=0.16, bb=0 0 720 720]{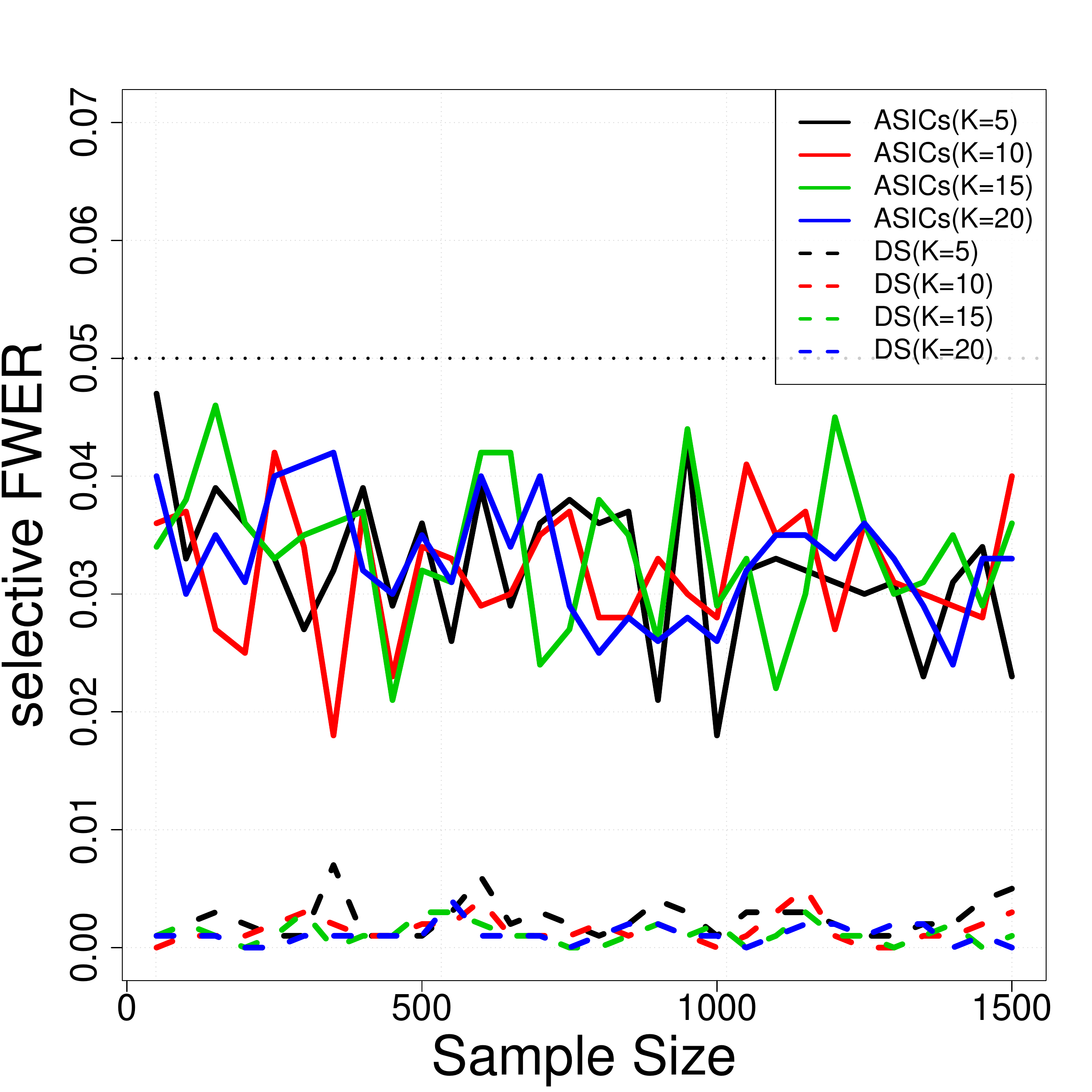}}
\subfloat[$d=500$]{\includegraphics[scale=0.16, bb=0 0 720 720]{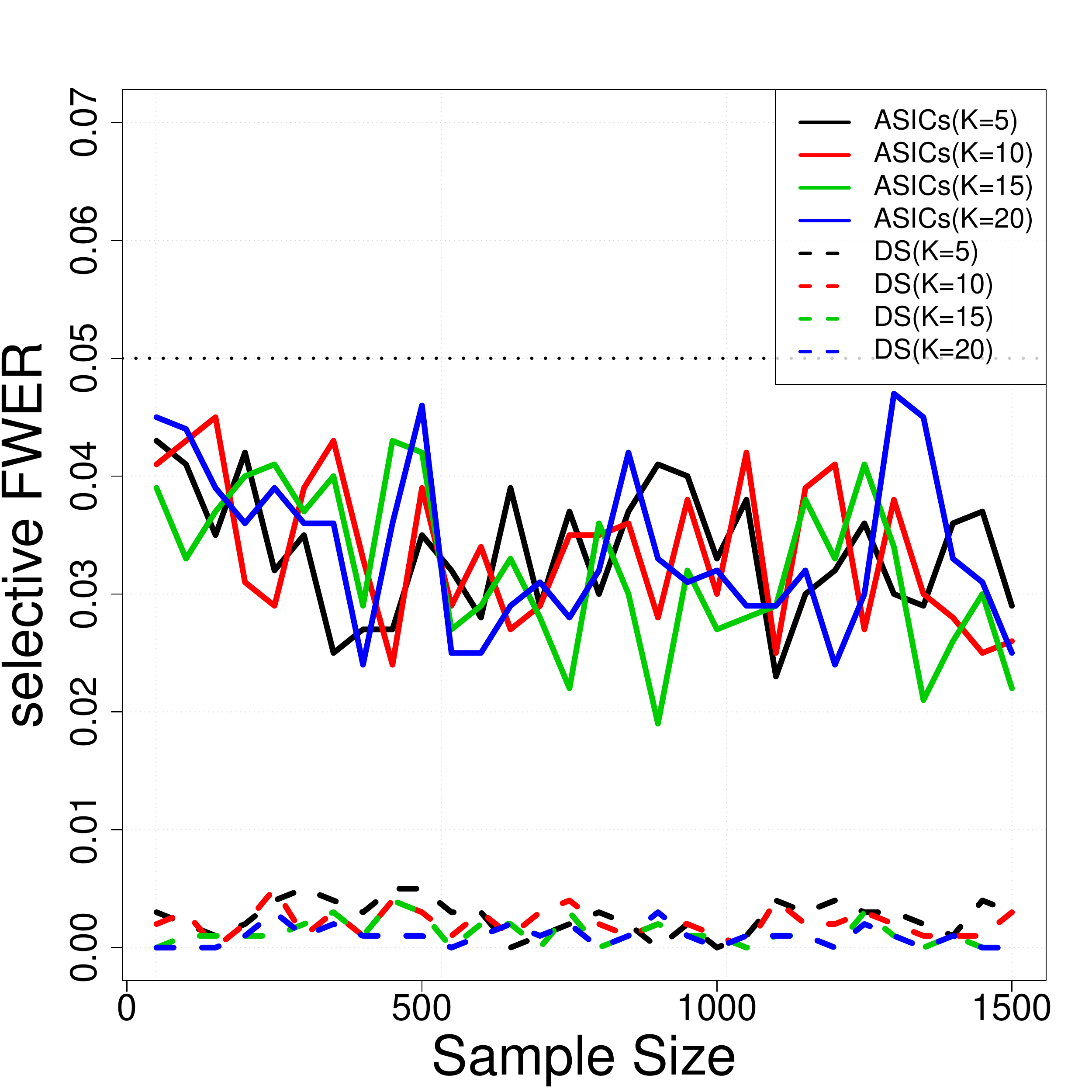}} 
\subfloat[$d=1,000$]{\includegraphics[scale=0.16, bb=0 0 720 720]{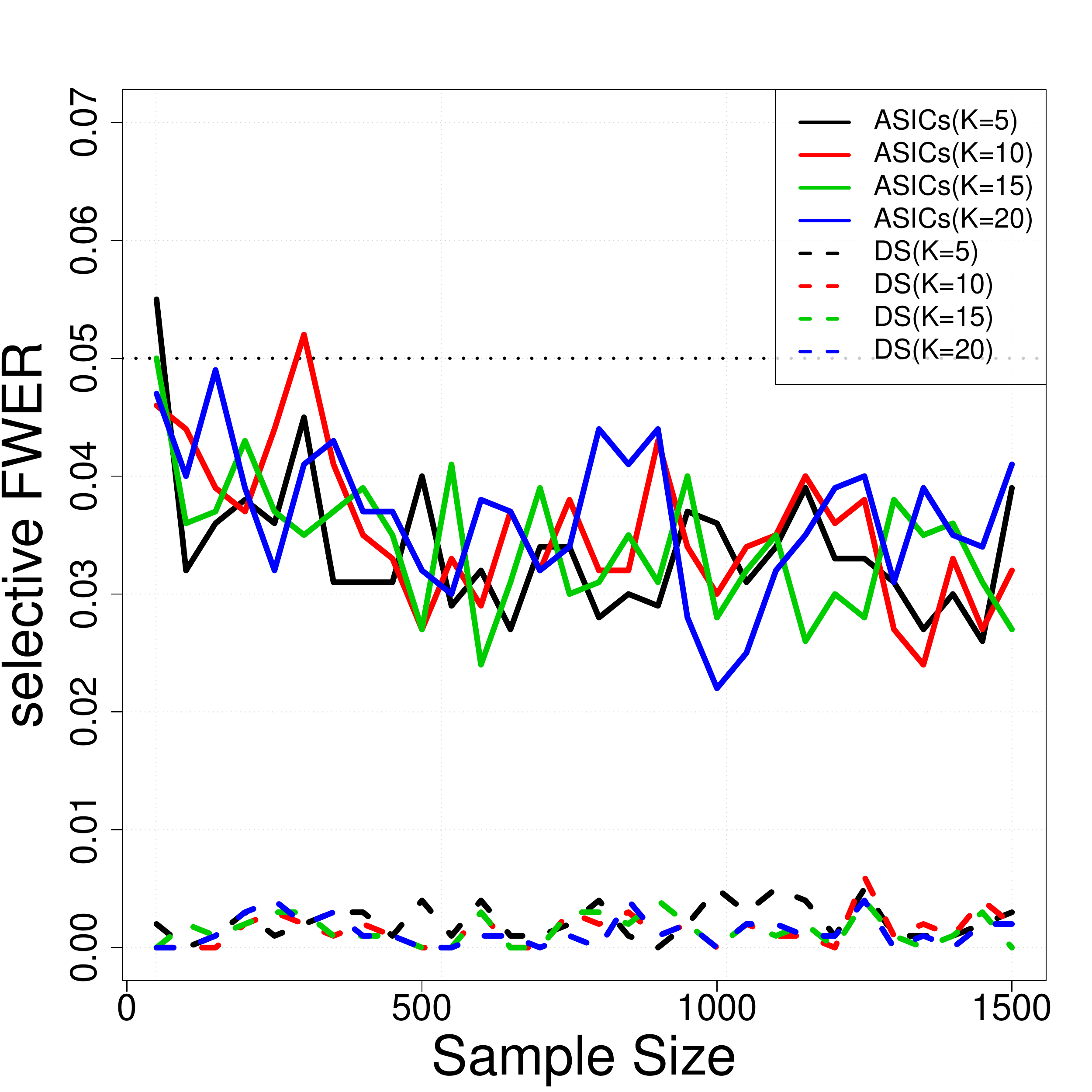}}
\caption{Method comparison using simulated data based on 1,000 Monte-Carlo runs.
The vertical and horizontal axes represent an average of (\ref{eq;FWER.est}) and sample size, respectively. 
The dotted line shows the significance level ($\alpha=0.05$).}
\label{fig:FWER}
\end{center}
\end{figure*}

Figure \ref{fig:FWER} shows the average (\ref{eq;FWER.est}) for each method.
ASICs and DS are both evaluated with respect to four values of $K$, thus eight lines are plotted in each graph.
Because of the randomness of simulation, some of the ASICs results are larger than 0.05 especially in small sample size and large variable dimension cases.
For both methods, it is clear that selective FWER tends to be controlled at the desired significance level, although DS is more conservative than ASICs.
To accord with our asymptotic theory, the number of selected variables must be $K={\rm o}(n^{1/3})$, which means that the normal approximation is not ensured in the case of $K=15$ and $20$.
However, we observe that selective FWER is correctly controlled even in these cases, which suggests that assumptions (C3) and (C4) can be relaxed.

\begin{figure*}[t]
\begin{center}
Case 1
\end{center}
\vspace{-35pt}
\begin{center}
\subfloat[$d=200$]{\includegraphics[scale=0.16, bb=0 0 720 720]{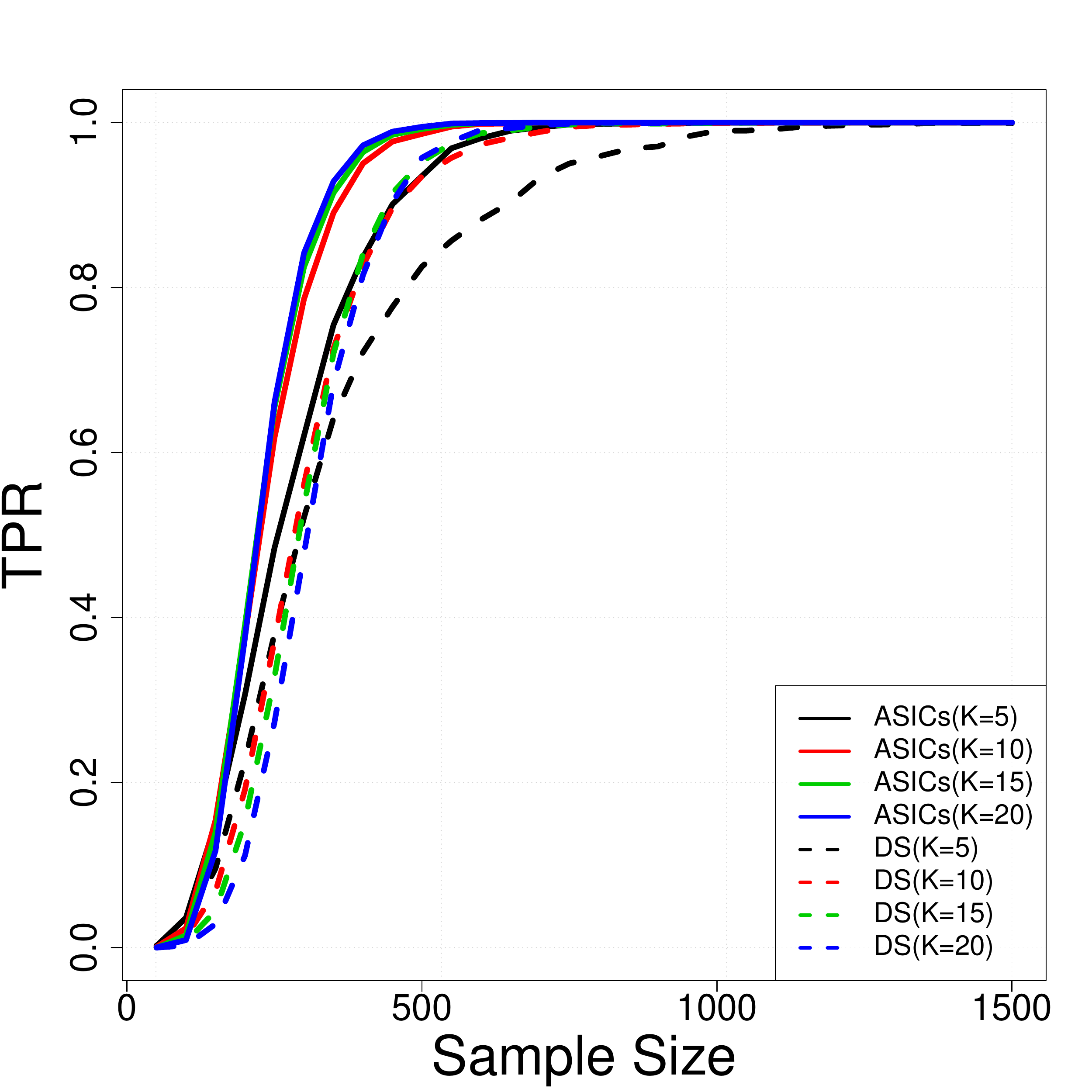}}
\subfloat[$d=500$]{\includegraphics[scale=0.16, bb=0 0 720 720]{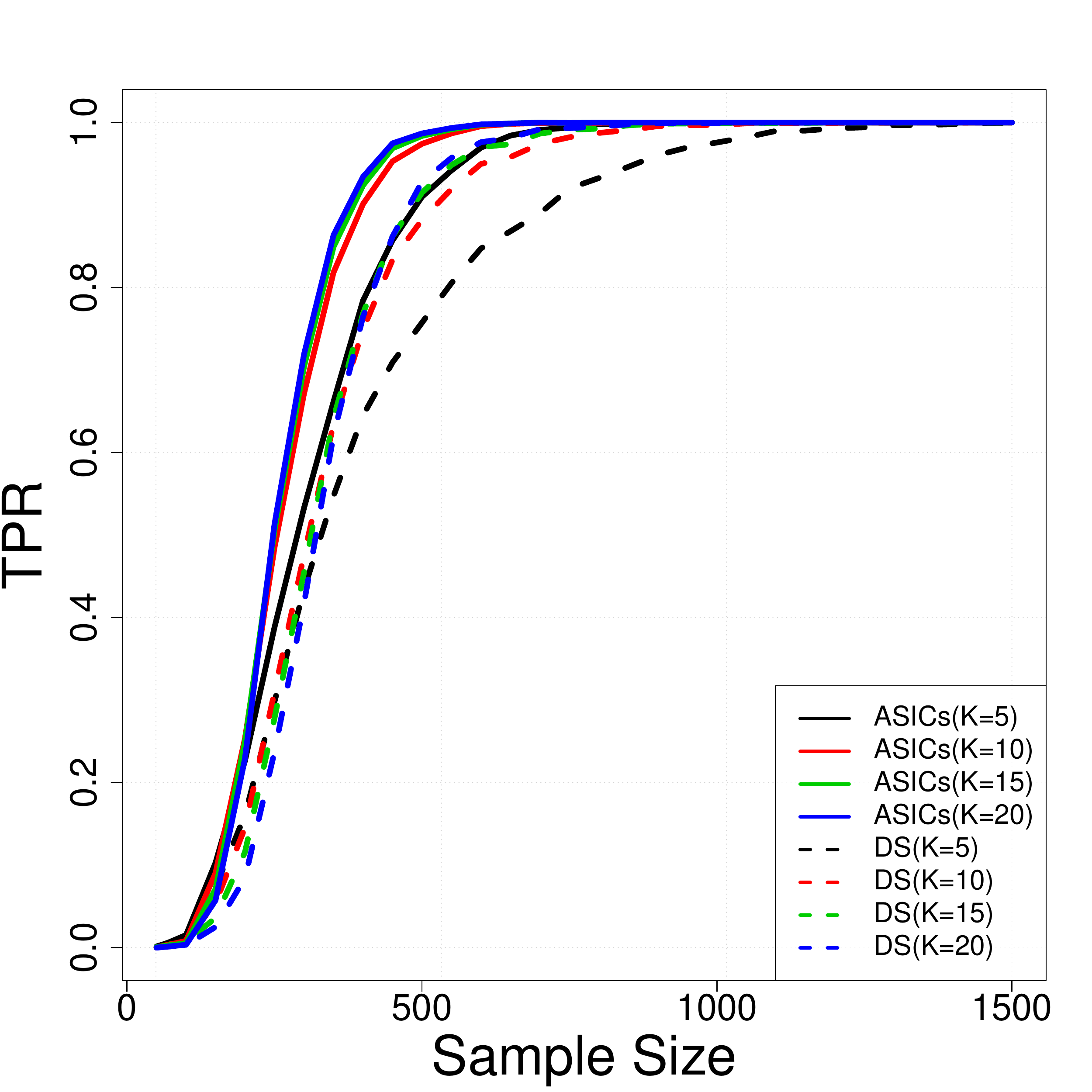}} 
\subfloat[$d=1,000$]{\includegraphics[scale=0.16, bb=0 0 720 720]{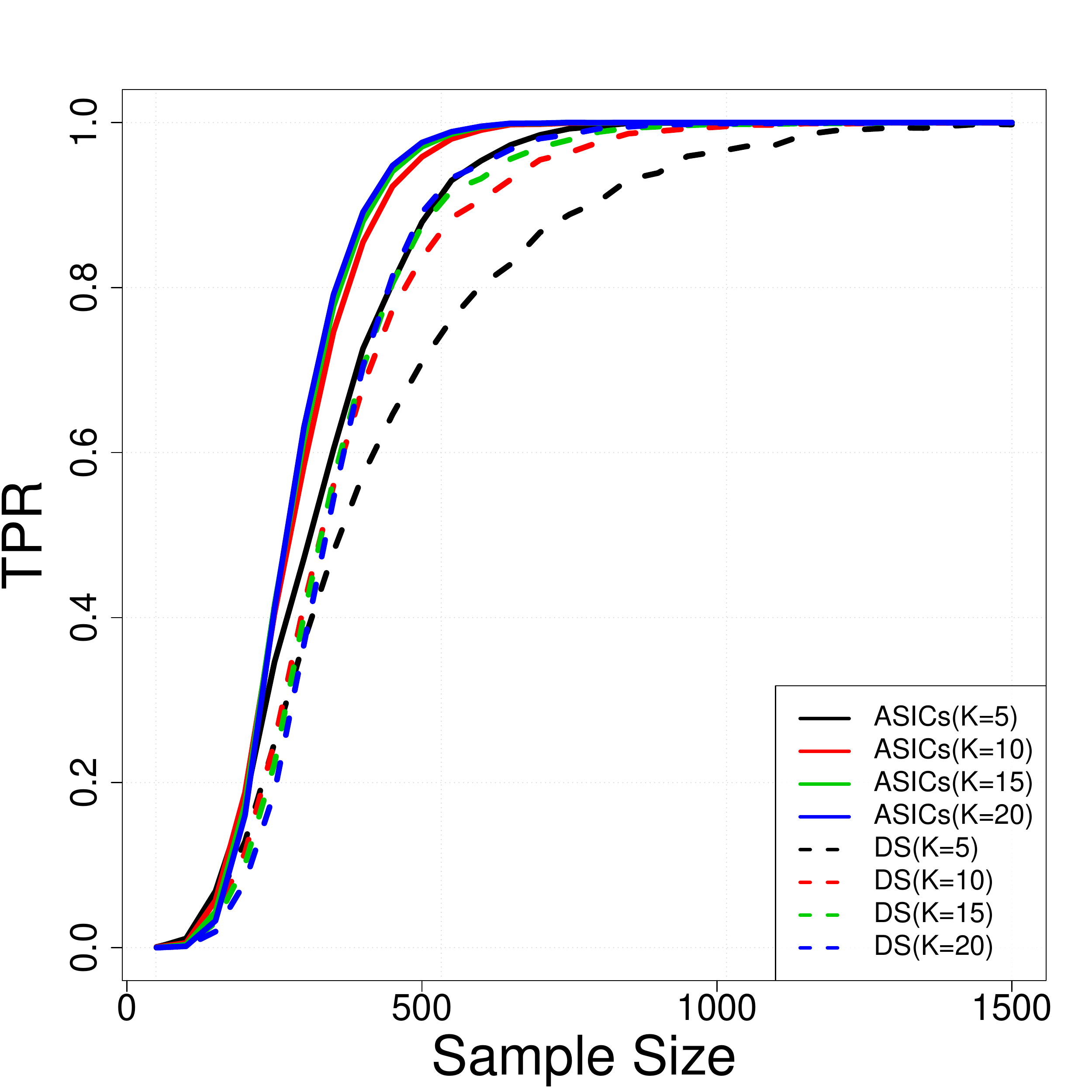}}
\end{center}
\begin{center}
Case 2
\end{center}
\vspace{-35pt}
\begin{center}
\subfloat[$d=200$]{\includegraphics[scale=0.16, bb=0 0 720 720]{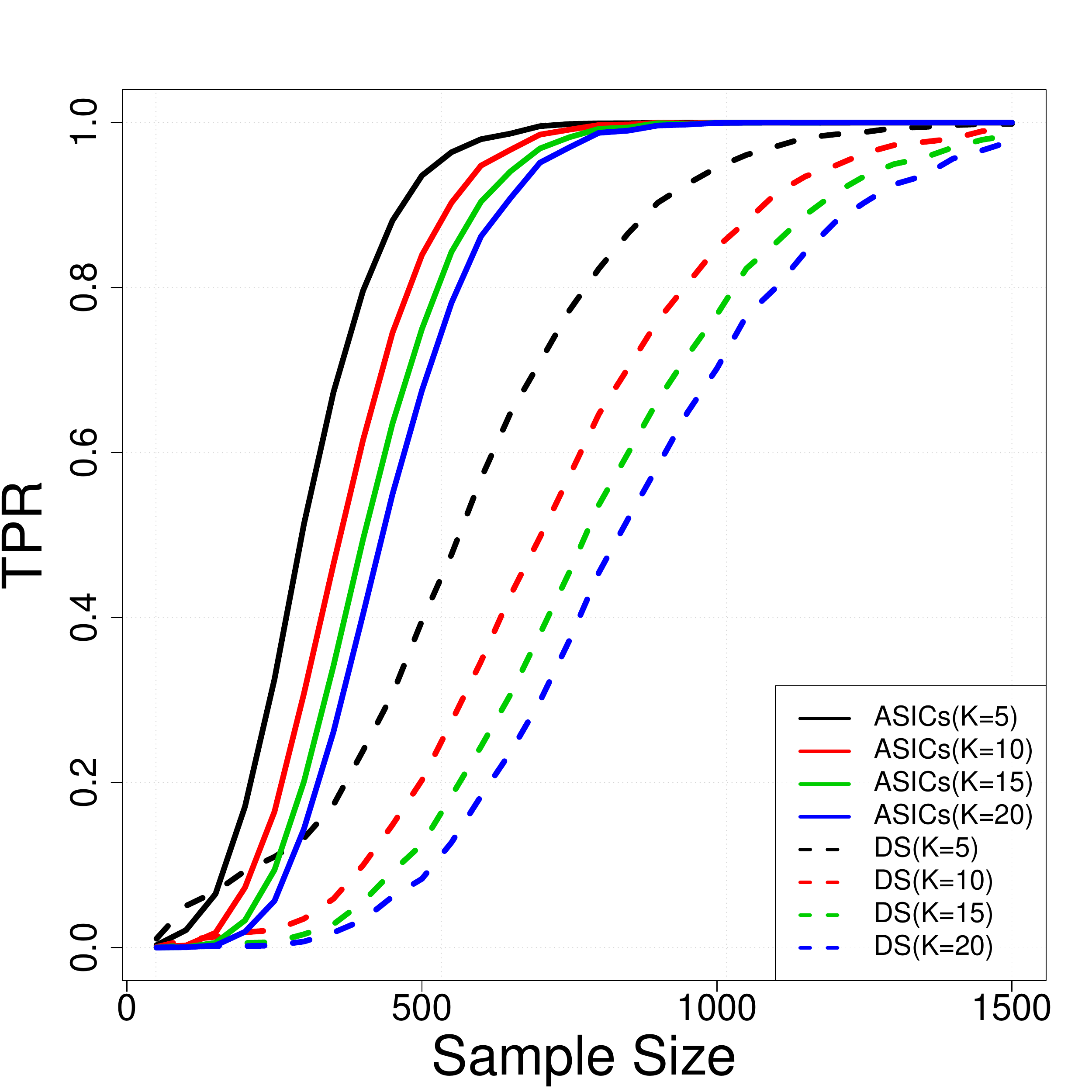}}
\subfloat[$d=500$]{\includegraphics[scale=0.16, bb=0 0 720 720]{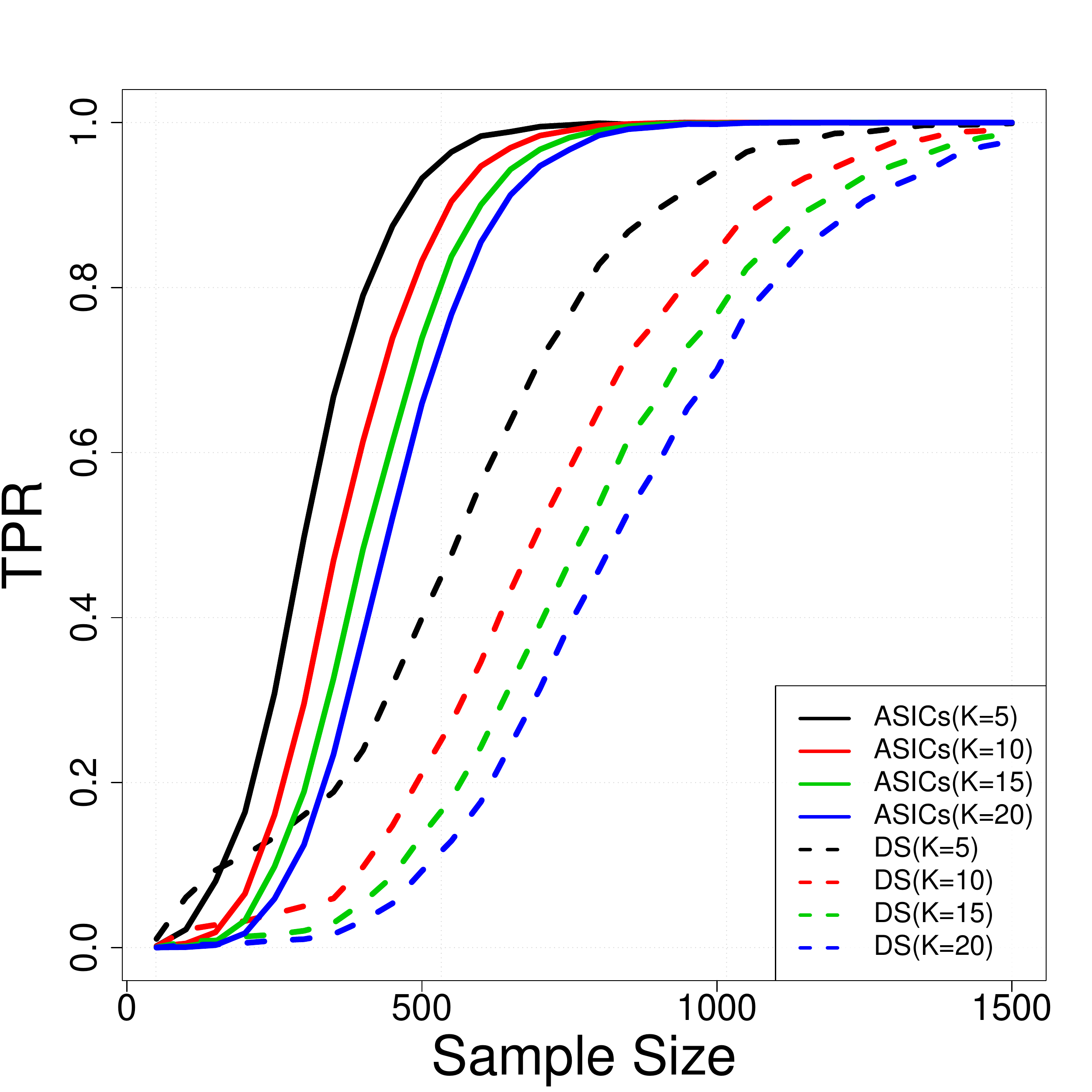}} 
\subfloat[$d=1,000$]{\includegraphics[scale=0.16, bb=0 0 720 720]{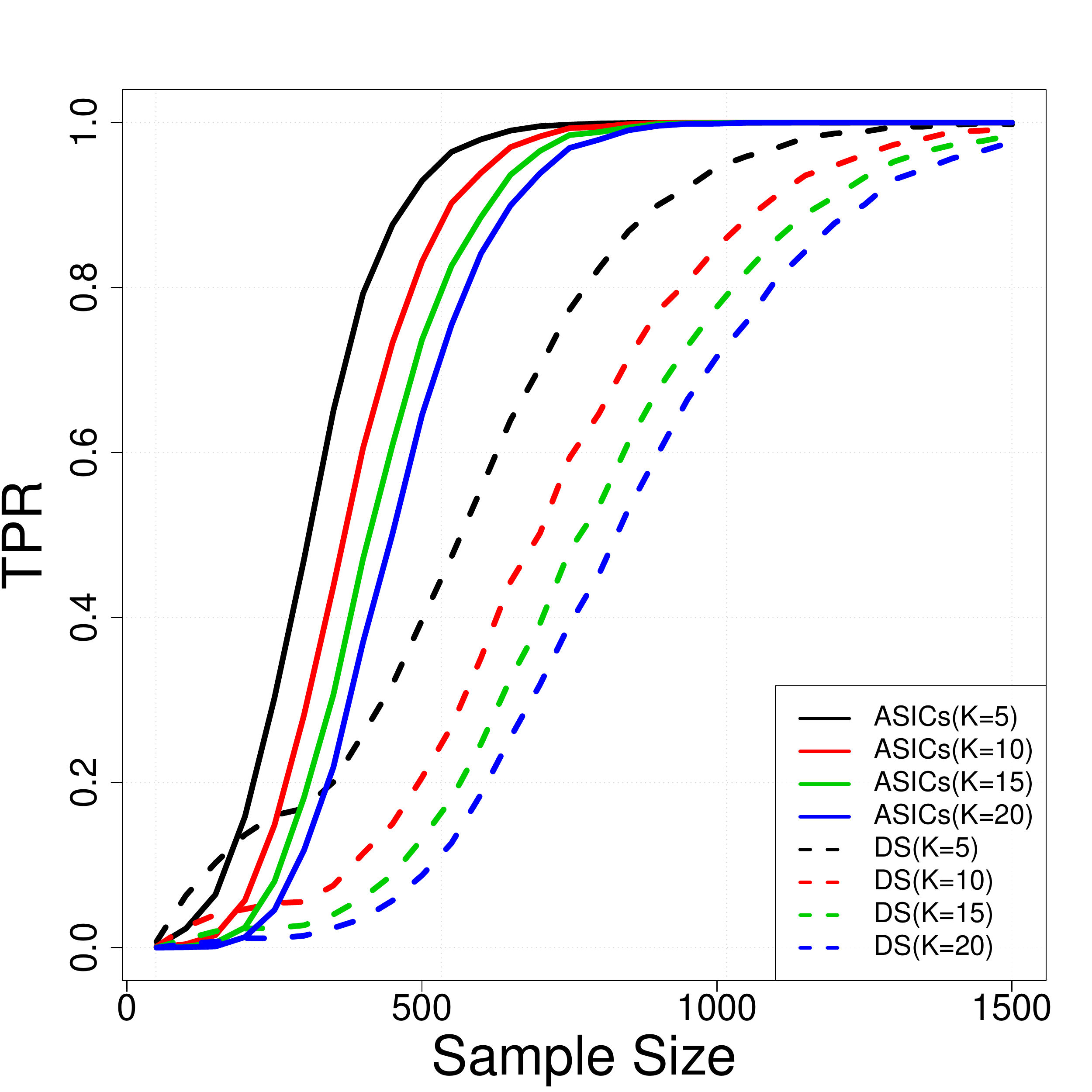}}
\caption{Method comparison using simulated data based on 1,000 Monte-Carlo runs.
The vertical and horizontal axes represent an average of (\ref{eq;TPR}) and sample size, respectively.}
\label{fig:Power1}
\end{center}
\end{figure*}

\begin{figure*}[t]
\begin{center}
Case 1
\end{center}
\vspace{-35pt}
\begin{center}
\subfloat[$d=200$]{\includegraphics[scale=0.16, bb=0 0 720 720]{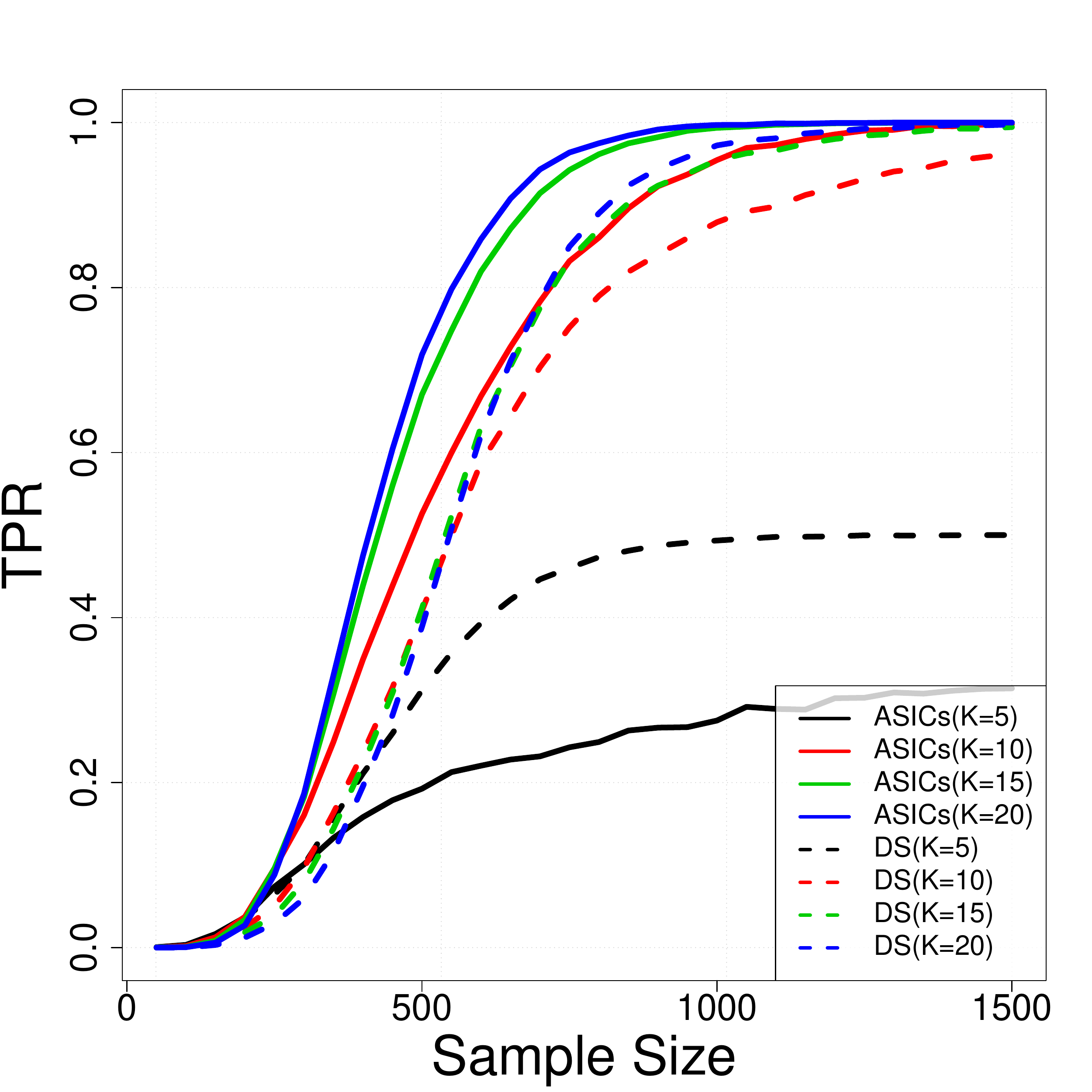}}
\subfloat[$d=500$]{\includegraphics[scale=0.16, bb=0 0 720 720]{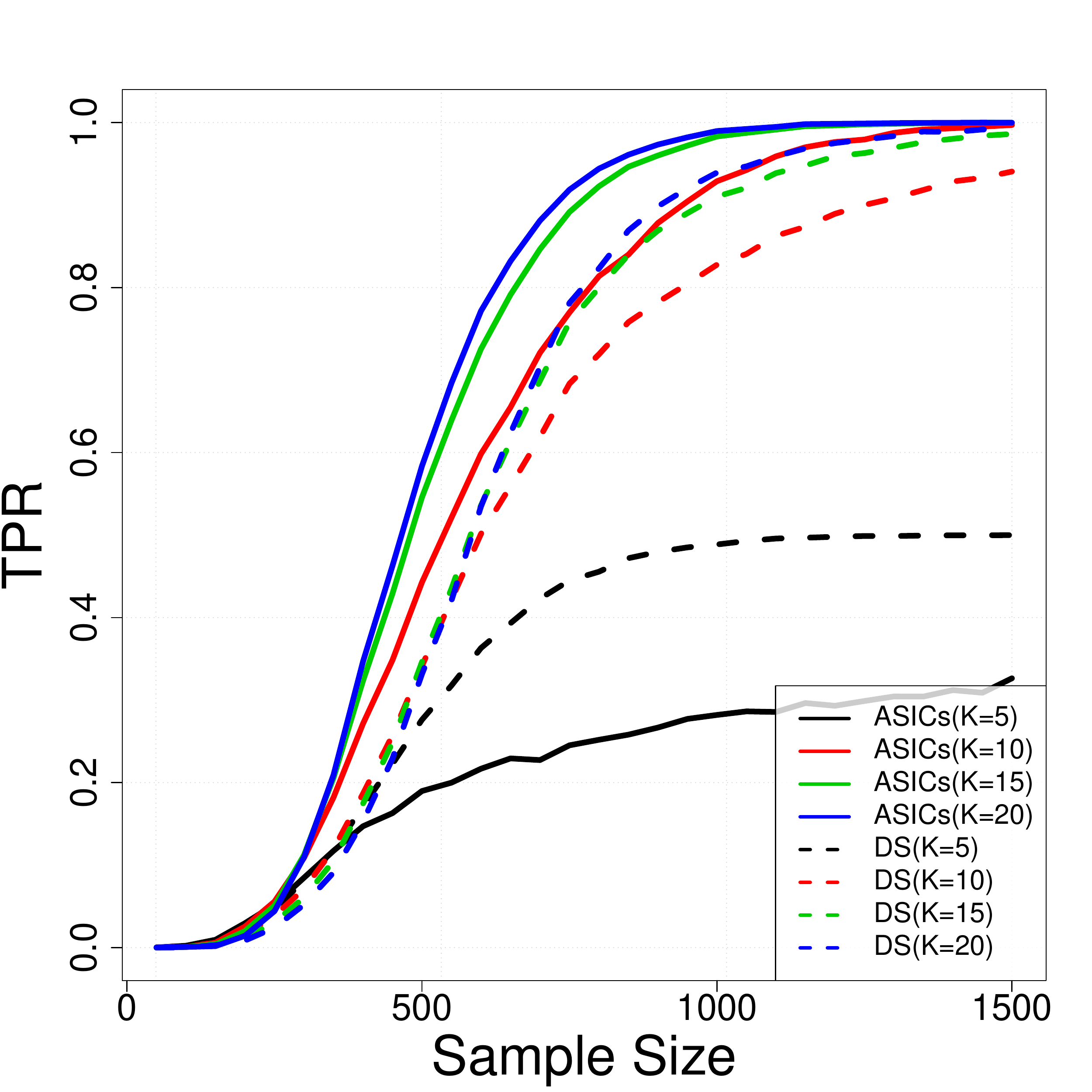}} 
\subfloat[$d=1,000$]{\includegraphics[scale=0.16, bb=0 0 720 720]{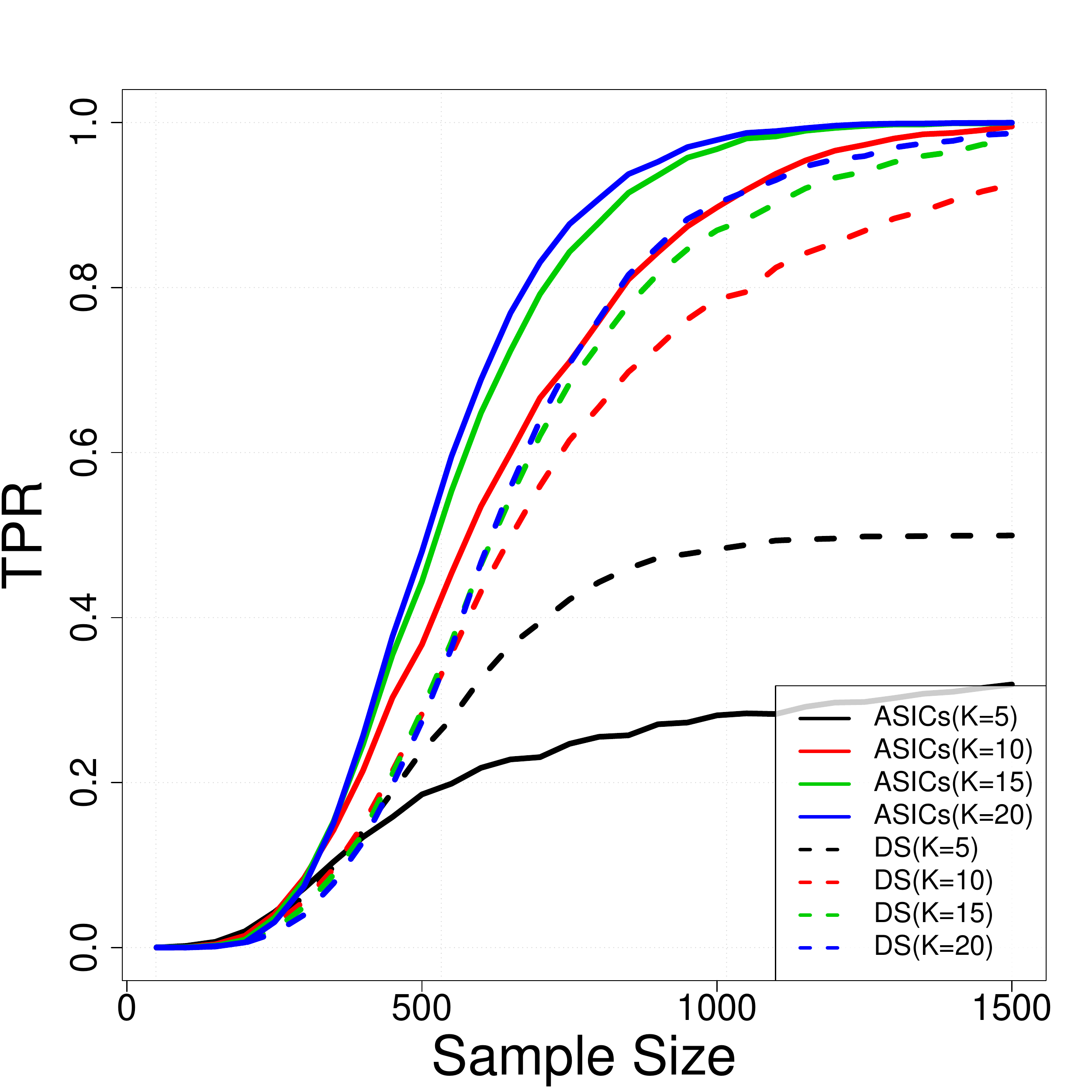}}
\end{center}
\begin{center}
Case 2
\end{center}
\vspace{-35pt}
\begin{center}
\subfloat[$d=200$]{\includegraphics[scale=0.16, bb=0 0 720 720]{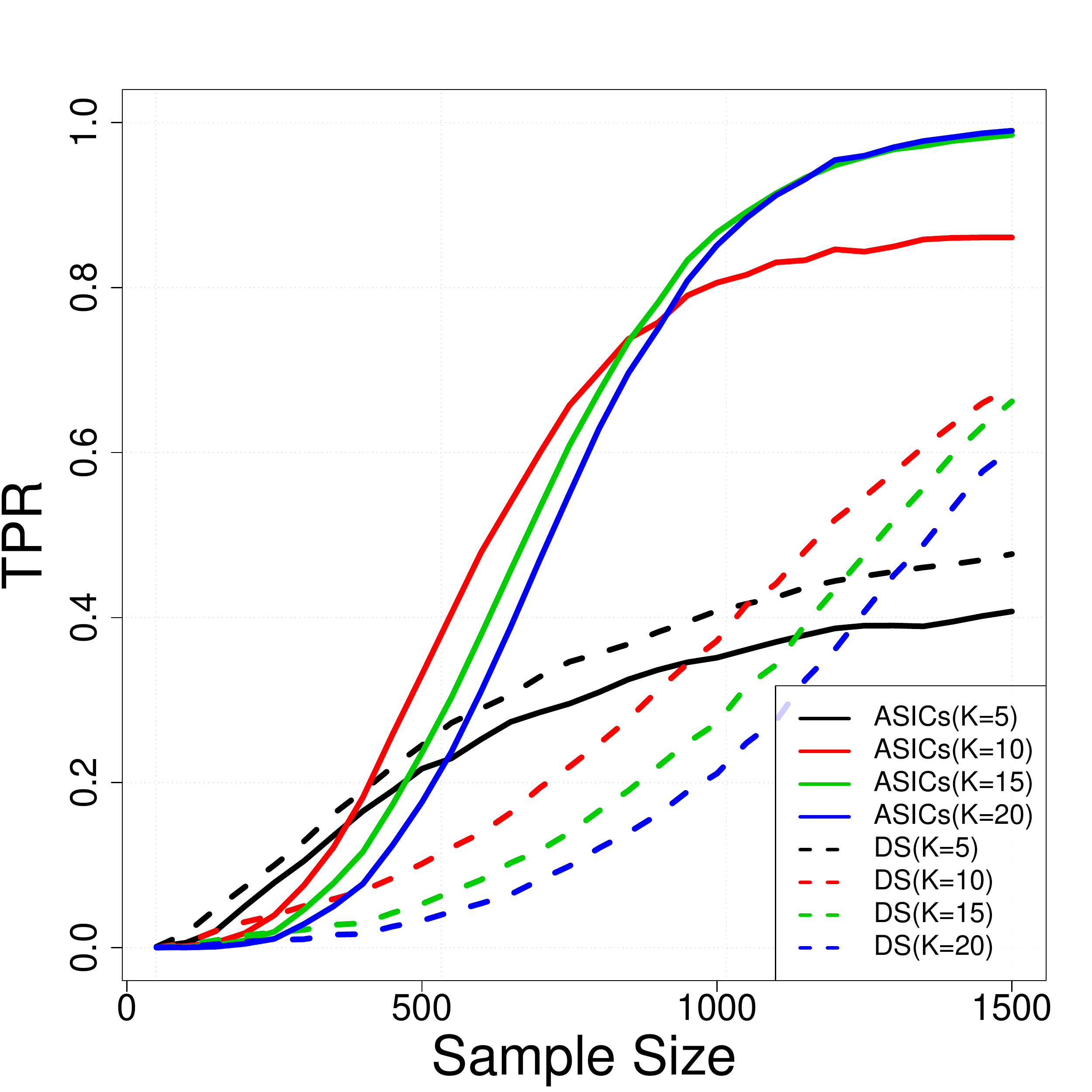}}
\subfloat[$d=500$]{\includegraphics[scale=0.16, bb=0 0 720 720]{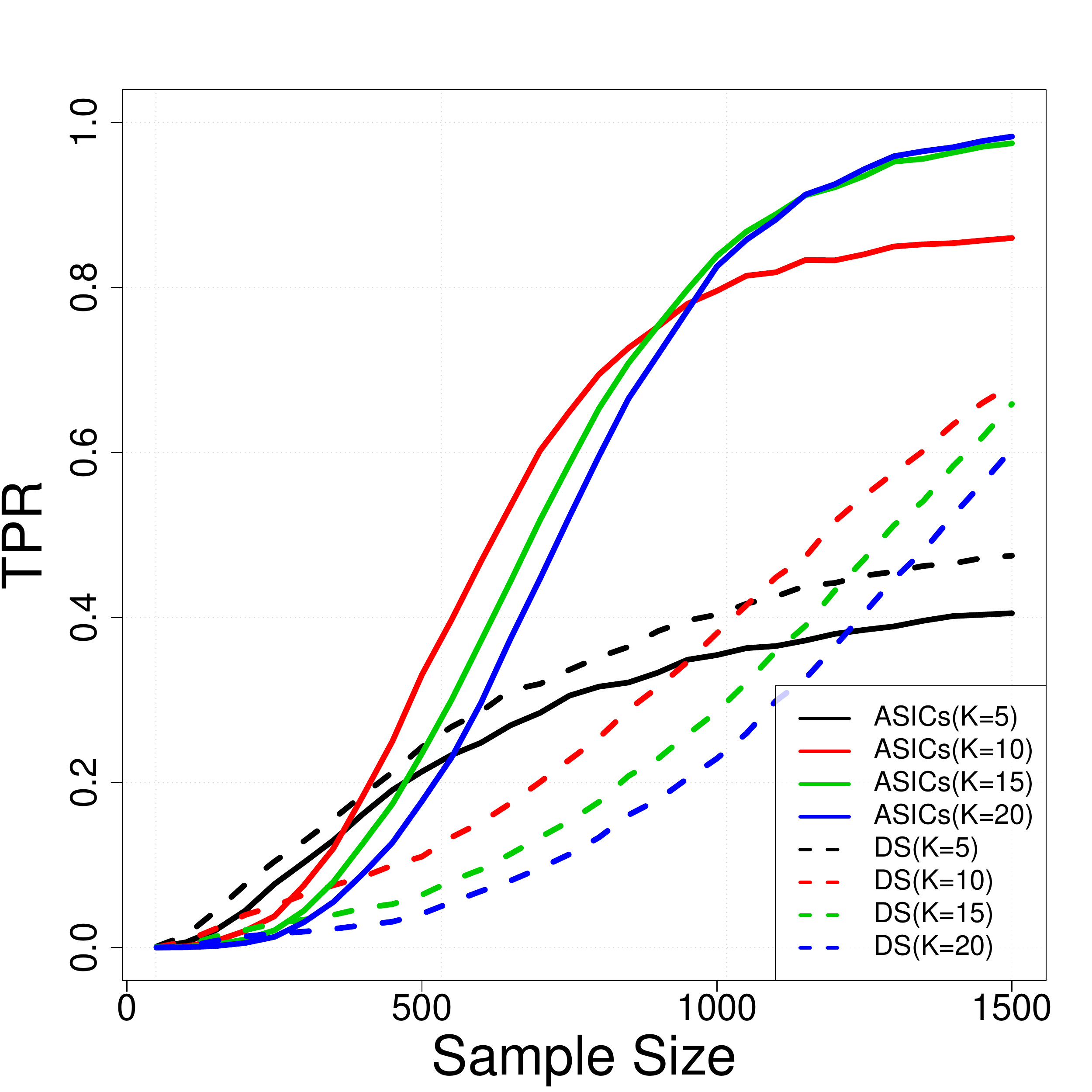}} 
\subfloat[$d=1,000$]{\includegraphics[scale=0.16, bb=0 0 720 720]{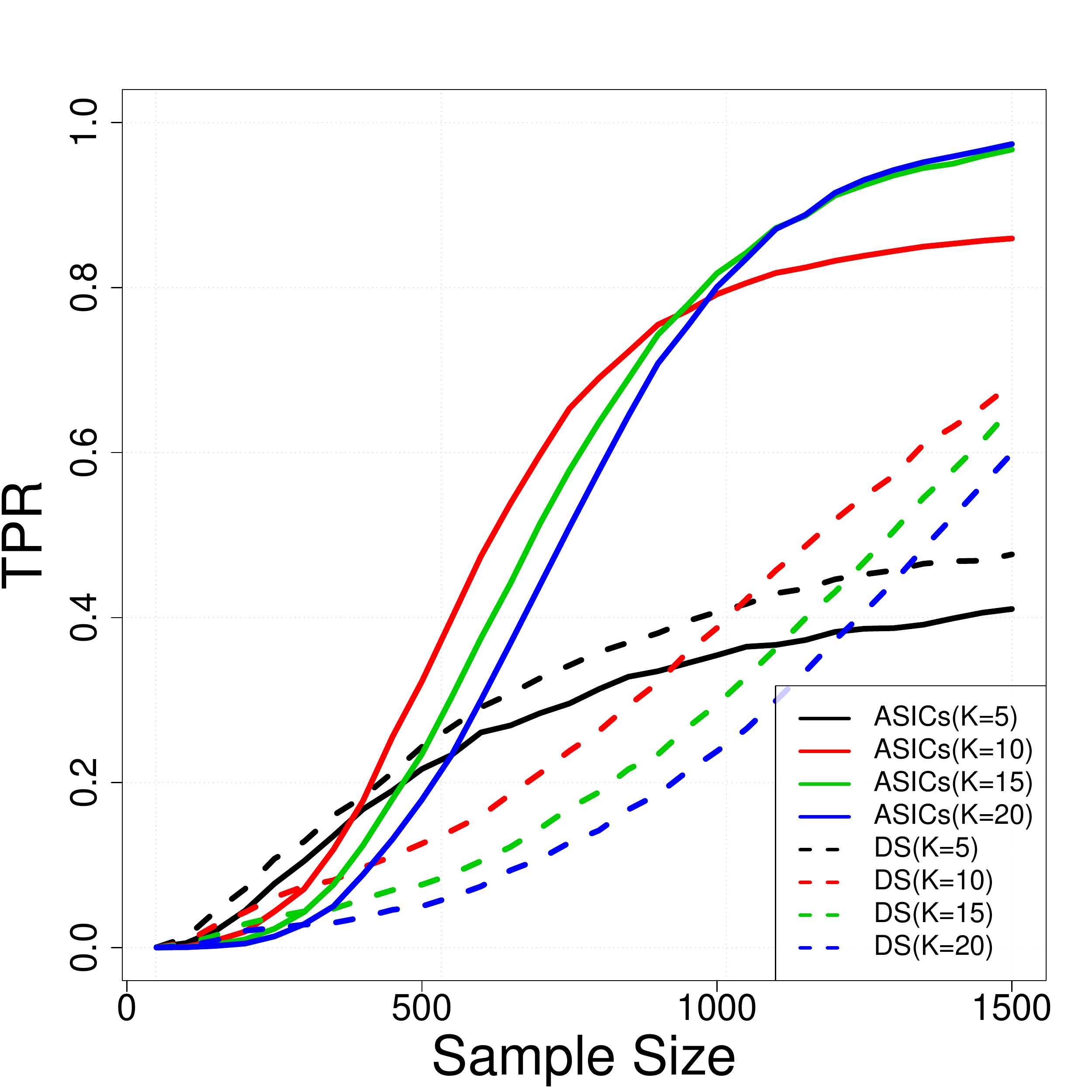}}
\caption{Method comparison using simulated data based on 1,000 Monte-Carlo runs.
The vertical and horizontal axes represent an average of (\ref{eq;TPR}) and sample size, respectively.}
\label{fig:Power2}
\end{center}
\end{figure*}

Figures \ref{fig:Power1} and \ref{fig:Power2} show the average of (\ref{eq;TPR}) for each method and settings in Model 1 and Model 2, respectively.
In Case 1 of Figure \ref{fig:Power1}, ASICs and DS have almost the same power for each $K$ and $d$.
In addition, ASICs is clearly superior to DS in Case 2. 
This is reasonable since DS uses only the half of the data for inference.
On the other hand, in all cases, the power of ASICs becomes higher as the number of selected variables $K$ decreases.
This can be explained by the condition (C3), that is, we need a much larger sample size when $K$ becomes large for assuring the asymptotic result in Theorem \ref{thm2}.
In Figure \ref{fig:Power2}, it is clear that the power of ASICs is superior in almost all settings.
However, neither AISCs nor DS appears to perform well when $K=5$.
In this case, the power of ASICs and DS cannot be improved by $50\%$ or more.
This is because we can only select at most $5$ true nonzero variables, while there are $10$ true nonzero variables.

\section{Empirical Applications}
\label{sec:real}
\begin{figure*}[h!]
\centering
{\bf Dexter Data Set} ($n=600, d=20,000$)
\vskip -10pt
\subfloat[$K=5$]{\includegraphics[scale=0.165, bb=0 0 1080 576]{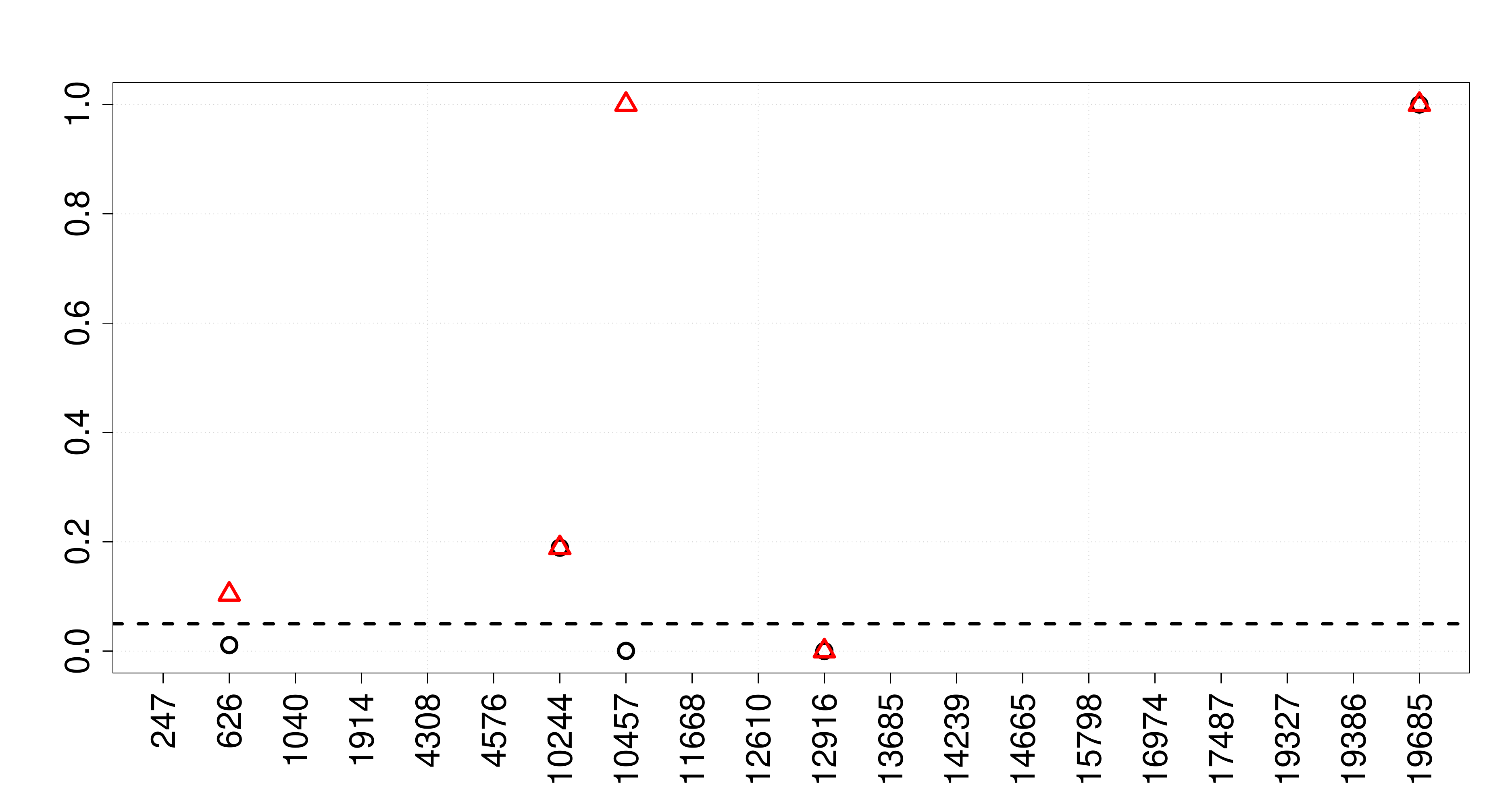}}
\subfloat[$K=10$]{\includegraphics[scale=0.165, bb=0 0 1080 576]{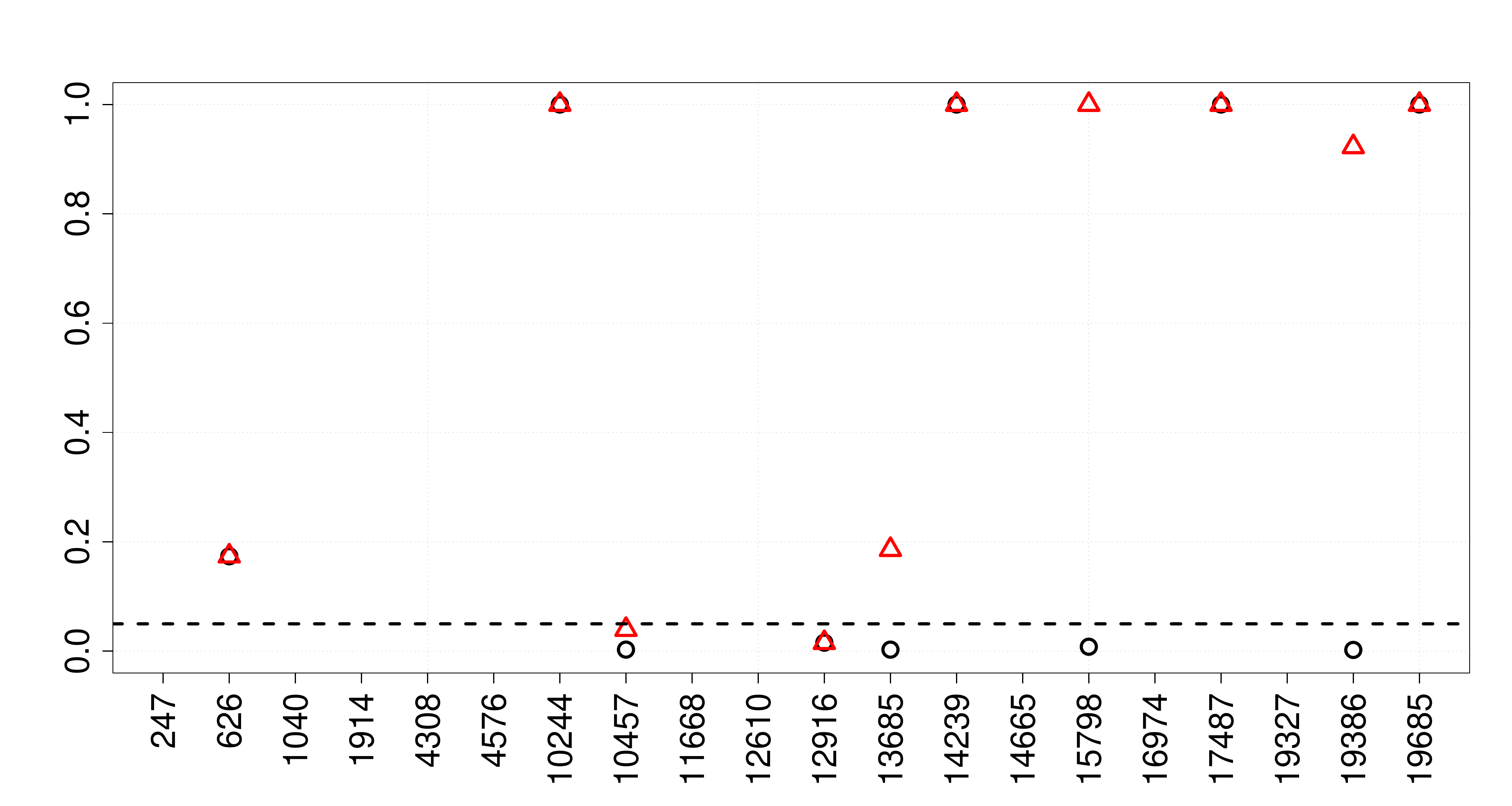}} \\
\subfloat[$K=15$]{\includegraphics[scale=0.165, bb=0 0 1080 576]{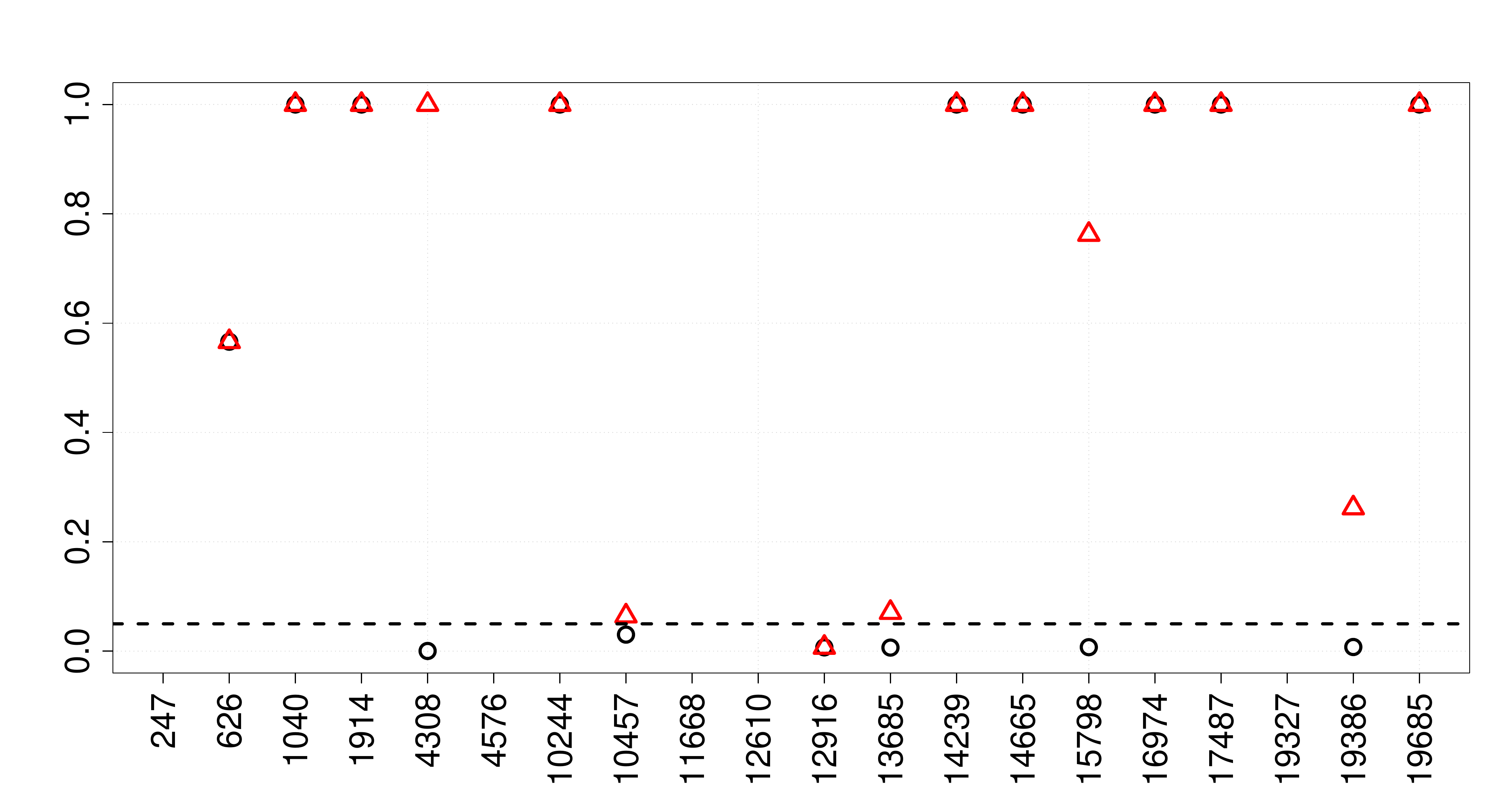}}
\subfloat[$K=20$]{\includegraphics[scale=0.165, bb=0 0 1080 576]{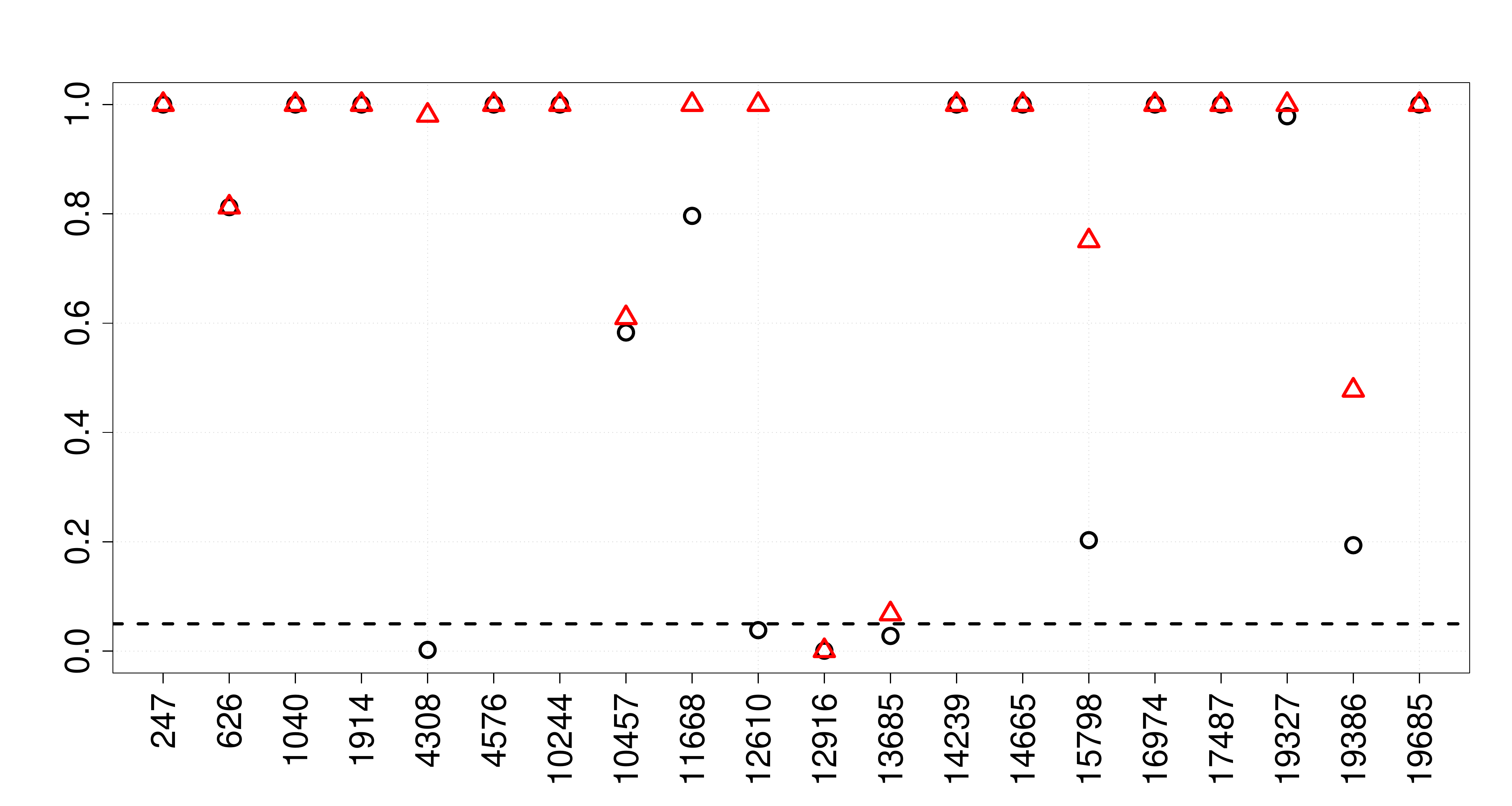}}
\caption{Comparison between adjusted selective $p$-values and nominal $p$-values.
The vertical and horizontal axes represent adjusted $p$-values and indices of selected variables, respectively, and the black dotted line shows the significance level ($\alpha=0.05$).
In each figure, black circles and red triangles respectively indicate adjusted nominal $p$-values and selective $p$-values.
}
\label{fig:real1}
\end{figure*}

\begin{figure*}[h!]
\centering
{\bf Dorothea Data Set} ($n=1,150, d=100,000$)
\vskip -10pt
\subfloat[$K=5$]{\includegraphics[scale=0.165, bb=0 0 1080 576]{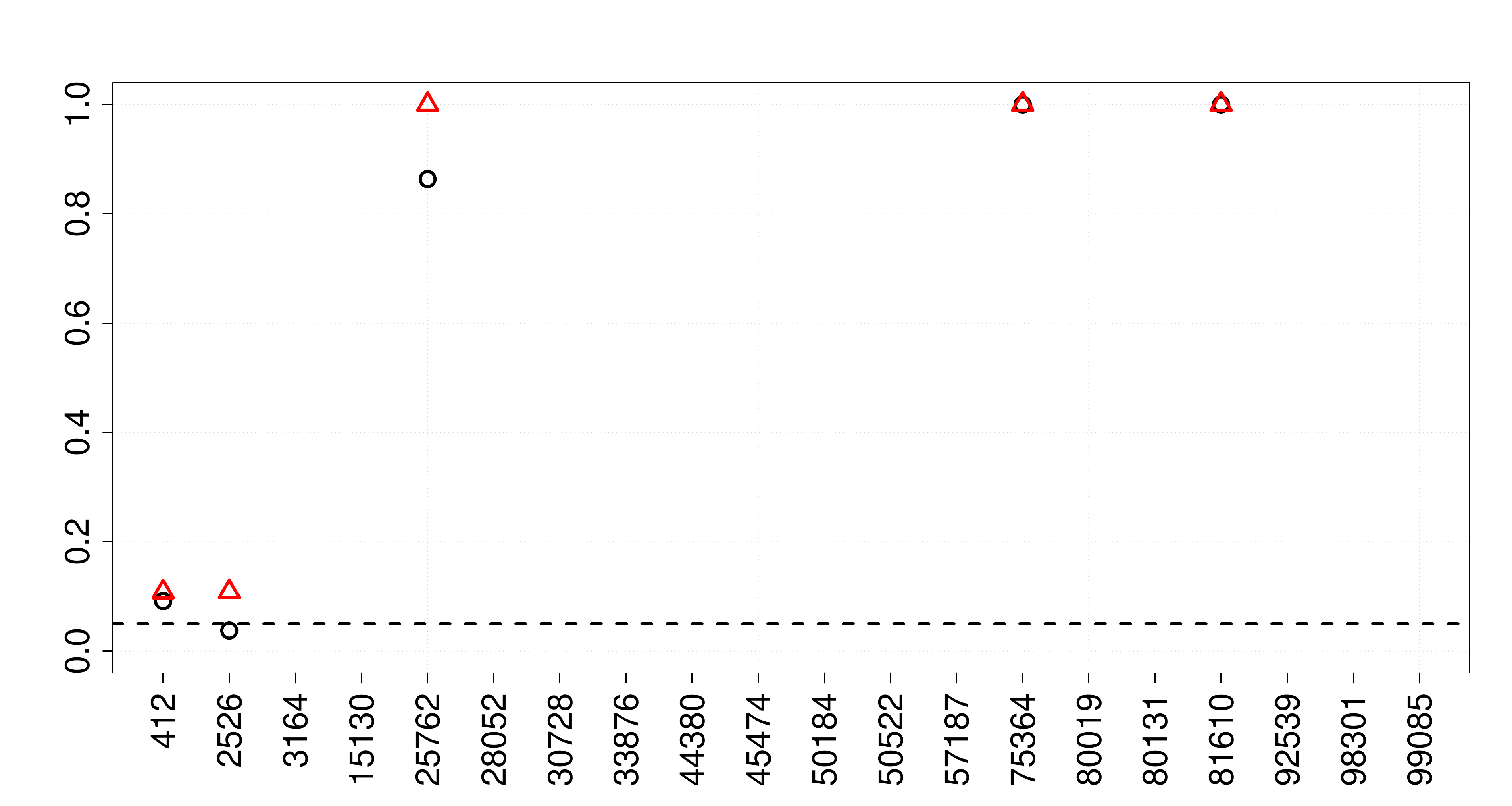}}
\subfloat[$K=10$]{\includegraphics[scale=0.165, bb=0 0 1080 576]{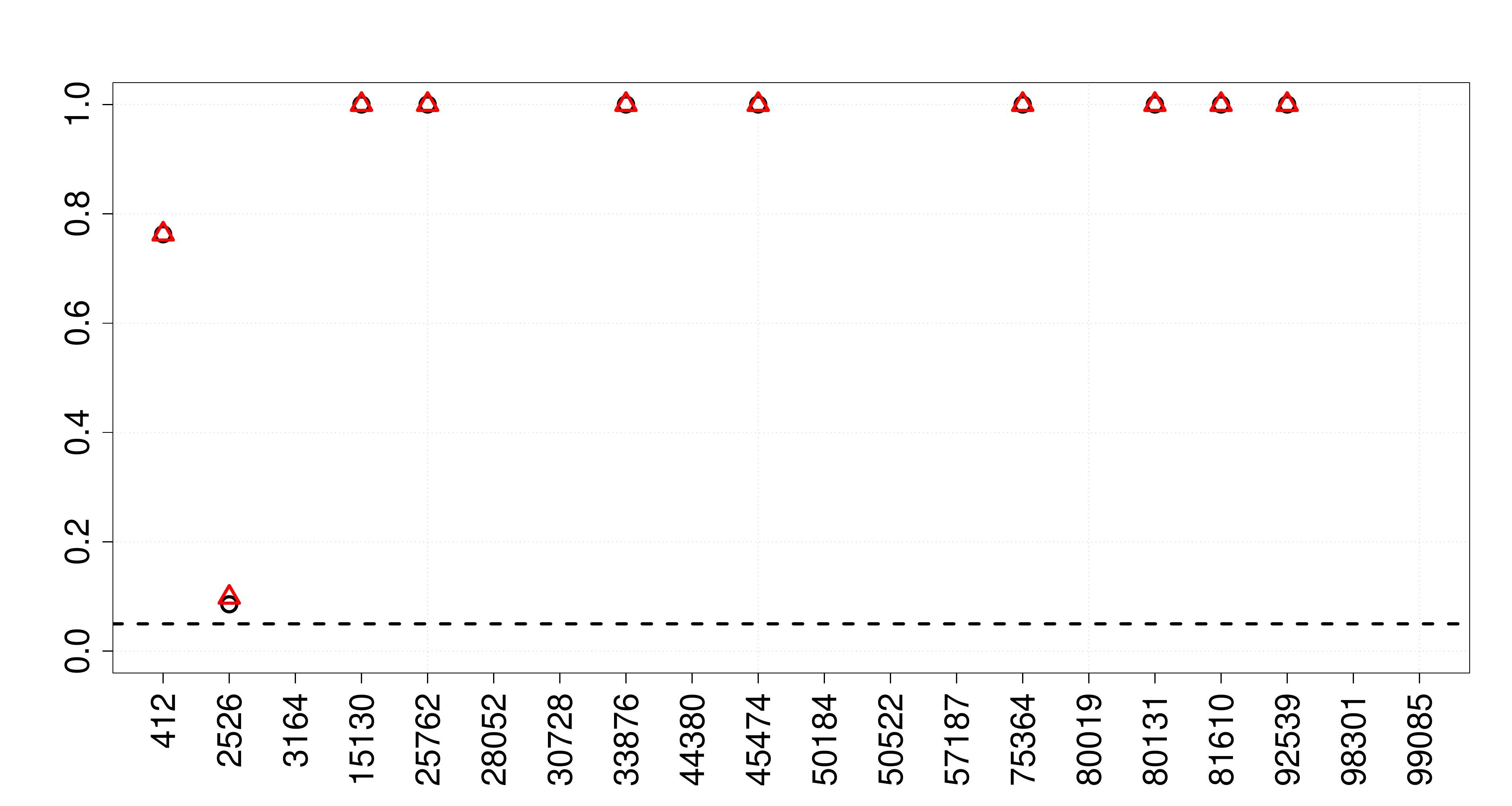}} \\
\subfloat[$K=15$]{\includegraphics[scale=0.165, bb=0 0 1080 576]{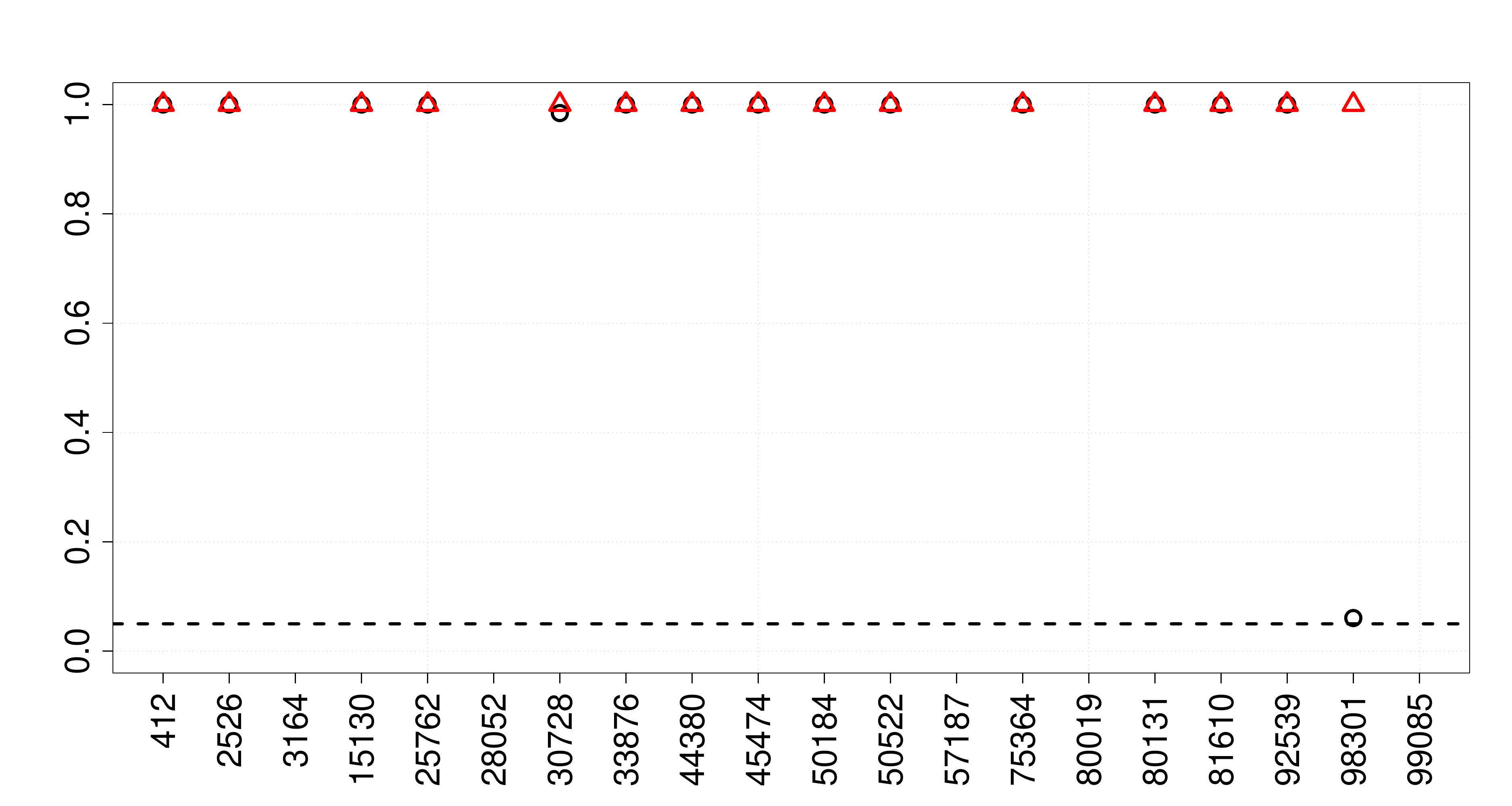}}
\subfloat[$K=20$]{\includegraphics[scale=0.165, bb=0 0 1080 576]{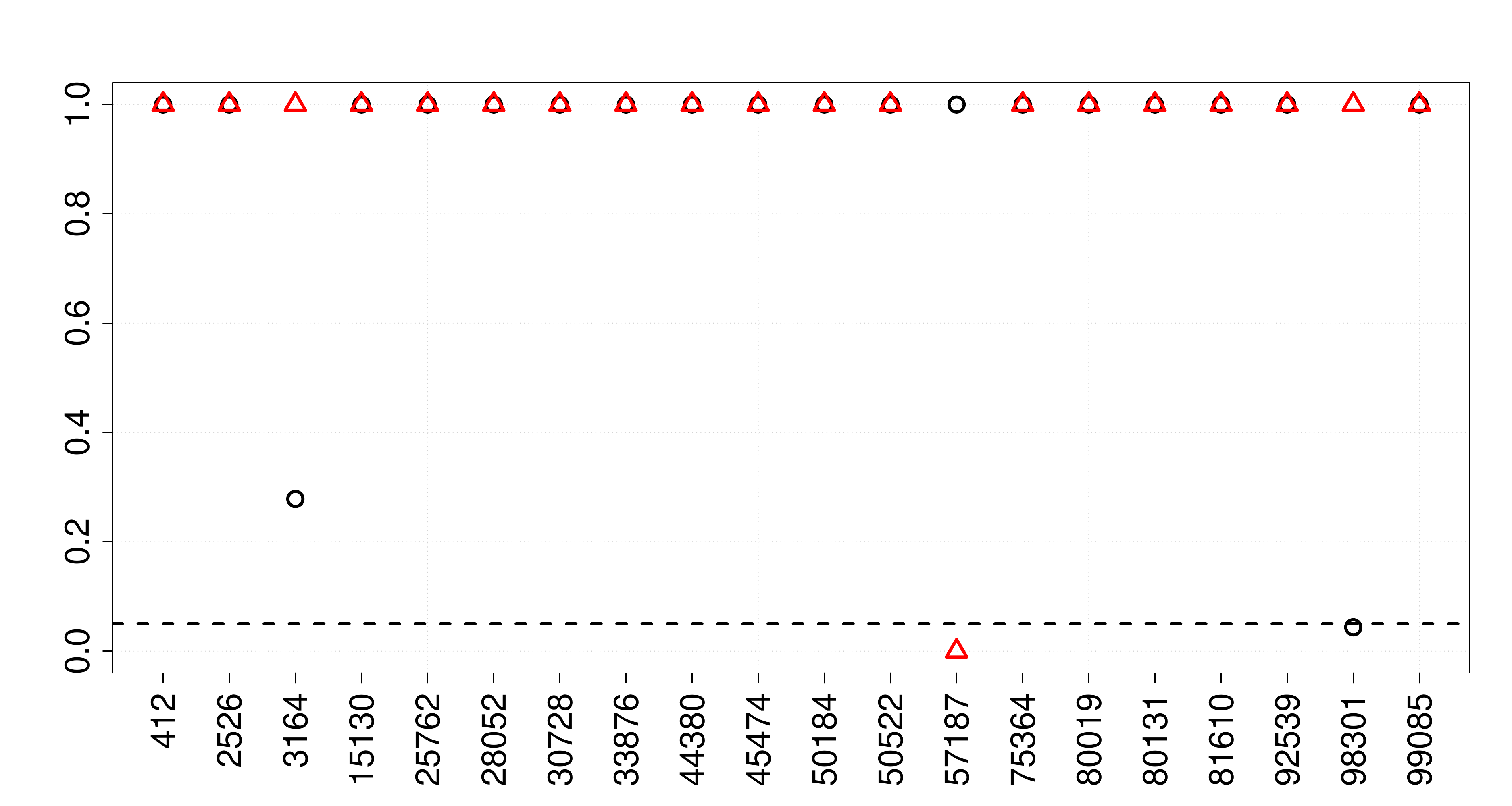}} \\
\vspace{20pt}
{\bf FarmAds Data Set} ($n=4,143, d=54,877$)
\vskip -10pt
\subfloat[$K=5$]{\includegraphics[scale=0.165, bb=0 0 1080 576]{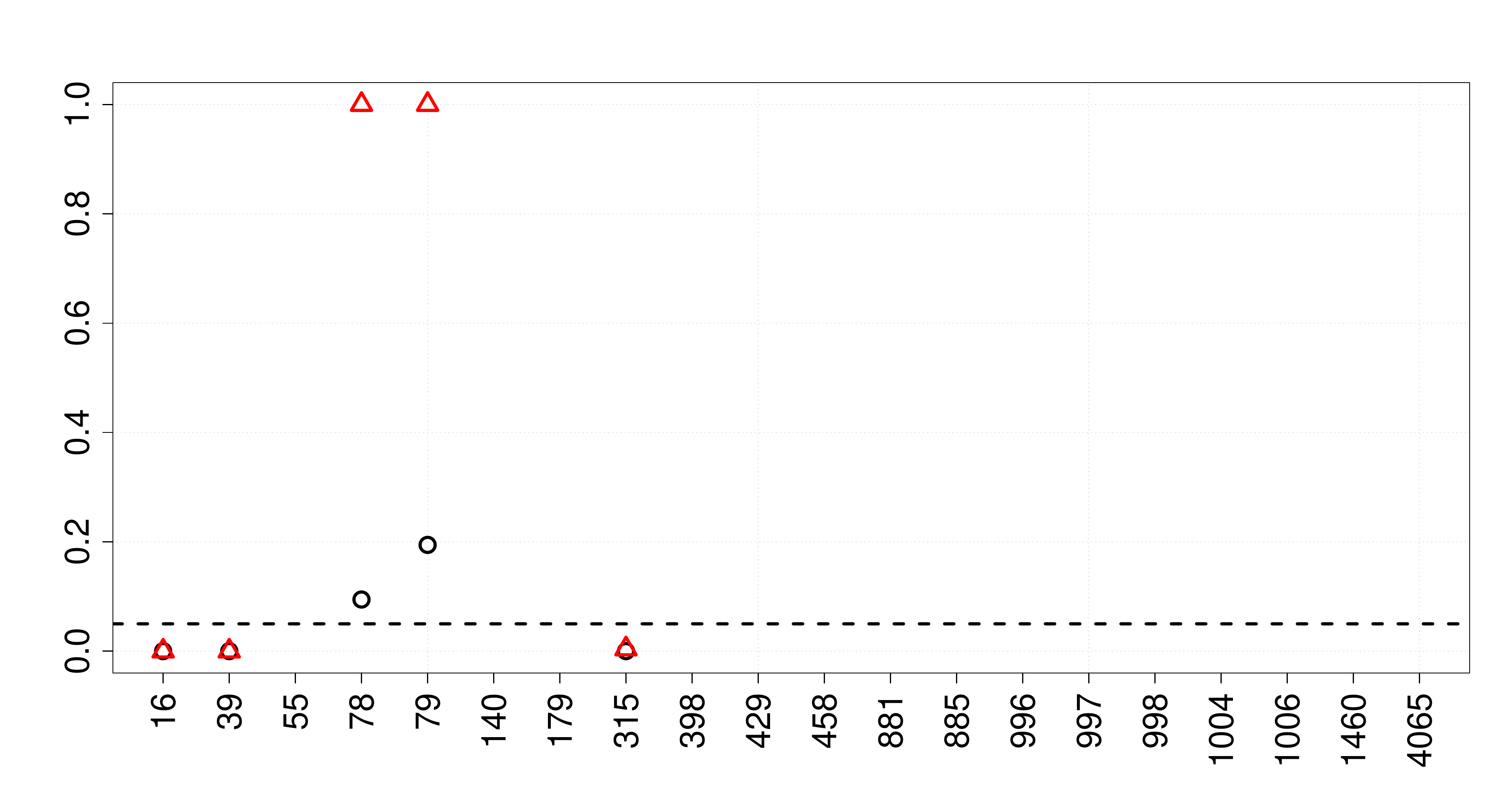}}
\subfloat[$K=10$]{\includegraphics[scale=0.165, bb=0 0 1080 576]{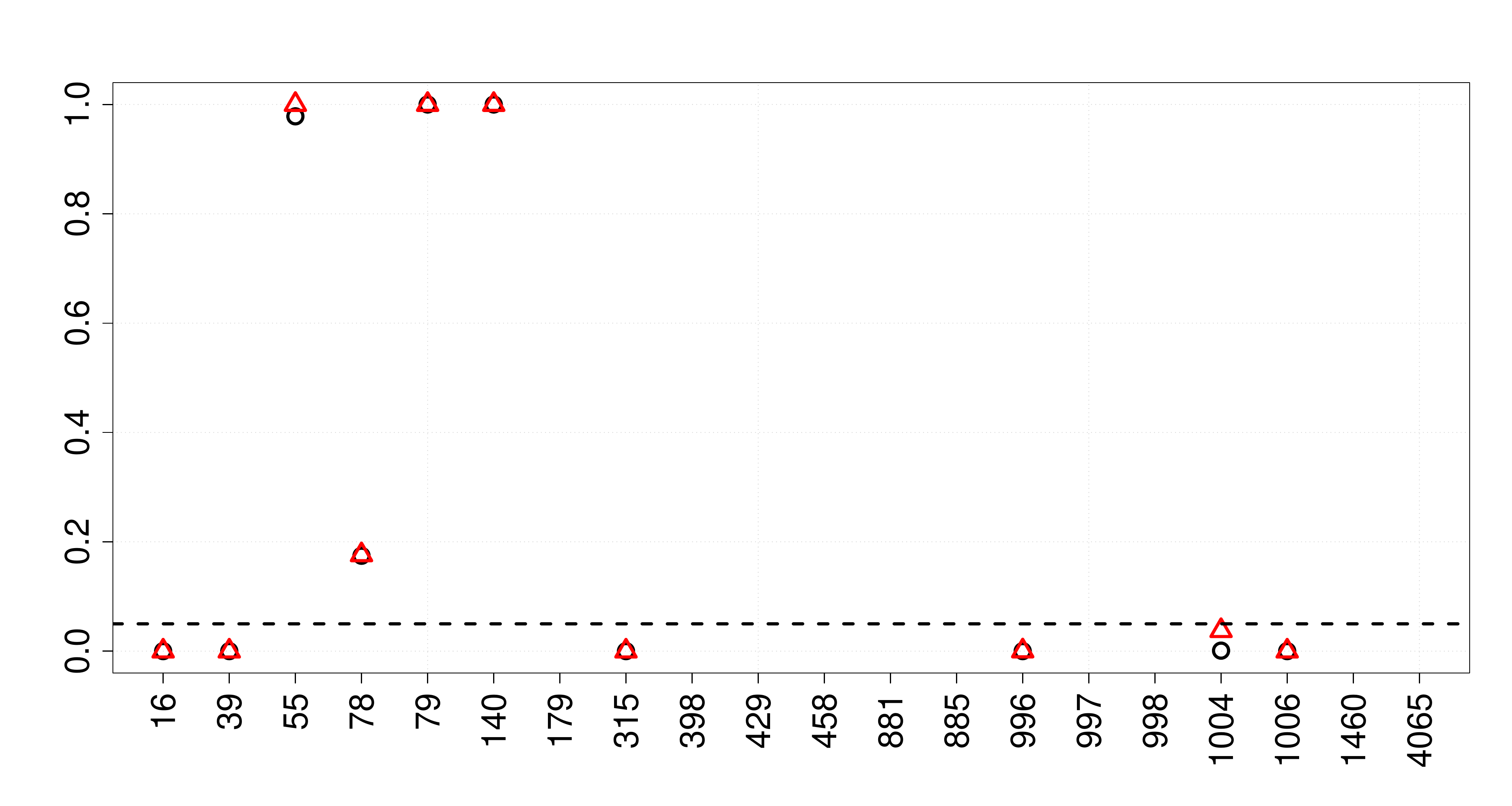}} \\
\subfloat[$K=15$]{\includegraphics[scale=0.165, bb=0 0 1080 576]{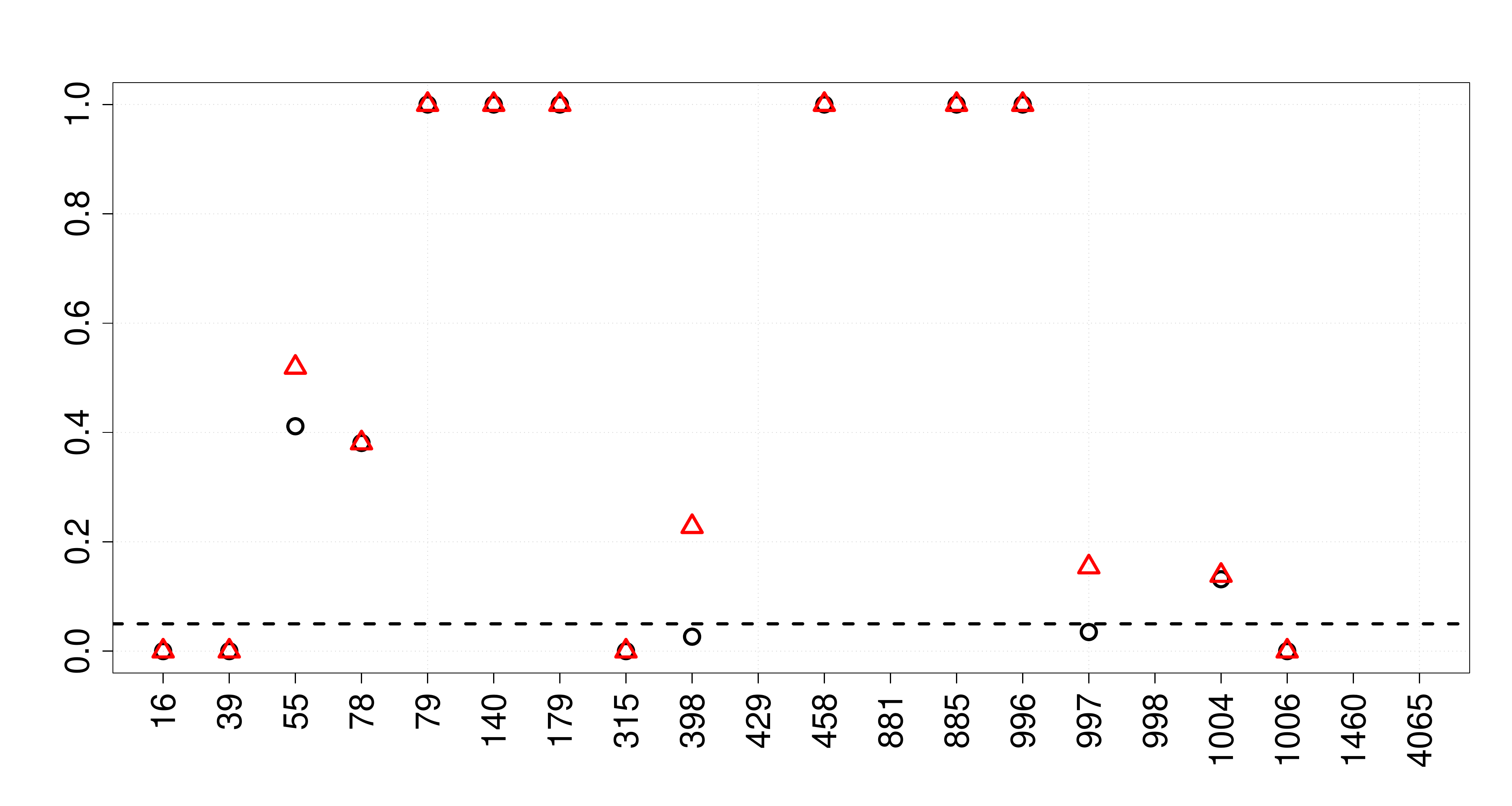}}
\subfloat[$K=20$]{\includegraphics[scale=0.165, bb=0 0 1080 576]{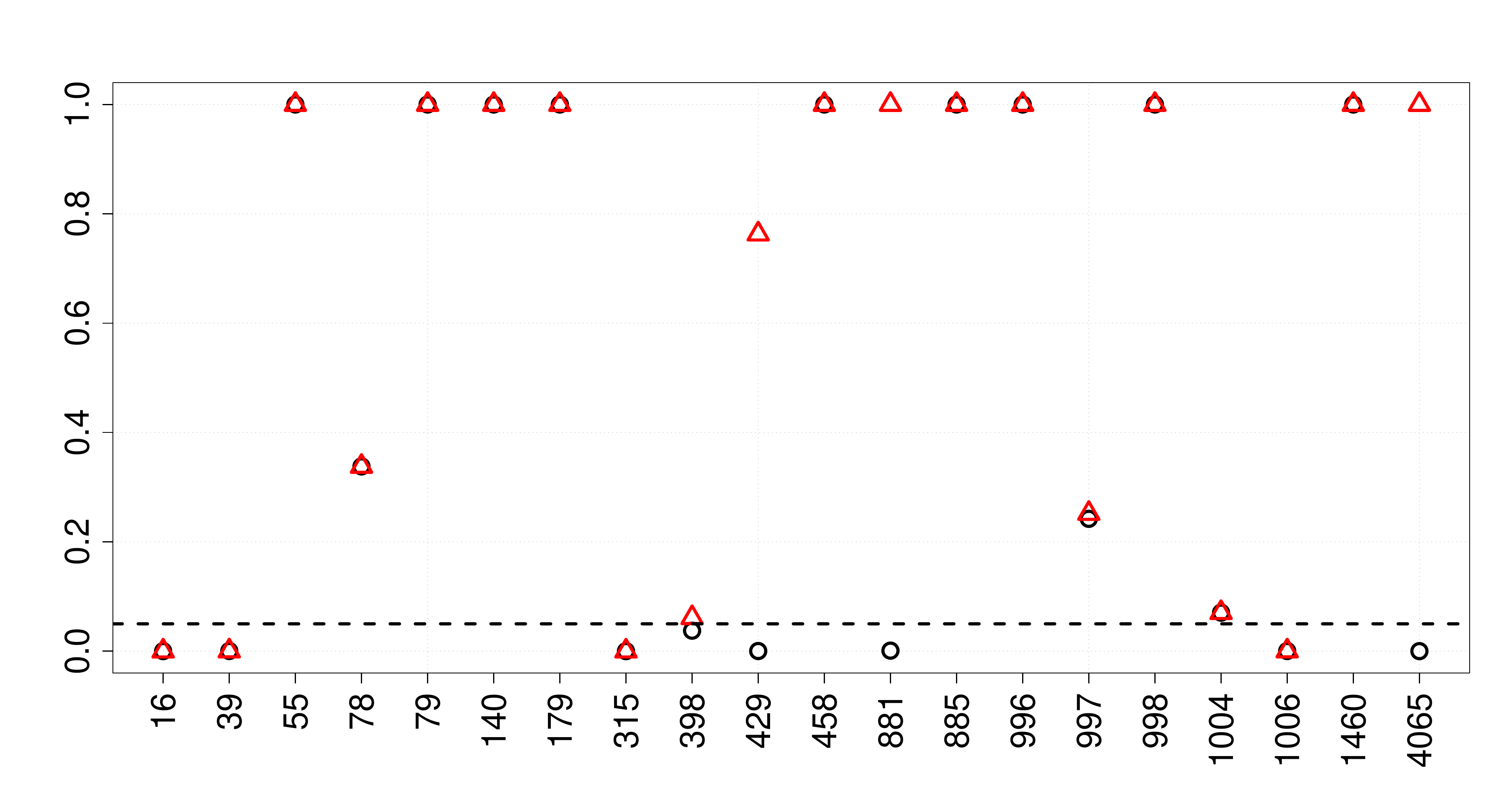}}
\caption{Comparison between adjusted selective $p$-values and nominal $p$-values.
The vertical and horizontal axes represent adjusted $p$-values and indices of selected variables, respectively, and the black dotted line shows the significance level ($\alpha=0.05$).
In each figure, black circles and red triangles respectively indicate adjusted nominal $p$-values and selective $p$-values.
}
\label{fig:real2}
\end{figure*}

\begin{figure*}[h!]
\centering
{\bf GISETTE Data Set} ($n=1,000, d=5,000$)
\vskip -10pt
\subfloat[$K=5$]{\includegraphics[scale=0.165, bb=0 0 1080 576]{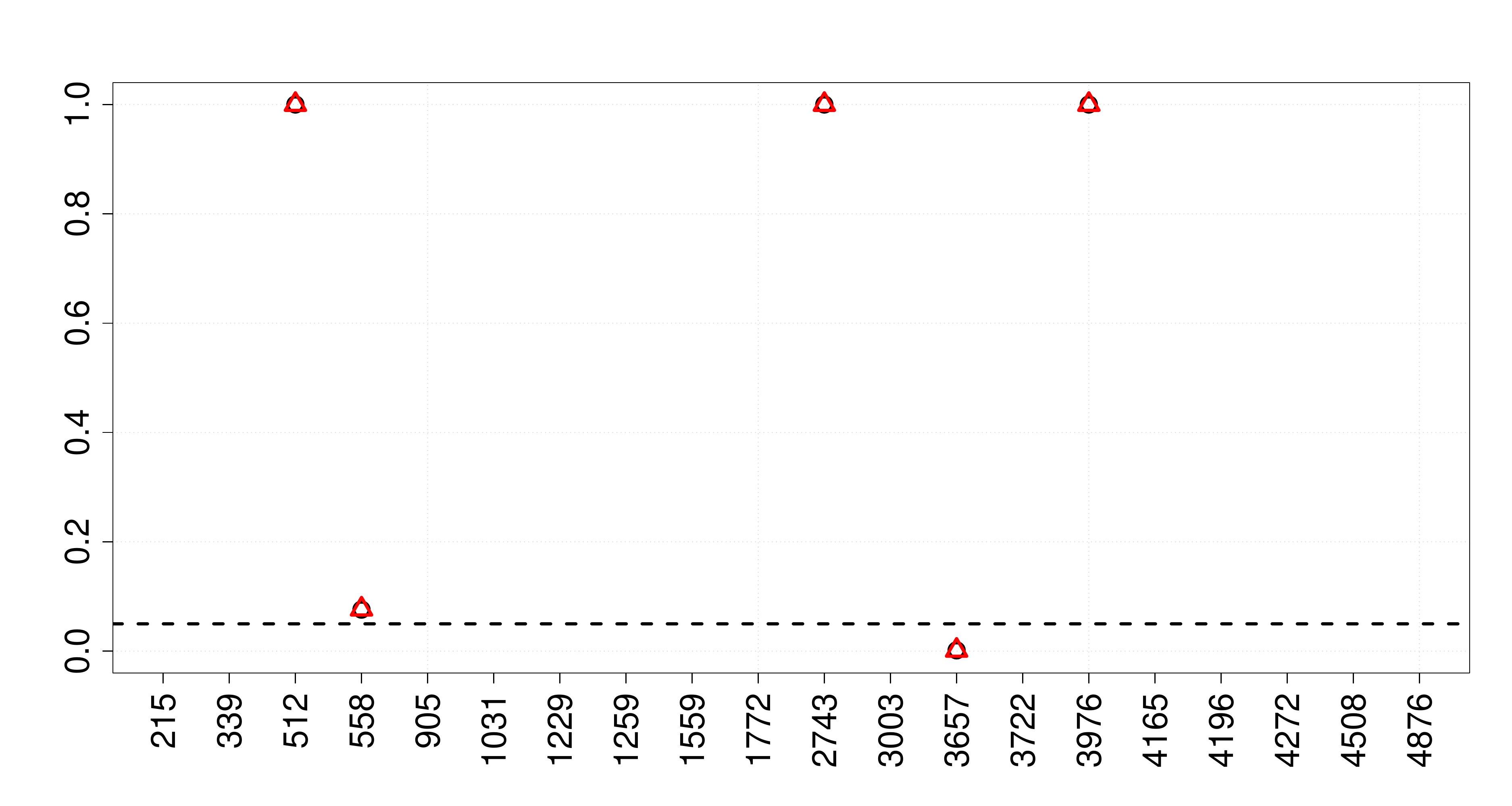}}
\subfloat[$K=10$]{\includegraphics[scale=0.165, bb=0 0 1080 576]{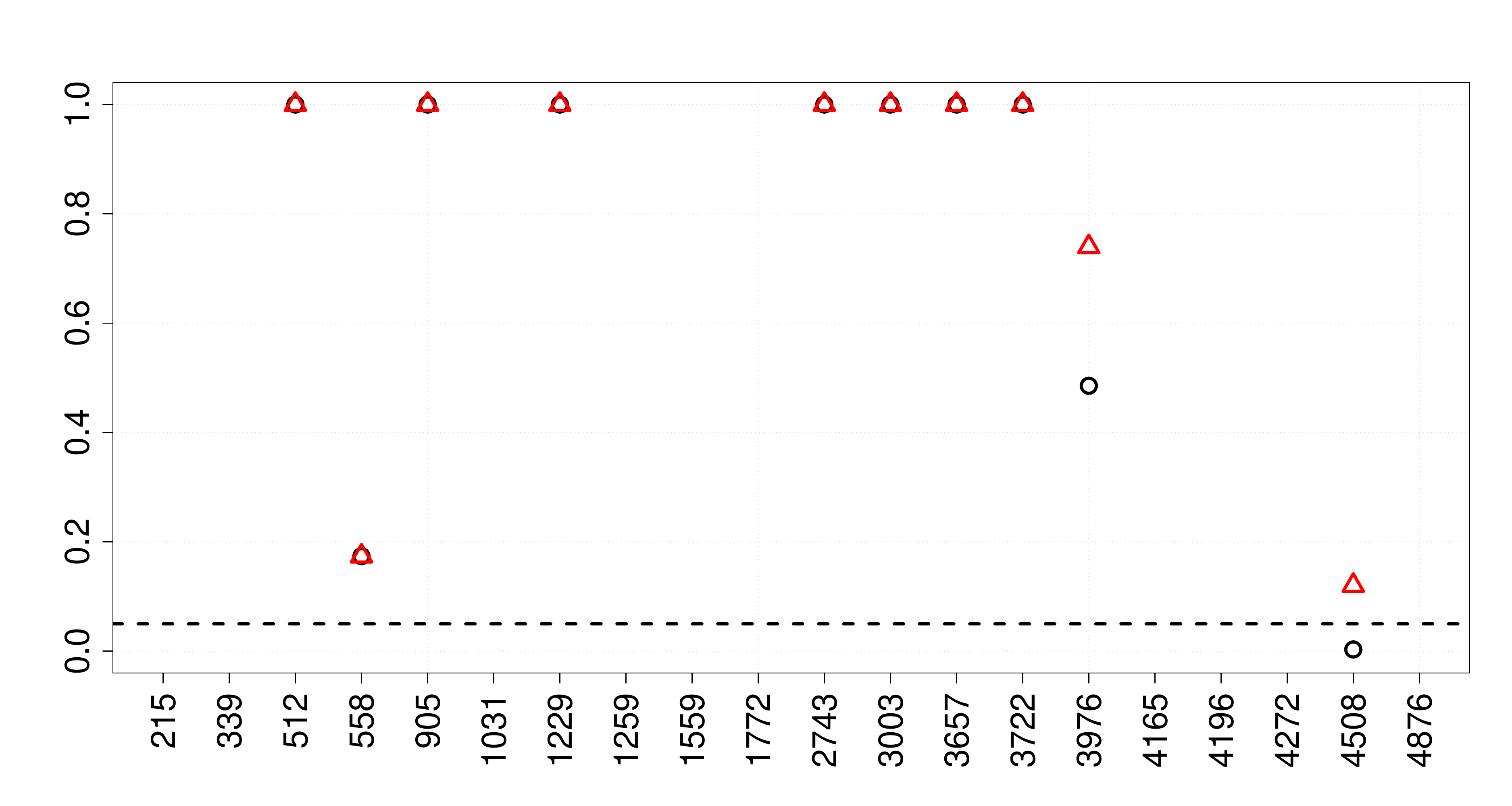}} \\
\subfloat[$K=15$]{\includegraphics[scale=0.165, bb=0 0 1080 576]{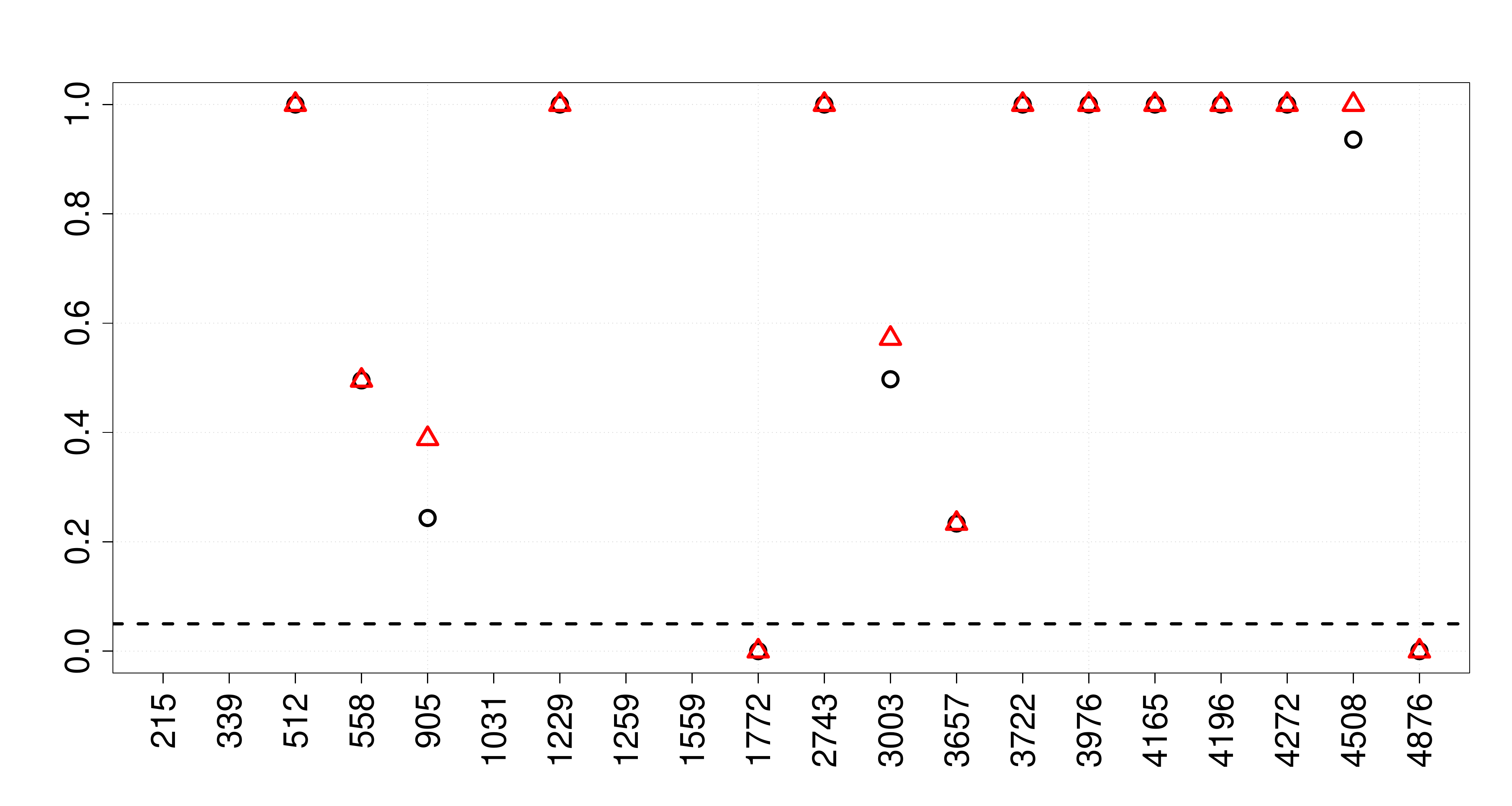}}
\subfloat[$K=20$]{\includegraphics[scale=0.165, bb=0 0 1080 576]{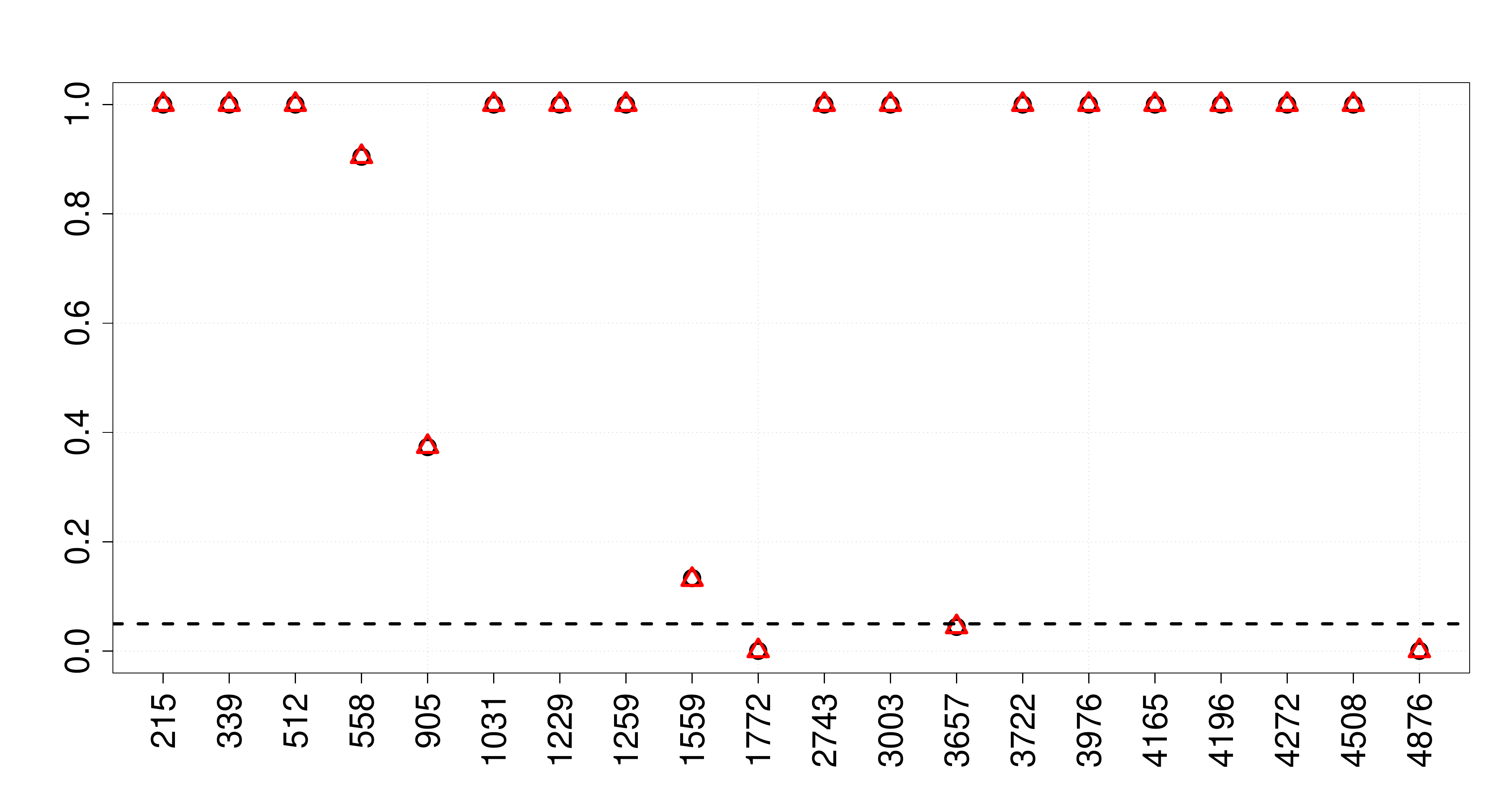}} \\
\vspace{20pt}
{\bf rcv1.Binary Data Set} ($n=20,242, d=47,236$)
\vskip -10pt
\subfloat[$K=5$]{\includegraphics[scale=0.165, bb=0 0 1080 576]{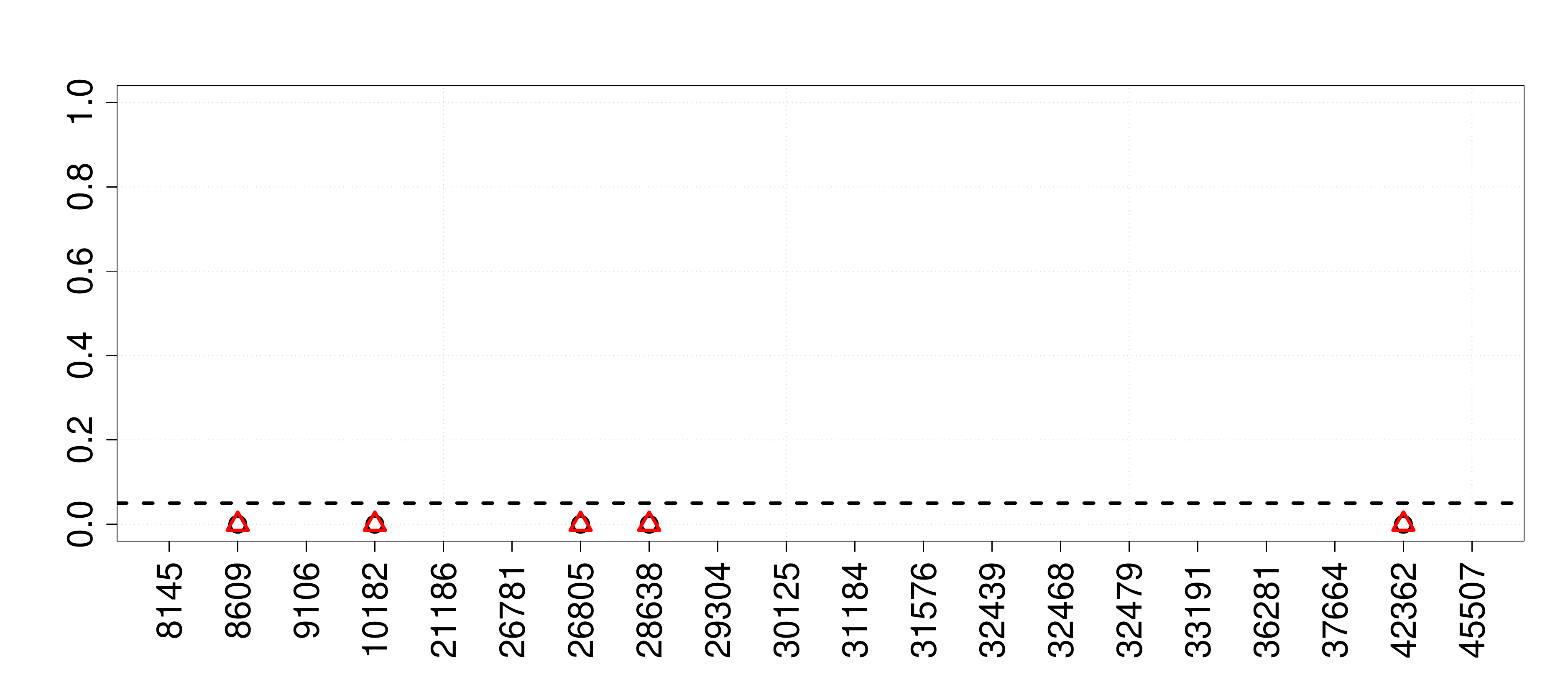}}
\subfloat[$K=10$]{\includegraphics[scale=0.165, bb=0 0 1080 576]{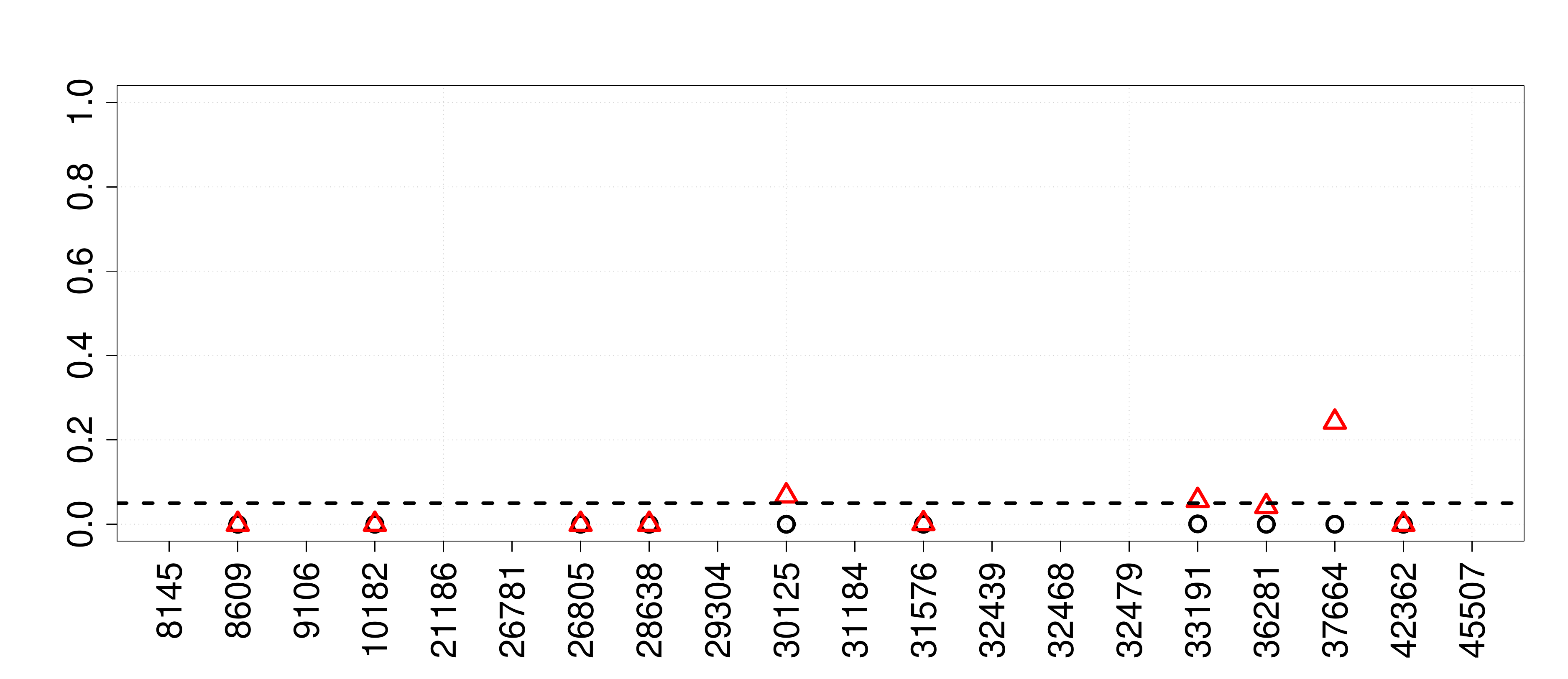}} \\
\subfloat[$K=15$]{\includegraphics[scale=0.165, bb=0 0 1080 576]{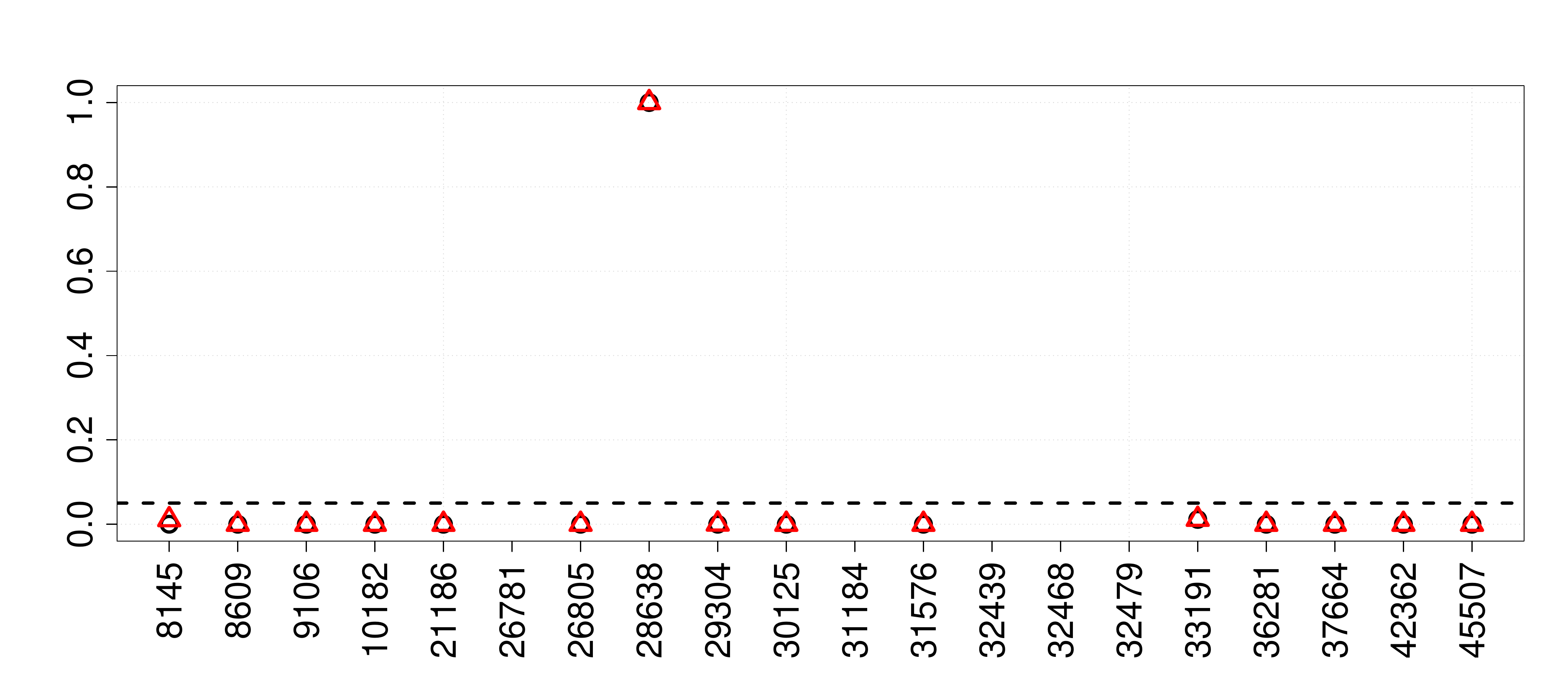}}
\subfloat[$K=20$]{\includegraphics[scale=0.165, bb=0 0 1080 576]{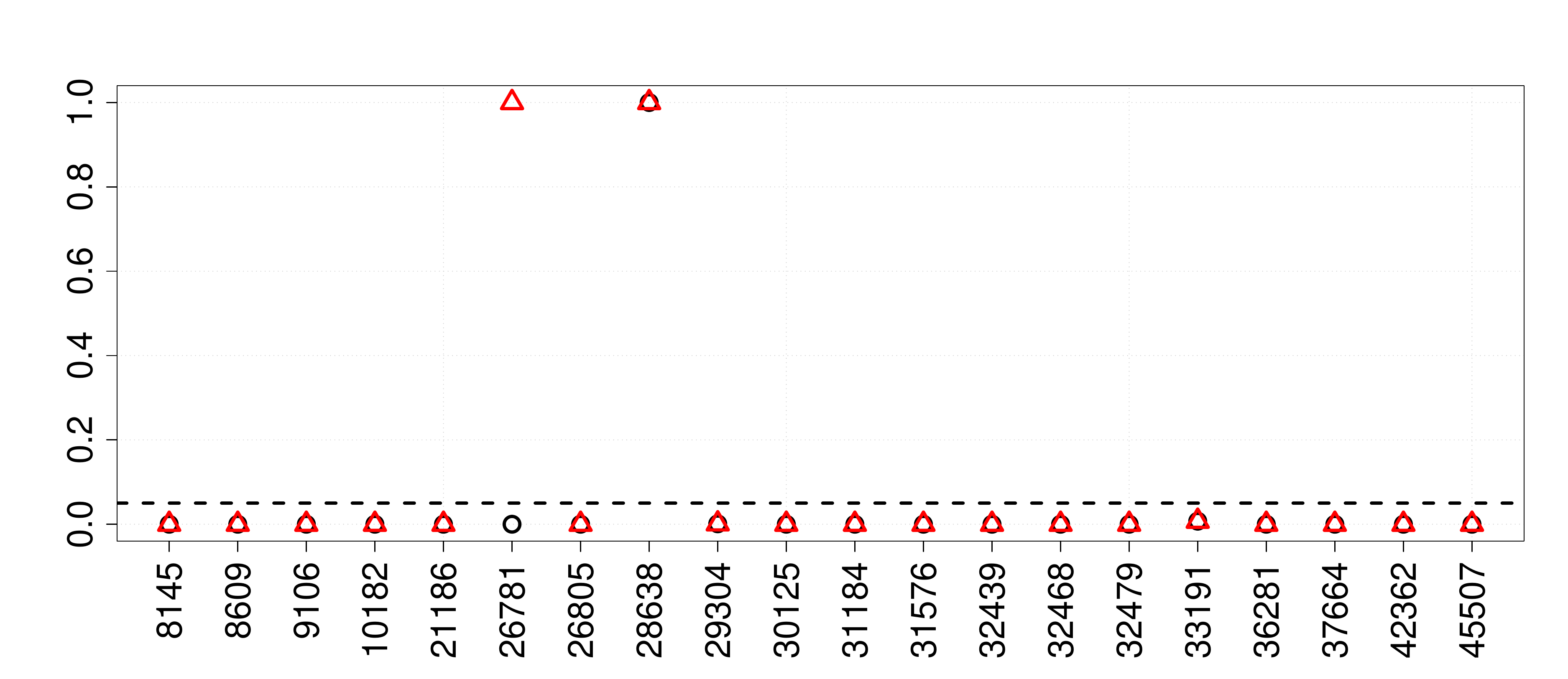}}
\caption{Comparison between adjusted selective $p$-values and nominal $p$-values.
The vertical and horizontal axes represent adjusted $p$-values and indices of selected variables, respectively, and the black dotted line shows the significance level ($\alpha=0.05$).
In each figure, black circles and red triangles respectively indicate adjusted nominal $p$-values and selective $p$-values.
}
\label{fig:real3}
\end{figure*}

We further explore the performance of the proposed method by applying it to several empirical data sets, all of which are available at LIBSVM\footnote{\url{https://www.csie.ntu.edu.tw/~cjlin/libsvmtools/datasets/}}.
In all experiments, we standardize the design matrix $X$ to make the scale of each variable the same.
We report adjusted selective $p$-values for selected variables.
To explore the selection bias, we also report naive adjusted $p$-values.
That is, we first compute $p$-values for selected variables based on NT, then we adjust these $p$-values by multiplying the number of selected variables.
The results are plotted in Figures \ref{fig:real1} -- \ref{fig:real3}.
The result shows that almost all adjusted nominal $p$-values are smaller than those of selective inference, and the difference between these $p$-values is interpreted as the effect of selection bias.

\section{Theoretical Analysis}
\label{sec:proof}
In this section, we provide proofs of the theoretical results derived herein.
We use the notation $p\lesssim q$, which means that, if for any $p,q\in\mathbb{R}$, there exists a constant $r>0$ such that $p\leq rq$, and $p\gtrsim q$ is defined similarly.
All proofs are based on fixed $S~(\supset S^*)$;
thus we simply denote $\hat{\bm{\beta}}_S$ and $X_S$ by $\hat{\bm{\beta}}$ and $X$, respectively.
This is because we need to verify several asymptotic condition before selections in the same way as in \cite{TiaTay15, Tay16}.

\subsection{Proof of (\ref{eq;consistency})}
\label{subsec:consistency}
Let $\alpha_n=\sqrt{K/n}$ and define a $K$ dimensional vector $\bm{u}$ satisfying $\|\bm{u}\|=C$ for a sufficiently large $C>0$.
The concavity of $\ell_n$ implies
\begin{align*}
{\rm P}\left(\|\hat{\bm{\beta}}-\bm{\beta}^*\|\leq \alpha_nC\right)
\geq {\rm P}\Bigl(\sup_{\|\bm{u}\|=C}\ell_n(\bm{\beta}^*+\alpha_n\bm{u})<\ell_n(\bm{\beta}^*)\Bigr),
\end{align*}
and thus we need to show that for any $\varepsilon>0$, there exists a sufficiently large $C>0$ such that
\begin{align}
{\rm P}\left(\sup_{\|\bm{u}\|=C}\ell_n(\bm{\beta}^*+\alpha_n\bm{u})<\ell_n(\bm{\beta}^*)\right)\geq 1-\varepsilon.
\label{eq;claim}
\end{align}
In fact, the  above inequality implies that $\hat{\bm{\beta}}\in\{\bm{\beta}+\alpha_n\bm{u};\;\|\bm{u}\|\leq C\}$, that is, $\|\hat{\bm{\beta}}-\bm{\beta}^*\|={\rm O}_{\rm p}(\alpha_n)$.

Observe that $|\psi'(\bm{x}_i^\top\bm{\beta})|,|\psi''(\bm{x}_i^\top\bm{\beta})|$ and $|\psi'''(\bm{x}_i^\top\bm{\beta})|$ are bounded uniformly with respect to $\bm{\beta}\in{\cal B}$ and $i$.
By using Taylor's theorem, we have
\begin{align*}
&\ell_n(\bm{\beta}^*+\alpha_n\bm{u})-\ell_n(\bm{\beta}^*) \\
&=\sum_{i=1}^n\Bigl[\alpha_ny_i\bm{x}_i^\top\bm{u}-\bigl\{\psi(\bm{x}_i^\top(\bm{\beta}^*+\alpha_n\bm{u}))-\psi(\bm{x}_i^\top\bm{\beta}^*)\bigr\}\Bigr] \\
&=\alpha_n\sum_{i=1}^{n}(y_i-\psi'(\bm{x}_i^\top\bm{\beta}^*))\bm{x}_i^\top\bm{u}
-\frac{\alpha_n^2}{2}\sum_{i=1}^{n}\psi''(\bm{x}_i^\top\bm{\beta}^*)(\bm{x}_i^\top\bm{u})^2
-\frac{\alpha_n^3}{6}\sum_{i=1}^{n}\psi'''(\theta_i)(\bm{x}_i^\top\bm{u})^3 \\
&\equiv I_1+I_2+I_3,
\end{align*}
where for $i=1,2,\ldots,n$, $\theta_i$ is in the line segment between $\bm{x}_i^\top\bm{\beta}^*$ and $\bm{x}_i^\top(\bm{\beta}^*+\alpha_n\bm{u})$.
From (C1) and (C2), we observe that
\begin{align*}
{\rm E}\left[\left\{\sum_{i=1}^{n}(y_i-\psi'(\bm{x}_i^\top\bm{\beta}^*))\bm{x}_i^\top\bm{u}\right\}^2\right] 
&=\sum_{i=1}^{n}{\rm E}\bigl[(y_i-\psi'(\bm{x}_i^\top\bm{\beta}^*))^2(\bm{x}_i^\top\bm{u})^2\bigr] \\
&=\sum_{i=1}^{n}\psi''(\bm{x}_i^\top\bm{\beta}^*)(\bm{x}_{i}^\top\bm{u})^2
\lesssim n\bm{u}^\top\Xi_n\bm{u}
\lesssim n\|\bm{u}\|^2,
\end{align*}
and thus we have $|I_1|={\rm O}_{\rm p}(\alpha_n\sqrt{n}\|\bm{u}\|)={\rm O}_{\rm p}(\sqrt{K}\|\bm{u}\|)$.
Next, by using (C1) again, $I_2$ can be bounded as
\begin{align*}
I_2
\lesssim -\alpha_n^2\sum_{i=1}^{n}(\bm{x}_i^\top\bm{u})^2
\lesssim -K\|\bm{u}\|^2
<0.
\end{align*}
Finally, for $I_3$, we have
\begin{align*}
|I_3|
&=\left|\frac{\alpha_n^3}{6}\sum_{i=1}^{n}\psi'''(\theta_i)(\bm{x}_i^\top\bm{u})^3\right| 
\lesssim \alpha_n^3\sum_{i=1}^{n}|\bm{x}_i^\top\bm{u}|^3
\leq n\alpha_n^3\bm{u}^\top\Xi_n\bm{u}\max_{1\leq i\leq n}|\bm{x}_i^\top\bm{u}| \\
&\lesssim n\alpha_n^3\sqrt{K}\|\bm{u}\|^3
={\rm O}\left(\frac{K^2}{\sqrt{n}}\|\bm{u}\|^3\right).
\end{align*}
Combining all the above, if $K^2/n\to0$ is satisfied, we observe that for sufficiently large $C$, $I_1$ and $I_2$ are dominated by $I_2~(<0)$.
As a result, we obtain (\ref{eq;claim}).
\begin{rem}
From (\ref{eq;consistency}) and (2), we have
\begin{align*}
|\bm{x}_i^\top\hat{\bm{\beta}}|
\leq |\bm{x}_i^\top(\hat{\bm{\beta}}-\bm{\beta}^*)|+|\bm{x}_i^\top\bm{\beta}^*| 
={\rm O}_{{\rm p}}(K/\sqrt{n})+\xi,
\end{align*}
and thus, with probability tending to 1, $\hat{\bm{\beta}}\in{\cal B}$ holds.
\end{rem}

\subsection{Proof of Theorem \ref{thm1}}
\label{subsec:thm1}
First, we prove that $\sqrt{n}(\hat{\bm{\beta}}-\bm{\beta}^*)$ is asymptotically equivalent to $\Sigma_n^{-1}\bm{s}_n$.
By using Taylor's theorem, we have
\begin{align}
\bm{0}
=\ell_n'(\hat{\bm{\beta}})
=\ell_n'(\bm{\beta}^*)+\ell''_n(\bm{\beta}^*)(\hat{\bm{\beta}}-\bm{\beta}^*)
+\frac{1}{2}\sum_{i=1}^n\psi'''(\tilde{\theta}_i)\bm{x}_i\{\bm{x}_i^\top(\hat{\bm{\beta}}-\bm{\beta}^*)\}^2,
\label{eq;taylor}
\end{align} 
where for $i=1,2,\ldots,n$, $\tilde{\theta}_i$ is in the line segment between $\bm{x}_i^\top\bm{\beta}^*$ and $\bm{x}_i^\top\hat{\bm{\beta}}$.
In addition, (\ref{eq;taylor}) can be rewritten as
\begin{align*}
\sqrt{n}(\hat{\bm{\beta}}-\bm{\beta}^*)
=\Sigma_n^{-1}\bm{s}_n+R_n,
\end{align*}
where
\begin{align*}
R_n
=-\frac{1}{2\sqrt{n}}\Sigma_n^{-1}\sum_{i=1}^n\psi'''(\tilde{\theta}_i)\bm{x}_i\{\bm{x}_i^\top(\hat{\bm{\beta}}-\bm{\beta}^*)\}^2.
\end{align*}
Noting that, from (C1),
\begin{align*}
\lambda_{{\rm min}}(\Sigma_n)
\gtrsim \lambda_{{\rm min}}(\Xi_n)
>C_1
>0,
\end{align*}
(C1), (C3) and (\ref{eq;consistency}) imply
\begin{align*}
\|R_n\|
&\lesssim \frac{1}{\sqrt{n}}\max_{1\leq i\leq n}|\bm{x}_i^\top(\hat{\bm{\beta}}-\bm{\beta}^*)|
\times\Bigl\|\sum_{i=1}^{n}\Sigma_n^{-1}\psi'''(\tilde{\theta}_i)\bm{x}_i\bm{x}_i^\top(\hat{\bm{\beta}}-\bm{\beta}^*)\Bigr\| \\
&\lesssim \frac{1}{\sqrt{n}}\max_{1\leq i\leq n}\|\bm{x}_i\|\|\hat{\bm{\beta}}-\bm{\beta}^*\|\times n\lambda_{{\rm max}}(\Xi_n)\|\hat{\bm{\beta}}-\bm{\beta}^*\| \\
&={\rm O}_{{\rm p}}\Bigl(\frac{K\sqrt{K}}{\sqrt{n}} \Bigr)
={\rm o}_{{\rm p}}(1).
\end{align*}

Now we can prove the asymptotic normality of $\sigma_n^{-1}\Sigma_n^{-1}\bm{s}_n$.
For any $K$ dimensional vector $\bm{\eta}$ with $\|\bm{\eta}\|<\infty$, define $\sigma_n^2=\bm{\eta}^\top\Sigma_n^{-1}\bm{\eta}$ and $\omega_n$ such that
\begin{align*}
\bm{\eta}^\top\Sigma_n^{-1}\bm{s}_n
=\frac{1}{\sqrt{n}}\sum_{i=1}^{n}\bm{\eta}^\top\Sigma_n^{-1}\bm{x}_i(y_i-\psi'(\bm{x}_i^\top\bm{\beta}^*))
=\sum_{i=1}^{n}\omega_{ni}.
\end{align*}
Then, since $S\supset S^*$, we observe that
\begin{align*}
\sum_{i=1}^{n}{\rm E}[\omega_{ni}]
 =\sum_{i=1}^{n}\frac{1}{\sqrt{n}}\bm{\eta}^\top\Sigma_n^{-1}\bm{x}_i{\rm E}[y_i-\psi'_i]
 =0,
\end{align*}
and
\begin{align*}
\sum_{i=1}^{n}{\rm V}[\omega_{ni}]
=\frac{1}{n}\sum_{i=1}^n\bm{\eta}^\top\Sigma_n^{-1}\bm{x}_i{\rm V}[y_i]\bm{x}_i^\top\Sigma_n^{-1}\bm{\eta}
=\sigma_n^2.
\end{align*}
To state the asymptotic normality of $\sigma_n^{-1}\Sigma_n^{-1}\bm{s}_n$, we check the Lindeberg condition for $\omega_n$: for any $\varepsilon>0$,
\begin{align}
\label{eq;Lindeberg}
\frac{1}{\sigma_n^2}\sum_{i=1}^{n}{\rm E}[\omega_{ni}^2I(|\omega_{ni}|>\sigma_n\varepsilon)]
={\rm o}(1).
\end{align}

For any $\varepsilon>0$, we have
\begin{align*}
&\frac{1}{\sigma_n^2}\sum_{i=1}^{n}{\rm E}[\omega_{ni}^2I(|\omega_{ni}|>\sigma_n\varepsilon)] \\
&=\frac{1}{\sigma_n^2}\cdot\frac{1}{n}\sum_{i=1}^{n}(\bm{\eta}^\top\Sigma_n^{-1}\bm{x}_i)^2{\rm E}[(y_i-\psi'_i)^2I(|\omega_{ni}|>\sigma_n\varepsilon)] \\
&\leq \frac{1}{\sigma_n^2}\max_{1\leq i\leq n}{\rm E}[(y_i-\psi'_i)^2I(|\omega_{ni}|>\sigma_n\varepsilon)]
\times\frac{1}{n}\sum_{i=1}^{n}(\bm{\eta}^\top\Sigma_n^{-1}\bm{x}_i)^2.
\end{align*}
By using the Cauchy-Schwarz inequality and (C1), 
\begin{align*}
\frac{1}{n}\sum_{i=1}^{n}(\bm{\eta}^\top\Sigma_n^{-1}\bm{x}_i)^2
\leq \frac{1}{n}\sum_{i=1}^{n}(\bm{\eta}^\top\Sigma_n^{-1}\bm{\eta})(\bm{x}_i^\top\Sigma_n^{-1}\bm{x}_i)
\lesssim\frac{1}{n}\sum_{i=1}^{n}\|\bm{x}_i\|^2
={\rm O}(K).
\end{align*}

Noting that each $y_i$ is distributed according to a Bernoulli distribution with parameter $\psi'$, ${\rm E}[(y_i-\psi'_i)^4]$ is uniformly bounded on ${\cal B}$ for any $i=1,\ldots,n$ by a simple calculation.
Thus, by using the Cauchy-Schwarz inequality and Chebyshev's inequality, we have
\begin{align*}
\max_{1\leq i\leq n}{\rm E}[(y_i-\psi'_i)^2I(|\omega_{ni}|>\sigma_n\varepsilon)]
&\leq \max_{1\leq i\leq n}{\rm E}[(y_i-\psi_i)^4]^{1/2}{\rm P}(|\omega_{ni}|>\sigma_n\varepsilon)^{1/2} \\
&\lesssim \frac{1}{\sigma_n}\max_{1\leq i\leq n}{\rm E}[\omega_{ni}^2]^{1/2} \\
&=\frac{1}{\sigma_n\sqrt{n}}\max_{1\leq i\leq n}|\bm{\eta}^\top\Sigma_n^{-1}\bm{x}_i|\sqrt{\psi'_i(1-\psi'_i)} \\
&\lesssim \frac{1}{\sqrt{n}}\max_{1\leq i\leq n}\|\bm{x}_i\|
={\rm O}\left(\frac{\sqrt{K}}{\sqrt{n}}\right).
\end{align*}
Finally, since
\begin{align*}
\sigma_n^2
=\bm{\eta}^\top\Sigma_n^{-1}\bm{\eta} 
\leq \lambda_{{\rm max}}(\Sigma_n^{-1})\|\bm{\eta}\|^2
=\frac{\|\bm{\eta}\|^2}{\lambda_{{\rm min}}(\Sigma_n)}
={\rm O}(1),
\end{align*}
we have
\begin{align*}
\frac{1}{\sigma_n^2}\sum_{i=1}^{n}{\rm E}[\omega_{ni}^2I(|\omega_{ni}|>\sigma_n\varepsilon)]
={\rm O}\left(\frac{\sqrt{K}}{\sqrt{n}}\cdot K \right).
\end{align*}
From (C3), this implies the Lindeberg condition (\ref{eq;Lindeberg}).

\subsection{Proof of Theorem \ref{thm2}}
\label{subsec:thm2}
First, we prove that, for any $K$ dimensional vector $\bm{\eta}$, the selection event can be expressed as an inequality with respect to $\bm{\eta}^\top\bm{T}_n$.
Let us define $\bm{w}=(I_K-\bm{c}\bm{\eta}^\top)\bm{T}_n$, where $\bm{c}=\Sigma_n^{-1}\bm{\eta}/\sigma_n^2$.
Then, since $\bm{T}_n=(\bm{\eta}^\top\bm{T}_n)\bm{c}+\bm{w}$, we have
\begin{align*}
\tilde{A}\bm{T}_n\leq \tilde{\bm{b}}
&\Leftrightarrow (\bm{\eta}^\top\bm{T}_n)\tilde{A}\bm{c}\leq \tilde{\bm{b}}-\tilde{A}\bm{w} \\
&\Leftrightarrow (\bm{\eta}^\top\bm{T}_n)(\tilde{A}\bm{c})_j\leq (\tilde{\bm{b}}-\tilde{A}\bm{w})_j,~\forall j \\
&\Leftrightarrow
\begin{cases}
\bm{\eta}^\top\bm{T}_n\leq (\tilde{\bm{b}}-\tilde{A}\bm{w})_j/(\tilde{A}\bm{c})_j,& j:(\tilde{A}\bm{c})_j>0 \\
\bm{\eta}^\top\bm{T}_n\geq (\tilde{\bm{b}}-\tilde{A}\bm{w})_j/(\tilde{A}\bm{c})_j,& j:(\tilde{A}\bm{c})_j<0 \\
0=(\tilde{\bm{b}}-\tilde{A}\bm{w})_j,& j:(\tilde{A}\bm{c})_j=0 \\
\end{cases}
\end{align*}
and this implies the former result in Theorem \ref{thm2}.

To prove the theorem, we need to verify asymptotic independency between $(L_n, U_n, N_n)$ and $\bm{\eta}^\top\bm{T}_n$.
By the definition of $\bm{w}$ and Theorem~\ref{thm1}, 
\begin{align*}
\left( \begin{array}{c}
\bm{\eta}^\top\bm{T}_n \\
\bm{w}
\end{array} \right)
=
\left( \begin{array}{c}
\bm{\eta}^\top \\
I_K-\bm{c}\bm{\eta}^\top
\end{array} \right)\bm{T}_n
\end{align*}
is asymptotically distributed according to a Gaussian distribution.
Thus, $\bm{w}$ and $\bm{\eta}^\top\bm{T}_n$ are asymptotically independent since
\begin{align*}
{\rm Cov}[\bm{w},\bm{\eta}^\top\bm{T}_n]
=(I_K-\bm{c}\bm{\eta}^\top){\rm E}[\bm{T}_n\bm{T}_n^\top]\bm{\eta}
=(I_K-\bm{c}\bm{\eta}^\top)\Sigma_n^{-1}\bm{\eta}
=\bm{0}.
\end{align*}
Now we only need to prove asymptotic independency between $\tilde{\bm{b}}$ and $\bm{\eta}^\top\bm{T}_n$.
Letting $\bm{\psi}'=\bm{\psi}'(\bm{\beta}^*)$, the definition of $\bm{T}_n$ and $\Sigma_n$ imply
\begin{align*}
X_S^\top\Bigl\{ (\bm{y}-\bm{\psi}')-\frac{1}{\sqrt{n}}\Psi X_S\bm{T}_n\Bigr\}
=\bm{0}, 
\end{align*}
and thus
\begin{align*}
\bm{y}=\bm{\psi}'+\frac{1}{\sqrt{n}}\Psi X_S\bm{T}_n
\end{align*}
Then, we observe that
\begin{align*}
\tilde{\bm{b}}
&=-\frac{1}{\sqrt{n}}A_SX_S^\top\bm{\psi}'-\frac{1}{\sqrt{n}}A_{S^\bot}X_{S^\bot}^\top\bm{y} \\
&=-\frac{1}{\sqrt{n}}AX^\top\bm{\psi}'-\frac{1}{\sqrt{n}}A_{S^\bot}X_{S^\bot}^\top\left(\bm{\psi}'+\frac{1}{\sqrt{n}}\Psi X_S\bm{T}_n\right) \\
&=-\frac{1}{\sqrt{n}}AX^\top\bm{\psi}'-\frac{1}{n}A_{S^\bot}X_{S^\bot}^\top\Psi X_S\bm{T}_n.
\end{align*}
Since $\tilde{\bm{b}}$ can be expressed as a linear combination of $\bm{T}_n$ as well as $\bm{w}$, the theorem holds when the covariance between $\tilde{\bm{b}}$ and $\bm{\eta}^\top\bm{T}_n$ converges to 0 as $n$ goes to infinity.
By noting that $\Sigma_n=X_S^\top\Psi X_S/n$, we have
\begin{align*}
{\rm Cov}[\tilde{\bm{b}},\bm{\eta}^\top\bm{T}_n]
&=-\frac{1}{n}A_{S^\bot}X_{S^\bot}^\top\Psi X_S{\rm E}[\bm{T}_n\bm{T}_n^\top]\bm{\eta} \\
&=-A_{S^\bot}(X_{S^\bot}^\top\Psi X_S)(X_S^\top\Psi X_S)^{-1}\bm{\eta}.
\end{align*}
In addition, letting $\bm{a}=(1,-1)^\top$, it is straightforward that
\begin{align*}
A_{S^\bot}
=\bm{1}_{K}\otimes
\left( \begin{array}{ccc}
0&\cdots&0 \\
\bm{a}&&O \\
&\ddots& \\
O&&\bm{a}
\end{array} \right)
=\bm{1}_{K}\otimes \tilde{J}
\end{align*}
by the definition of the selection event, where $\tilde{J}=(\bm{0}_{d-K}, I_{d-K}\otimes\bm{a}^\top)^\top$.
This implies $A_{S^\bot}^\top A_{S^\bot}=2KI_{d-K}$.
Finally, (C1), (C3), and (C4) imply
\begin{align*}
\|{\rm Cov}[\tilde{\bm{b}},\bm{\eta}^\top\bm{T}_n]\|^2
=2K\|(X_{S^\bot}^\top\Psi X_S)(X_S^\top\Psi X_S)^{-1}\bm{\eta}\|^2 
\lesssim K\Bigl\|\frac{1}{n}X_{S^\bot}^\top X_S\Bigr\|^2 
={\rm O}(K^3/n),
\end{align*}
and this proves the asymptotic independency between $\tilde{\bm{b}}$ and $\bm{\eta}^\top\bm{T}_n$.

\section{Concluding Remarks and Future Research}
\label{sec:conclusion}
Recently, methods for data driven science such as selective inference and adaptive data analysis have become increasingly important as described by \cite{Bar16}.
Although there are several approaches for carrying out post-selection inference, we have developed a selective inference method for high dimensional classification problems, based on the work in \cite{Lee16}.
In the same way as that seminal work, the polyhedral lemma (Lemma \ref{lem1}) plays an important role in our study.
By considering high dimensional asymptotics concerning sample size and the number of selected variables, we have shown that a similar result to the polyhedral lemma holds even for high dimensional logistic regression problems.
As a result, we could construct a pivotal quantity whose sampling distribution is represented as a truncated normal distribution which converges to a standard uniform distribution.
In addition, through simulation experiments, it has been shown that the performance of our proposed method, in almost all cases, superior to other methods such as data splitting.

As suggested by the results from the simulation experiments, conditions might be relaxed to accommodate more general settings.
In terms of future research in this domain, while we considered the logistic model in this paper, it is important to extend the results to other models, for example, generalized linear models.
Further, higher order interaction models are also crucial in practice.
In this situation, the size of the matrix in the selection event becomes very large, and thus it is cumbersome to compute truncation points in the polyhedral lemma.
\cite{suzumura2017selective} have shown that selective inference can be constructed in such a model by utilizing a pruning algorithm.
In this respect, it is desirable to extend their result not only to linear regression modeling contexts but also to other models.

\bibliographystyle{econ}
\bibliography{myref}

\end{document}